\documentclass[twocolumn,aps,prd,showpacs,nofootinbib,superscriptaddress,preprintnumbers,floatfix,10pt]{revtex4-1}

\usepackage{amsmath}
\usepackage{amsfonts}
\usepackage{amssymb}
\usepackage{epsfig}
\usepackage{graphicx}
\usepackage{color}
\usepackage{slashed}
\usepackage{epstopdf}
\usepackage[utf8]{inputenc}
\usepackage[markup=nocolor]{changes}

\usepackage{hyperref}

\newcommand{\be}{\begin{equation}}
\newcommand{\ee}{\end{equation}}
\newcommand{\bea}{\begin{eqnarray}}
\newcommand{\eea}{\end{eqnarray}}

\begin{document}

\preprint{}

\title {Universal behavior of coupled order parameters below three dimensions} 

\author{Julia Borchardt}
\email{julia.borchardt@uni-jena.de}
\affiliation{Theoretisch-Physikalisches Institut, %
Friedrich-Schiller-Universit\"at Jena, Max-Wien-Platz 1, D-07743 Jena, Germany}
\author{Astrid Eichhorn}
\email[]{a.eichhorn@imperial.ac.uk} 
\affiliation{Blackett Laboratory, Imperial College, London SW7 2AZ, United Kingdom}
\preprint{Imperial/TP/2016/AE/03}

\begin{abstract}
 We explore universal critical behavior in models with two competing order parameters, and an O($N$) $\oplus$ O($M$) symmetry 
 for dimensions $d\leq 3$. In $d=3$, there is always exactly one stable Renormalization Group fixed point, corresponding to bicritical or tetracritical behavior.
Employing novel, pseudo-spectral techniques to solve functional Renormalization Group equations in a two-dimensional field space, we uncover a more intricate structure of fixed points in $d<3$, where two additional  bicritical fixed points play a role.
Towards $d=2$, we discover ranges of $N=M$ with several simultaneously stable fixed points, indicating the coexistence of several universality classes. 
\end{abstract}


\maketitle

\section{Introduction}
\label{sec:intro}

Two competing order parameters, parameterized by an O($N$) and O($M$) symmetry, respectively, feature multicritical points in their phase diagram \cite{Fisher:1974zz,Kosterlitz:1976zza, FisherLiu}: As a function of two external parameters, the separate symmetries are spontaneously broken across second-order phase transition lines. These lines meet at a multicritical point.  
If it is bicritical, there are three phases adjacent to it - the two broken phases and a phase of unbroken symmetry. 
For a tetracritical point, an additional mixed phase with two spontaneously broken symmetries exists. The universality class of the multicritical point is encoded in an infrared (IR) Renormalization Group (RG) fixed point of the O($N$) $\oplus$ O($M$) model \cite{PhysRevLett.33.813}.

Examples of systems with two competing order parameters include, e.g., anisotropic antiferromagnets in an external magnetic field, with $N=1$, $M=2$ \cite{Rohrer1975, Rohrer1977,KingRohrer,Oliveira, Butera,Ohgushi,beccera88,basten80}. Models of high-$\rm T_c$ superconductors could also fall into this category \cite{Zhang1997}, as well as certain properties of graphene \cite{Roy:2011pg,Classen:2015ssa,Classen:2015mar}. These systems have been explored theoretically in great detail in $d=3$ dimensions \cite{Aharony:2002,Aharony:2002zz,Aharony:1976book,Aharony:1973zz,Calabrese:2002bm,Calabrese:2002bq,Vicari:2007ma,Folk2008,Eichhorn:2013zza,Eichhorn:2014asa,Boettcher:2015pja,Eichhorn:2015woa}. Two dimensions are a special case for continuous phase transitions. O$(N)$ models with $N > 2$ cannot exhibit universal critical behavior due to the Mermin Wagner theorem \cite{PhysRevLett.17.1133}.  $N=2$ features
a Kosterlitz-Thouless  (BKT) phase transition with algebraic order \cite{Kosterlitz:1973xp}. Experimentally, this was observed, e.g., in liquid-helium films \cite{Bishop:1978zz,PhysRevLett.47.1533} and atomic gases \cite{Nature05851,PhysRevLett.105.230408,2012NatPh...8..645D,PhysRevLett.115.010401}.
Hence, the universal critical behavior in $d=2$ in the vicinity of the multicritical point for $N=M=1$, corresponding to two coupling Ising models
could be determined by the Onsager solution for the simple Ising model. In this case, the pertinent RG fixed point encoding the universal behavior would be the decoupled fixed point (DFP), at which all mixed interactions between the O($N$) and the O($M$) sector vanish. On the other hand, multicritical models also feature fixed points exhibiting an enhanced O($N+M$) symmetry. For two coupled Ising models, this would suggest BKT type physics as $d$ approaches 2.
Besides, the model also features an additional, biconical fixed point which could provide another candidate for a universality class in $d=2$ for $N=M=1$.
Finally, additional fixed points may exist below $d=3$, which could become relevant for the physics of two coupled Ising models in $d=2$.

To investigate the physics of coupled order parameters in $d=2$, we employ the functional RG. It has previously been shown to give reliable results for the physics of coupled order parameters in $d=3$ \cite{Eichhorn:2013zza,Eichhorn:2014asa,Boettcher:2015pja,Eichhorn:2015woa}, and for O($N$) models in $2\leq d \leq 3$ dimensions, see, e.g., \cite{Morris:1994jc,Canet:2003qd,Bervillier:2007rc,Litim:2010tt,Codello:2012ec,Codello:2012sc,Delamotte:2015aaa}, including the BKT phase transition.
It can be efficiently captured with the help of the functional RG, formulated in terms of the O(2) scalar field, without introducing any explicit notion of vortex-antivortex pairs \cite{Grater:1994qx,VonGersdorff:2000kp,Jakubczyk:2014isa,Jakubczyk:2016sul}.

On the technical side, we advance the recently introduced pseudospectral tool \cite{Borchardt:2015rxa,Borchardt:2016pif} to solve functional RG flows. We show that this method is applicable to systems with more than one field, where global information on the potential is required, including
examples from high-energy physics, such as, e.g., stability-properties of the Higgs potential \cite{Gies:2013fua,Gies:2014xha,Eichhorn:2015kea,Borchardt:2016xju} to cases with additional fields \cite{Eichhorn:2014qka}.

Our main results lie in the derivation of a ``phase diagram" for the system, which shows which of the various O($N$) $\oplus$ O($M$) fixed points is the stable one as a function of $N=M$ and $d$. In particular, we
discover within truncated RG flows that several simultaneously stable fixed points underlie different possibilities for the universal critical behavior of the system. We find that besides the long-range degrees of freedom, symmetries and dimensionality of the system, additional information 
is required to determine which of the possible universality classes is realized in the IR.

\section{Competing orders in $d \leq 3$ dimensions}
The model is given by an O($N$) vector $\phi^I$, $I=1,...,N$ and an $O(M)$ vector $\chi^J$, $J=1,...,M$
describing order parameters of the separate O($N$) and O($M$) symmetry: If $\phi^I$ assumes a nonvanishing vacuum expectation value, the O($N$) symmetry is broken spontaneously to an O($N-1$) symmetry, and similarly for $\chi$.
The effective potential for the model
depends on
the invariants
\be
\bar{\rho}_{\phi} = \frac{\sum_{I=1}^N\phi^I \phi^I}{2}, \quad \bar{\rho}_{\chi} = \frac{\sum_{J=1}^M \chi^J \chi^J}{2},
\ee
and can be expanded as
\be
U(\bar{\rho}_{\phi}, \bar{\rho}_{\chi}) = \sum_{i,j} \frac{\bar{\lambda}_{i,j}}{i! j!} \left(\bar{\rho}_{\phi}- \bar{\kappa}_{\phi}\right)^i\left( \bar{\rho}_{\chi}- \bar{\kappa}_{\chi}\right)^j.\label{eq:dimfullpot}
\ee

For the symmetry-broken regime, it is advantageous to choose the expansion points $\bar{\kappa}_{\phi/\chi}$ as the non trivial minima. 
In the main part of this work, we consider the full potential as a general functional of the two fields.

Universal critical behavior is encoded in potentials which are fixed points of the RG flow. In $d=3$, these can be studied within the $\epsilon$ expansion around $d=4$ dimensions, however, this becomes challenging already for the O($N$) model in $d=2$, see, e.g., \cite{Orlov:2000wn}, as well as for two coupled order parameters \cite{2007PhRvB..76b4436P}. Thus, we focus on the nonperturbative, functional RG approach, which allow us to evaluate the RG flow with respect to the momentum-scale $k$ in general dimensions $d$.

\section{Fixed-point properties}
\label{sec:fixedpointstructure}

\subsection{Stability of fixed points and nature of the multicritical point}

This section provides an overview of the fixed-point content of the model for dimensions $2\leq d\leq 3$.
We are interested in the \emph{stable} fixed point, which features not more than two relevant directions\footnote{Note that stability of the fixed-point potential is an unrelated question. As fixed-point potentials which are not stable, i.e., unbounded from below, do not define viable quantum field theories, we do not consider such solutions here, and exclusively focus on potentials which are bounded from below.}. 
These correspond to parameters that require tuning in a given experimental situation, in order to observe the universal scaling behavior associated to the fixed-point solution.  
Typically, there is one tunable parameter for each of the order parameters, e.g., the temperature and the magnetic field for an anisotropic antiferromagnet.

The critical exponents are derived from the beta functions, for details see Sec.~\ref{sec:FRG},
\be
\beta_{\lambda_{i,j}} := k\, \partial_k\, \lambda_{i,j} (k),
\ee
of the dimensionless couplings
\be
\lambda_{i,j} = k^{-d_{\bar{\lambda}_{i,j}}} \bar{\lambda}_{i,j},\label{eq:dimlesscouplings}
\ee
where $k$ is the RG scale and $[\bar{\lambda}_{i,j}] = d_{\bar{\lambda}_{i,j}}$ the canonical dimensionality of the couplings $\bar{\lambda}_{i,j}$. 
The critical exponents are 
minus the eigenvalues of the stability matrix
\be
\theta= - {\rm eig} \,\mathcal{M}_{n,p}.
\label{eq:critExp}
\ee
The stability matrix is defined as follows
\be
\mathcal{M}_{n,p} = - \frac{\partial \beta_{\lambda_n}}{\partial \lambda_{p}}\Big|_{\lambda_{i,j} = \lambda_{i,j}^{\ast}},
\ee
where $ \lambda_{i,j}^{\ast}$ are the fixed-point values. Here, we have summarized the two indices $i,j$ in one, such that $\lambda_p$, $p=1,...$ runs through all couplings $\lambda_{i,j}$.
A positive critical exponent is associated to a relevant direction. As each of the two fields $\phi^I$ and $\chi^J$ comes with a mass-like parameter, we always have $\theta_{1,2}>0$. Thus the third critical exponent, $\theta_3$, determines the stability of a fixed point.

An important property of the fixed point is encoded in the field-dependent parameter $\Delta$, which is related to the determinant of the matrix consisting of the second derivatives of the potential, \cite{Eichhorn:2013zza}. 
The general definition is given by
\be
\Delta = u^{(2,0)} u^{(0,2)}- \left(u^{(1,1)}\right)^2,
\ee
where we have introduced the dimensionless potential 
\be
u = k^{-d} U\left(\bar{\rho}_{\phi}(\rho_{\phi}), \bar{\rho}_{\chi}(\rho_{\chi})\right)
\ee
and the dimensionless, renormalized field variables
\be
\rho_{\phi} = Z_{\phi}\frac{\sum_I\phi^I \phi^I}{2 \,k^{d-2}}, \quad \rho_{\chi} = Z_{\chi}\frac{\sum_J\chi^J \chi^J}{2 \,k^{d-2}}
\ee
with the wave function renormalizations $Z_\phi$ and $Z_\chi$.
We denote derivatives by the shorthand
\be
\frac{\delta^{n_1}}{\delta\rho_{\phi}^{n_1}}\frac{\delta^{n_2}}{\rho_{\chi}^{n_2}}u[\rho_{\phi},\rho_{\chi}] = u^{(n_1,n_2)}[\rho_{\phi}, \rho_{\chi}].\label{eq:notation2}
\ee
If we choose an expansion point for the effective potential, as in \eqref{eq:dimfullpot}, and evaluate $\Delta$ at that point
\be
\Delta= \lambda_{2,0} \lambda_{0,2} - \lambda^2_{1,1}.\label{eq:Delta}
\ee
If the first derivatives of $u$ vanish at the expansion point, the expansion point corresponds to a saddle point for $\Delta \leq 0$, whereas it corresponds to a minimum for $\Delta>0$.
Thus, for $\Delta>0$, the minimum of the fixed-point potential lies at nonvanishing expectation values for both order-parameter fields. The fixed point is called tetracritical, as a mean-field analysis relates it to a multicritical point which is bordered by a mixed phase \cite{FisherLiu}. 
On the other hand, for $\Delta<0$, there is no mixed phase, and the minimum of the bicritical fixed-point potential lies on one of the axes.
As the sign of $\Delta$ for the fixed-point solutions we consider does not depend on whether it is evaluated at the extremum/saddle point or the origin in field space,
we typically extract $\Delta$ at the origin in field-space. 

\begin{figure}[!t]
\includegraphics[width=\linewidth]{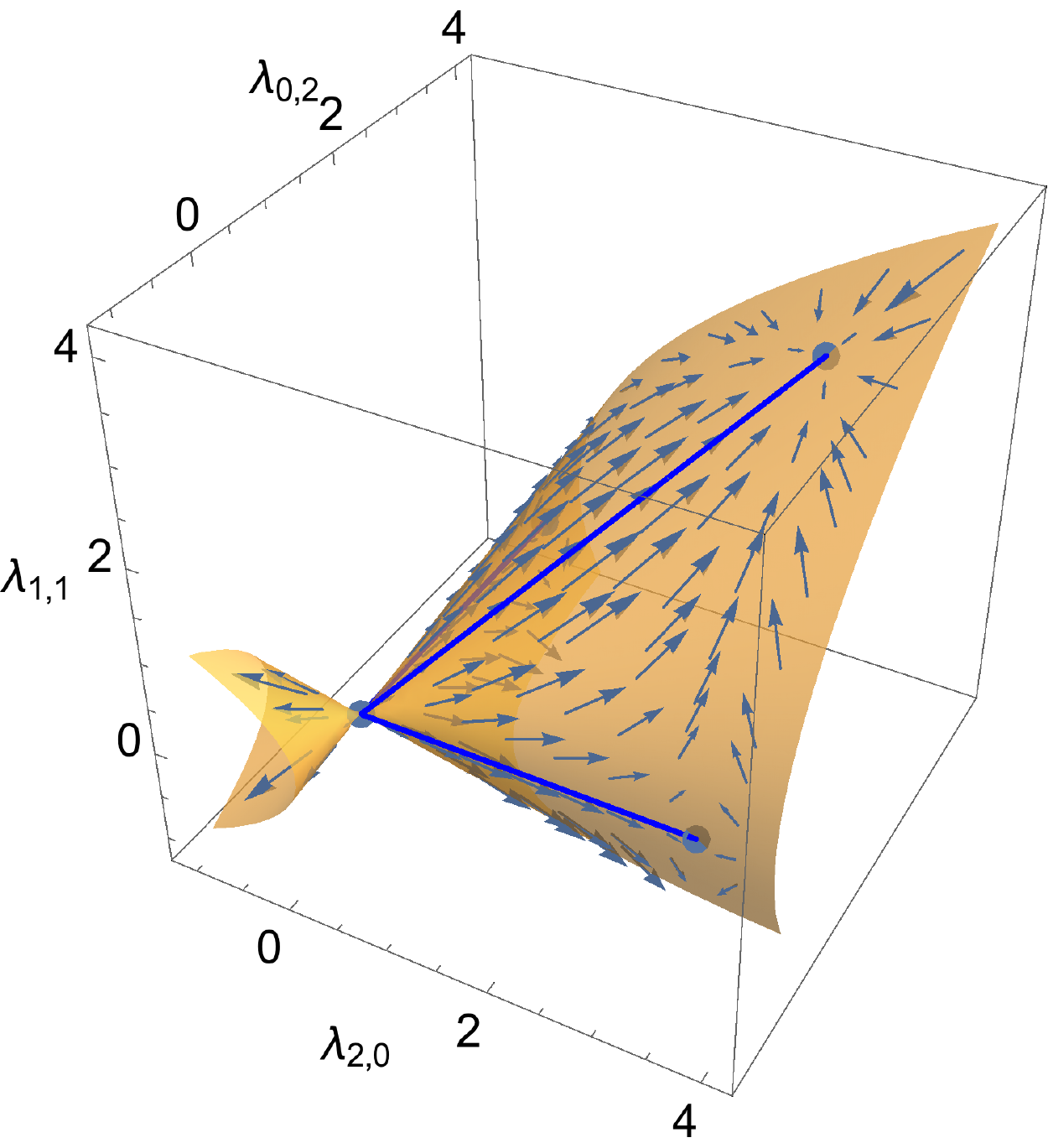}
\caption{\label{fig:Deltasurface} We show the 3-dimensional theory space spanned by the quartic couplings. The orange surface is defined by $\Delta=0$, and blue arrows show the RG flow towards the IR. Blue dots denote different fixed points that lie in this surface.}
\end{figure}

If $\Delta=0$ at every point in field space, the symmetry is enhanced to an O$(N+M)$ symmetry. 
An enhanced O$(N+M)$ symmetry requires that the potential has a flat direction everywhere in field space, i.e., the Hessian must have vanishing determinant. 
Hence, the RG flow cannot cross the hypersurface defined by $\Delta=0$ as a \emph{global} criterion. That hypersurface also contains several separatrices between fixed points, cf.~Fig.~\ref{fig:Deltasurface}.  
Note that this does not necessarily imply that $\Delta=0$, if it is imposed only locally in field space, is preserved during the flow. Within a local expansion up to fourth order in the fields, one can show that the flow of $\Delta$ is proportional to $\Delta$, using scale-dependent redefinitions of the field. These correspond to deforming ``elliptical" potentials such that the symmetry- enhancement is obvious.
We assume in the following that the sign of $\Delta$ evaluated at the extremum/saddle point $\kappa_\phi=\kappa_\chi \neq 0$ of the potential does not change under the flow, if that flow ends at a fixed point in the IR. In particular, it might be possible that additional separatrices outside the global surface $\Delta=0$ connect these fixed points, if they have appropriate attractive directions perpendicular to that surface. These separatrices would serve to separate the theory space.
To comprehensively uncover the structure of the theory space and its separate regions, global flows have to be considered that do not rely on a choice of expansion point.

\subsection{Fixed-point content}

Besides the trivial scaling solution, the Gaussian fixed point (GFP), with canonical critical exponents ($[\bar{\lambda}_{n,m}] = d - (n+m)(d-2)$) and thus infinitely many relevant couplings in $d=2$, there exist fixed points which are candidates for the stable fixed point.

\subsubsection{Decoupled fixed point} \label{sec:DFP}

\begin{figure}[!t]
\includegraphics[width=0.45\linewidth]{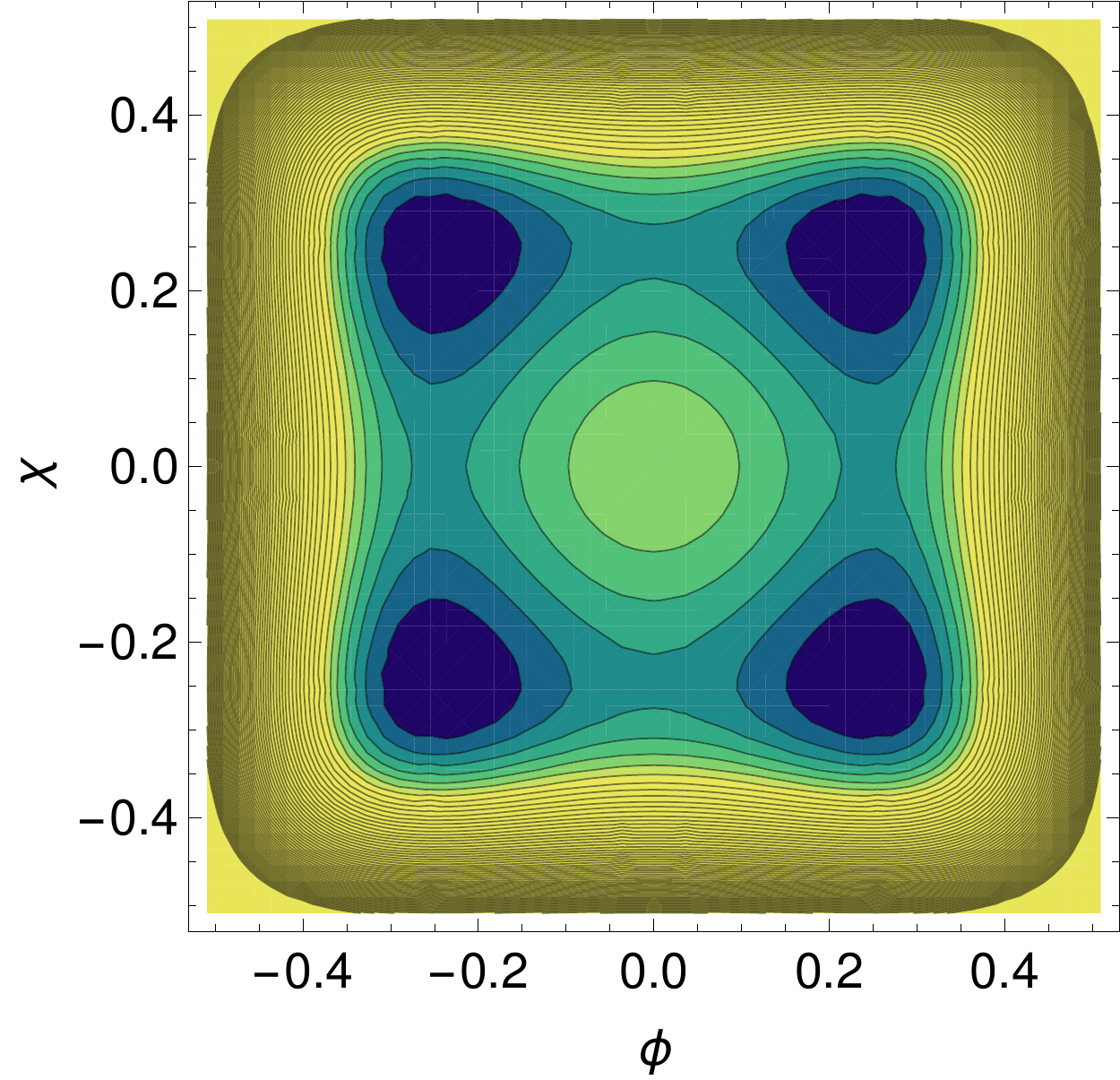} \quad \includegraphics[width=0.45\linewidth]{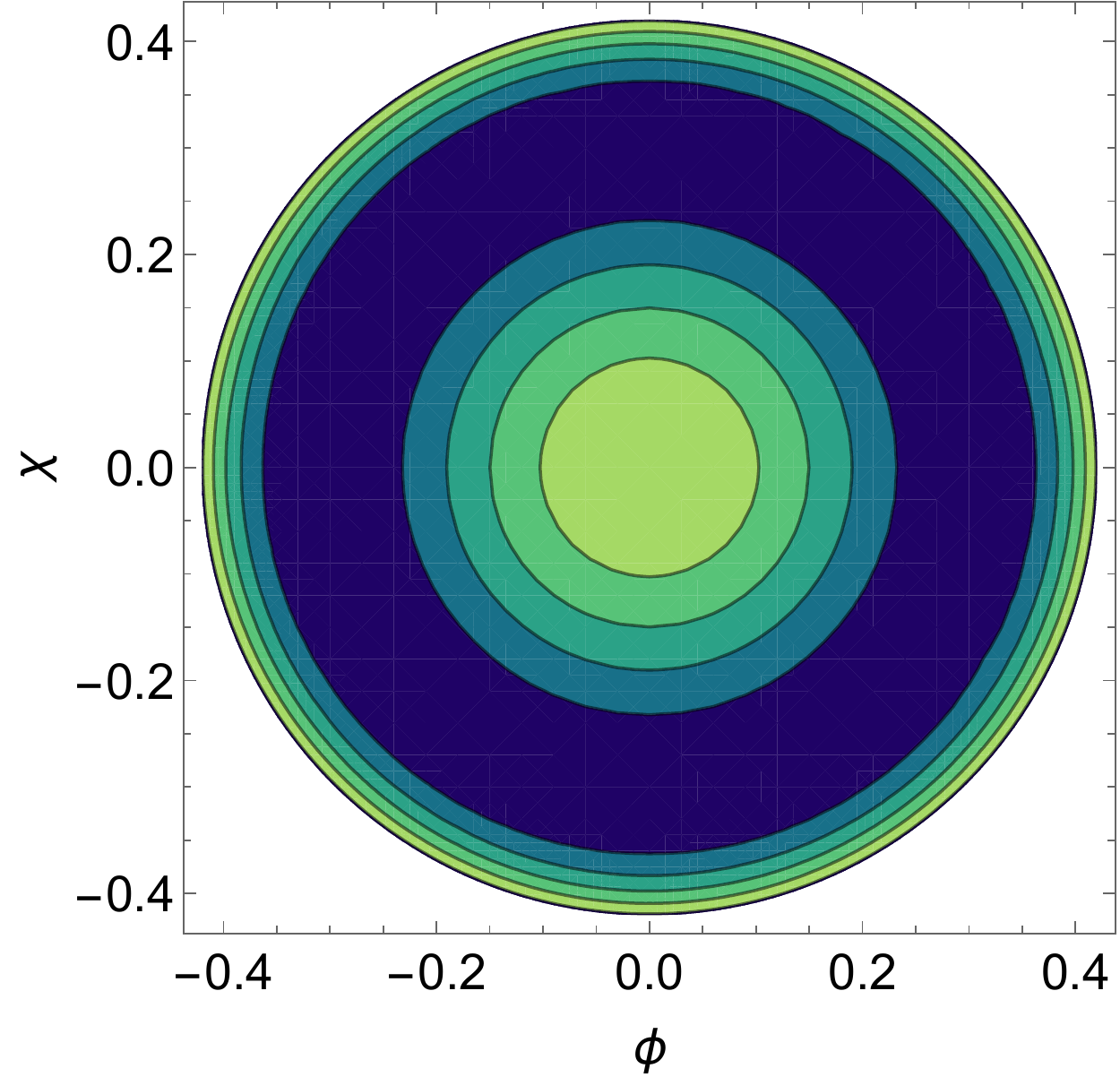}
\caption{\label{plot:DFPpot} Fixed-point potential at the DFP (left panel) and IFP (right panel) in $d=3$ and $N=M=1$. The decoupled fixed point is clearly tetracritical, i.e., $\Delta>0$, as the minima lie between the axes. The isotropic fixed point features a symmetry enhancement to an O(2) rotational symmetry.}
\end{figure}
At the decoupled fixed point (DFP), the fixed-point potential in \eqref{eq:dimfullpot} decouples into two independent O($N$) and O($M$) Wilson-Fisher fixed points, if the mixed couplings vanish, $\bar{\lambda}_{n,m}=0$ for $n,m>0$. Thus $\Delta>0$, i.e., the fixed point corresponds to tetracritical behavior, cf.~Fig.~\ref{plot:DFPpot}. 
In addition to one relevant direction from each Wilson-Fisher solution,  the vanishing mixed couplings $\bar{\lambda}_{n,m}$ are associated to non trivial critical exponents.
Hence, the fixed point can be stable, i.e., feature two relevant directions, depending on the values of those exponents.
The third critical exponent is related to the inverse Wilson-Fisher correlation length critical exponents, $\theta_I = \frac{1}{\nu_I}$  by Aharony's scaling relation \cite{Aharony:2002,Aharony:2002zz,Aharony:1976book,Aharony:1973zz}
\begin{equation}
 \label{eq:scalerelation}
 \theta_3 = \theta_1 + \theta_2 - d.
\end{equation}
The scaling relation \eqref{eq:scalerelation} is 
satisfied to any order in the $\epsilon$-expansion \cite{Calabrese:2002bm} and thus expected to be exact.

Additionally, there are two fixed points at which one sector is trivial and the other assumes a Wilson-Fisher fixed point, the decoupled Gaussian fixed points (DGFP). The separatrices between those fixed points and the GFP lie in the hypersurface $\Delta=0$,  cf.~Fig.~\ref{fig:Deltasurface}.
These fixed points, however, do not play any role for the stability trading.

\subsubsection{Fixed point with symmetry enhancement} 

There is a solution with $\lambda_{n,m}= \lambda_{c,0}=\lambda_{0,c}$ with $n+m =c$, 
which corresponds to an enhancement of the O($N$) $\oplus$ O($M$) symmetry to an O($N+M$) symmetry. This symmetry-enhanced fixed point features the coordinates of the single-field O($N+M$) Wilson-Fisher fixed point. We call it the isotropic fixed point (IFP), cf.~Fig.~\ref{plot:DFPpot}.

\subsubsection{Biconical fixed point}
\begin{figure}[!t]
\includegraphics[width=0.45\linewidth]{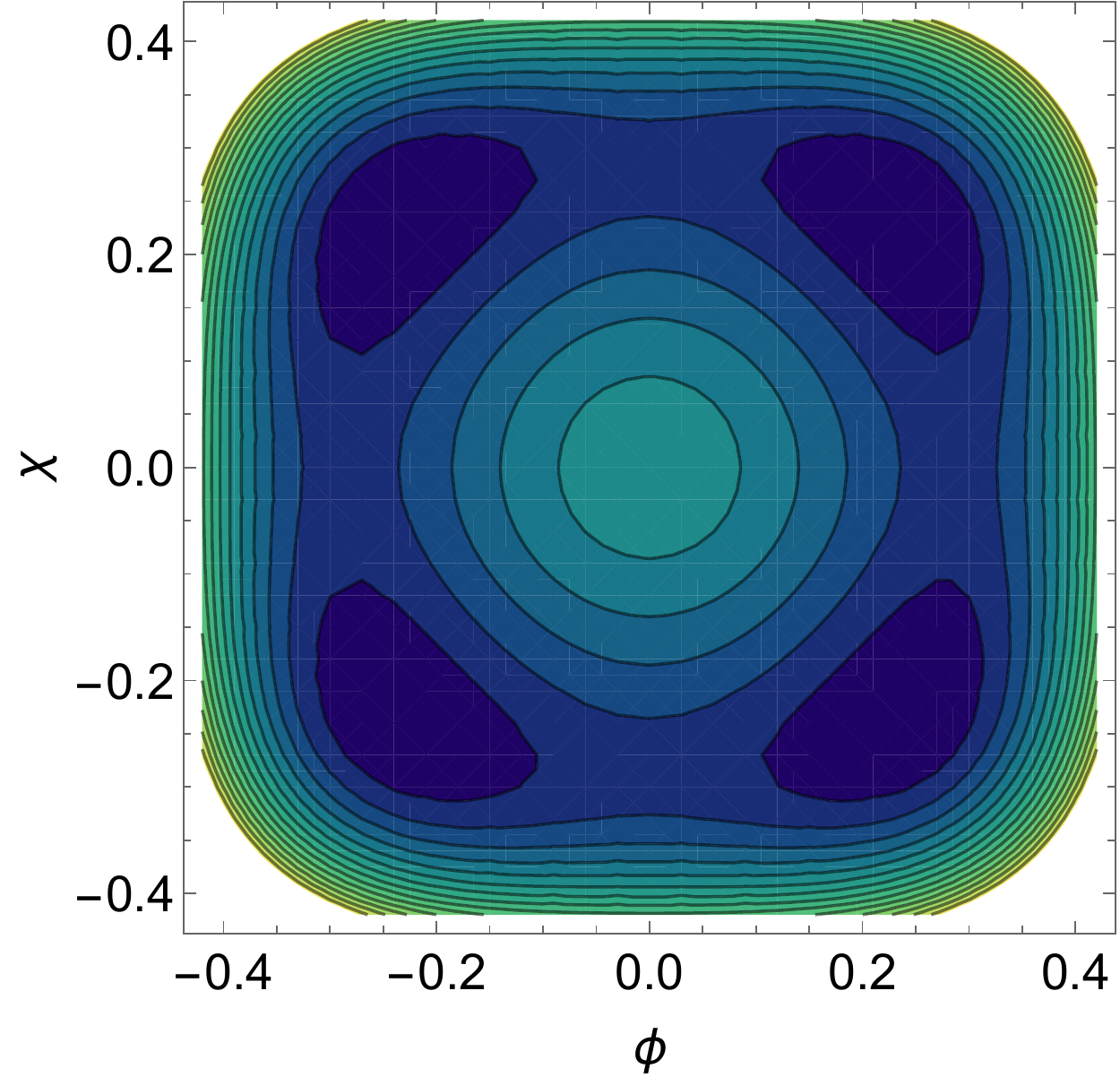}\quad
\includegraphics[width=0.45\linewidth]{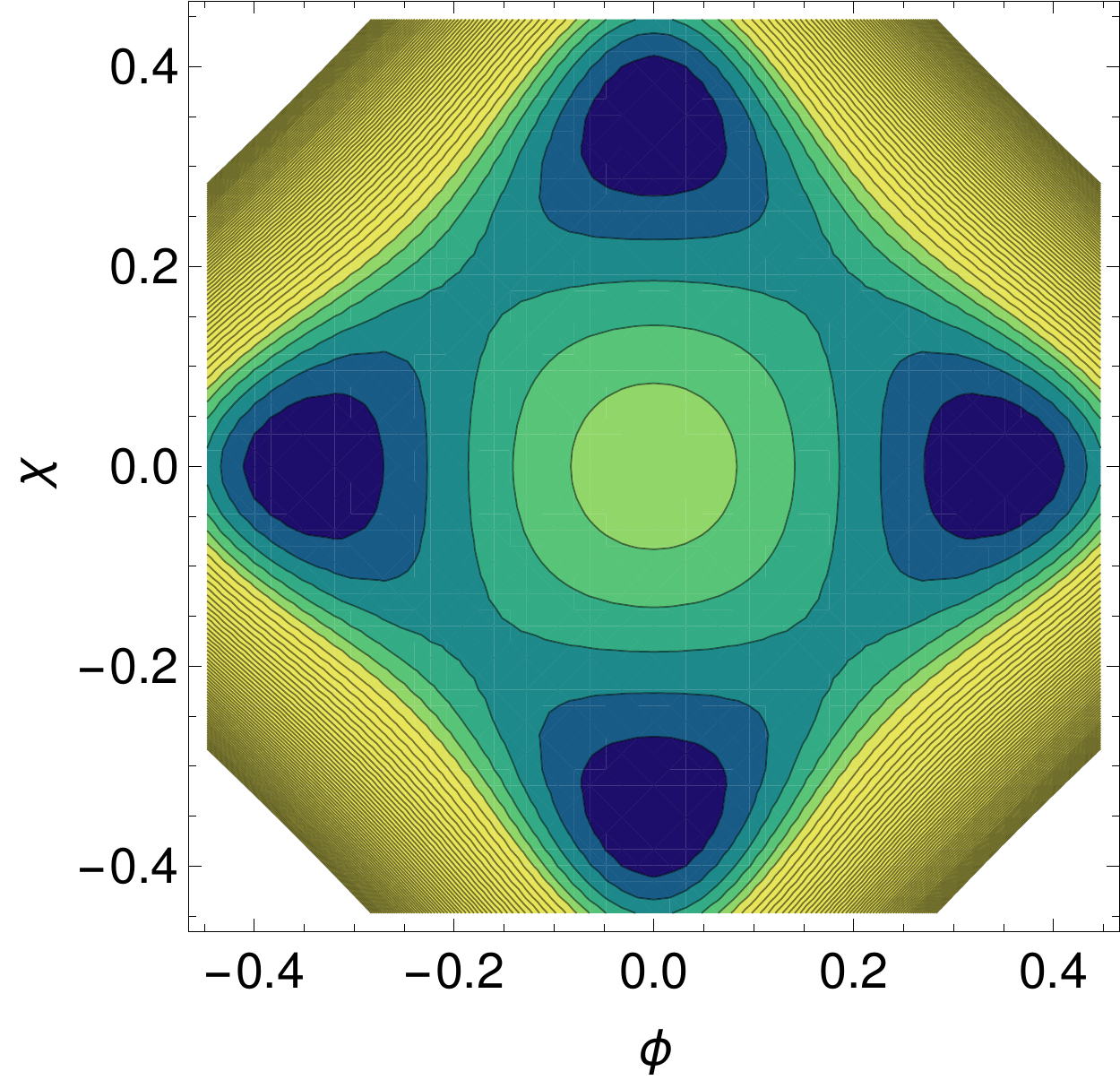}
\caption{\label{plot:BFPpot} Fixed-point potential at the BFP in $d=3$ and $N=M=1.2$ (left panel) within a local potential approximation (LPA). The fixed point features global minima between the axes, i.e., $\Delta>0$ and corresponds to tetracritical behavior. For $N=M=1$ (right panel) the minima lie on the axes, and accordingly $\Delta<0$.}
\end{figure}

There is one scaling solution for which both sectors are coupled non trivially,  i.e., $\lambda_{i,j}\neq 0$, without continuous symmetry-enhancement\footnote{The existence of new fixed points from non trivial interactions occurs more generally in models of several coupled sectors \cite{Eichhorn:2014asa,Eichhorn:2015woa}, as well as for coupled gauge theories \cite{Esbensen:2015cjw},  and provides a template for the possible fixed-point structure of asymptotically safe gravity and matter, see, e.g., \cite{Eichhorn:2012va,Eichhorn:2016esv}.}. 
This biconical fixed point (BFP) occurs in a tetracritical or bicritical form for different values of $N$, cf.~Fig.~\ref{plot:BFPpot}.

\subsubsection{Bicritical fixed points} 
\label{sec:bicriticalFP}
At $N=M=1$, the fixed-point equation has an accidental exchange symmetry under $\phi \leftrightarrow \chi$. Then, additional solutions can be deduced from fixed-point solutions that preserve that symmetry. For instance, the DGFPs do not, but the DFP does.
A solution $u(\phi,\chi)_{\ast}$ (expressed in dimensionless variables) to the fixed-point equation in $\phi, \chi$, also solves
the fixed-point equation in $\phi', \chi'$, where
\bea
\phi&=& \frac{1}{\sqrt{2}}\left(\phi' + \chi' \right),\quad
\chi= \frac{1}{\sqrt{2}}\left( -\phi' + \chi' \right),\label{eq:coordrot}
\eea
see App.~\ref{rotatedsolutions} for details.
This property is, e.g., present in the $\epsilon$ expansion to $\mathcal{O}(\epsilon)$
\cite{Kosterlitz:1976zza,Folk2008}
\bea
\beta_{\lambda_{2,0}}&=&- \epsilon \lambda_{2,0} + (64+8\,N)\, \lambda_{2,0}^2 + 8 M \lambda_{1,1}^2,\\
\beta_{\lambda_{0,2}}&=&- \epsilon \lambda_{0,2} + (64+8\,M)\, \lambda_{0,2}^2 + 8 N \lambda_{1,1}^2,\\
\beta_{\lambda_{1,1}}&=&- \epsilon \lambda_{1,1}+ 32 \lambda_{1,1}^2 + (16+8 N)\, \lambda_{2,0}\,\lambda_{1,1}\nonumber\\
&{}&+ (16+8 M)\, \lambda_{0,2} \lambda_{1,1}.
\eea
Here, we have adopted a convenient redefinition of the couplings, corresponding to a potential of the form
\be
u_{\epsilon}[\phi, \chi] = \lambda_{2,0}\phi^4 + 2 \lambda_{1,1}\phi^2 \chi^2 + \lambda_{0,2} \chi^4,
\ee
where we have specialized to the case $N=M=1$. We now implement the rotation \eqref{eq:coordrot}, which leaves us with
\bea
u_{\epsilon} &=& \frac{\phi'^4 +\chi'^4}{4}(\lambda_{2,0}+\lambda_{0,2} +2 \lambda_{1,1})\nonumber\\
&{}&+ \left(\phi'^3\, \chi'+ \phi'\,\chi'^3 \right) \left(\lambda_{2,0}- \lambda_{0,2} \right) \nonumber\\
&{}&+ \frac{ \phi'^2 \, \chi'^2 }{2}\left(3(\lambda_{2,0}+\lambda_{0,2}) - 2\lambda_{1,1}\right).
\eea
For this to be a viable fixed-point potential, we demand
\bea
0 &=& \lambda_{2,0}- \lambda_{0,2}, \quad \lambda'_{2,0}= \frac{\lambda_{2,0}}{2}+ \frac{\lambda_{1,1}}{2}, \nonumber\\
\lambda'_{1,1}&=&\frac{3}{2}\lambda_{2,0}- \frac{\lambda_{1,1}}{2}.
\eea
The first requirement corresponds to an exchange symmetry $\phi \leftrightarrow \chi$.
The DFP fulfills this requirement. Its coordinates are $\lambda_{2,0} = \lambda_{0,2} = \frac{1}{72}$. Upon the rotation \eqref{eq:coordrot}, this gives a fixed-point potential with coordinates $\lambda_{2,0}= \lambda_{0,2}= \frac{1}{144}$ and $\lambda_{1,1}= \frac{1}{48}$. 
The rotated counterpart of the DFP, called SFP in  \cite{Eichhorn:2013zza}, see also \cite{Bornholdt:1994rf,Bornholdt:1995rn,Tissier:2002zz}, must be bicritical. 
We refer to it as the RDFP. In $d=3$, the BFP equals the rotated DFP at $N=1$, which is also visible by comparing
the left panel in Fig.~\ref{plot:DFPpot} and the right panel in Fig.~\ref{plot:BFPpot}.
Below $d=3$, that degeneracy is lifted, and the BFP exists independently of the RDFP at $N=1$.
When the BFP is tetracritical at $N=1$, it implies the existence of another bicritical scaling solution, which we call the RBFP.

\begin{table}[!top]
\begin{tabular}{c| c c c c c c}
 & $\theta_1$ & $\theta_2$ & $\theta_3$ & $\theta_4$ & $\theta_5$ & $\theta_6$ \\ \hline
 \multicolumn{7}{c}{$d=2.8$}\\ \hline
 DFP & 1.408 & 1.408 & 0.0165 & -0.753 & -0.753 & -2.145\\
 RDFP & 2 & 1.408 & 0.0165 & 0 & -0.753 & -2 \\ \hline
 BFP & 1.498 & 1.329 & -0.0098 & -0.734 & -0.759 & -2.126\\
 RBFP & 1.944 & 1.329 & -0.0098 & -0.154 & -0.759 & -2.077\\ \hline
 \multicolumn{7}{c}{$d=2.7$}\\ \hline
 DFP & 1.334 & 1.334 & -0.033 & -0.800 & -0.800 & -2.166\\
 RDFP & 2 & 1.334 & 0 & -0.033 & -0.800 & -2  \\ \hline
\end{tabular}
\caption{\label{tab:critExpLPA} We show the five largest critical exponents of the DFP and BFP and the rotated counterparts in LPA (see Sec.~\ref{sec:FRG}) for $N=M=1$.
The dimension $d=2.8$ is chosen as a representative. For the RDFP we additionally give the values for $d=2.7$ to clarify
that $-2,0,2,\ldots$ are always present in the spectrum at $N=1$, independent of the dimension.}
\end{table}

Some of the eigenperturbations of the rotated fixed-point solutions are related to their unrotated counterparts, for details see App.~\ref{rotatedsolutions}.
For instance, the instability of the DFP implies instability of the RDFP, cf.~Tab.~\ref{tab:critExpLPA}.
By contrast, if the DFP is stable, the RDFP features an additional marginal direction, as we obtain $\theta_i=2,0,-2,...$ as additional critical exponents\footnote{Note that $\theta_3=0$ presumably implies the existence of another fixed point which the RDFP collides with at that point. In this work, we do not search for such a fixed-point solution.}, independent of $d$, at least within the approximation without anomalous dimensions. 

\section{Synopsis: Key results}

\begin{figure}[!t]
 \includegraphics[width=\columnwidth]{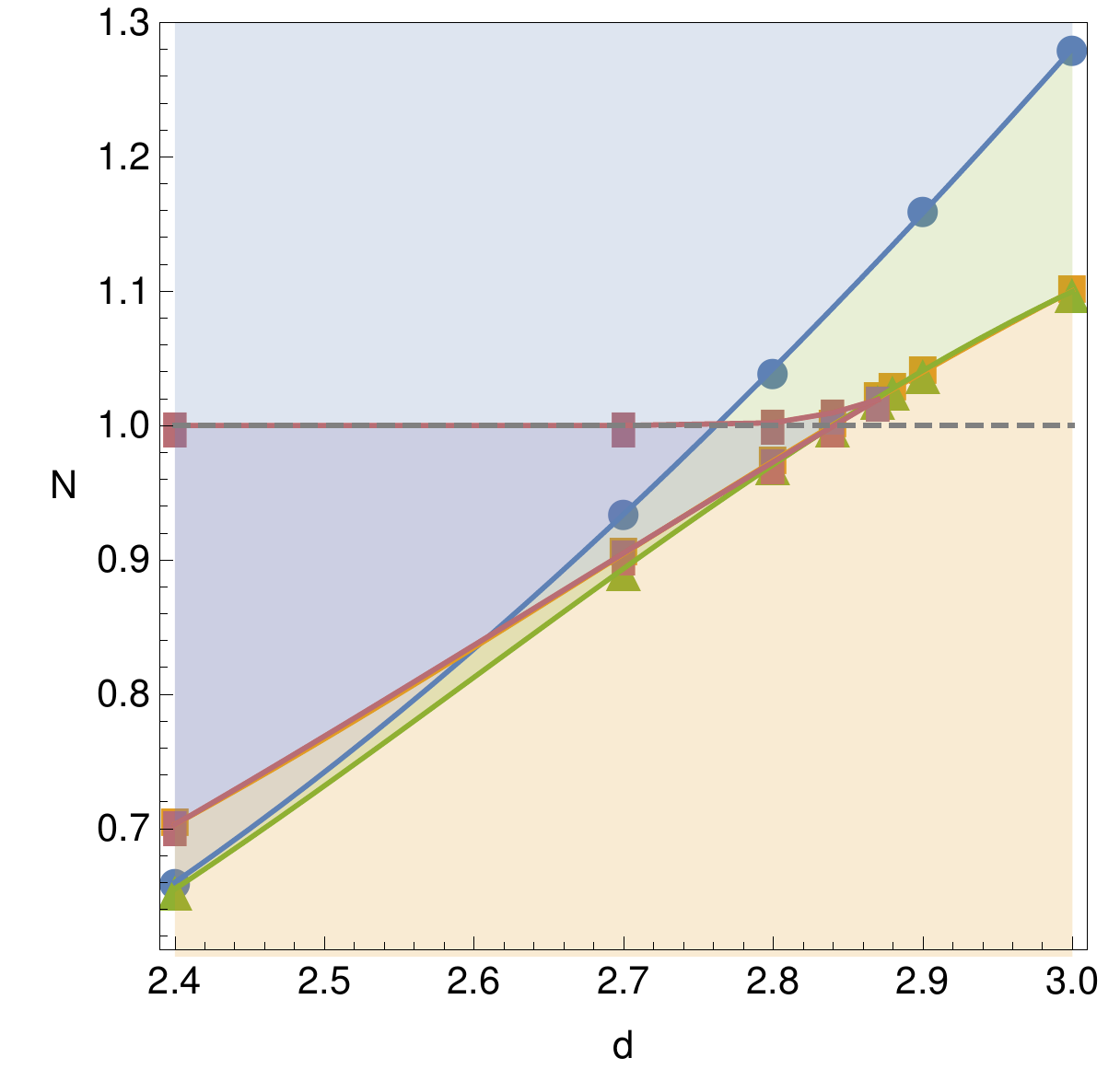}
 \includegraphics[width=\columnwidth]{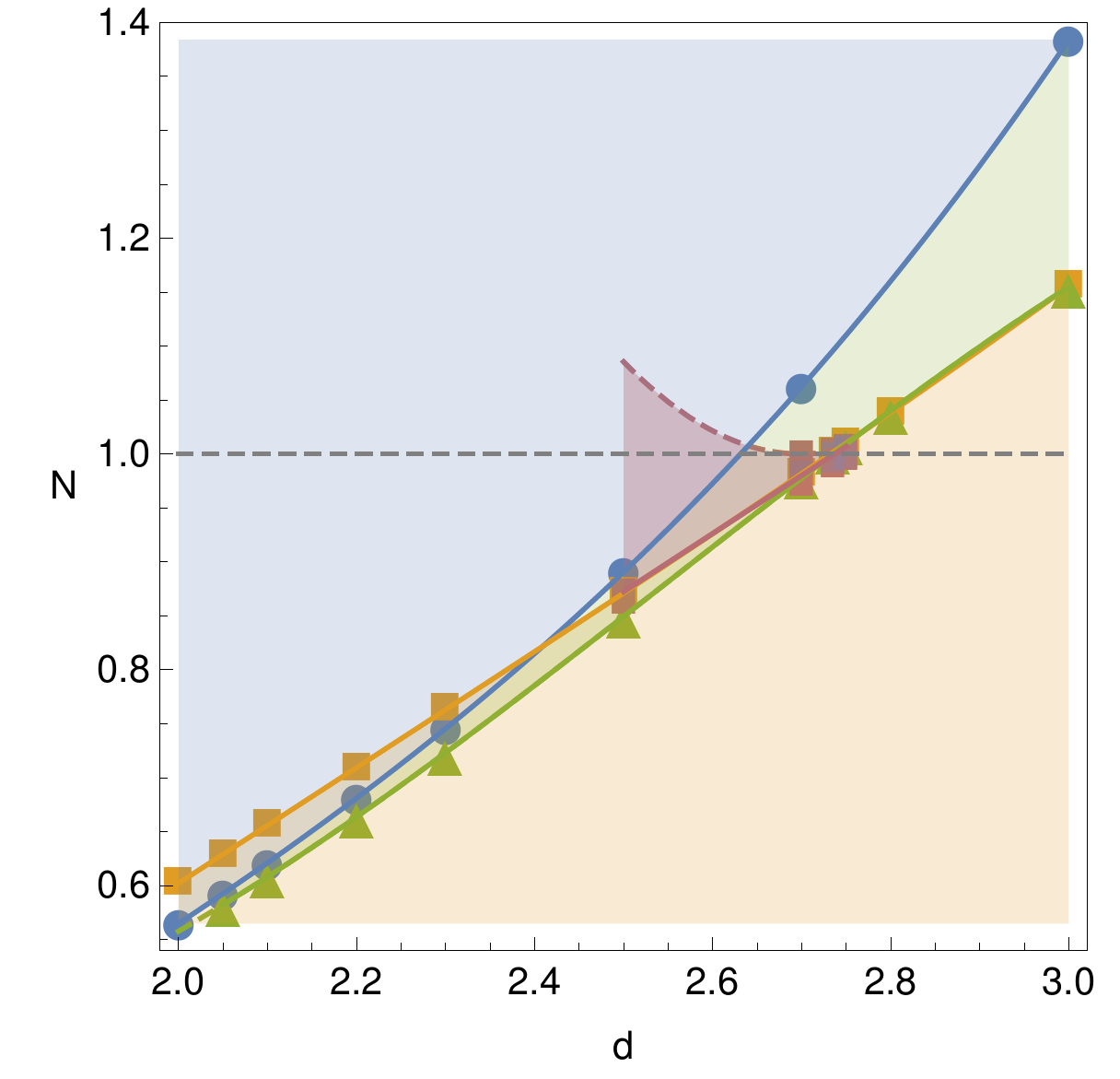}
 \caption{Stability regions of the DFP (blue circles), IFP (orange squares), BFP (green triangles)  and the R(B/D)FP (mauve rectangles) as a function of $N=M$ and the dimension. The DFP is stable in the blue region (between blue line and large $N=M$), the IFP in the light-orange region (between orange line and small $N=M$), the BFP in the light green region (between green and blue line) and the R(B/D)FP in the mauve region (between the mauve lines). The upper panel is for LPA, the lower for LPA' (see Sec.~\ref{sec:FRG}). In LPA', we can only give an estimate for the upper stability bound of the R(B/D)FP (dashed line). In fact, it extends to $d=2$, which is not depicted here. The lower bound continues on the orange line.
 \label{fig:stabilityregions}}
\end{figure}

Here, we present the main results of our study, before discussing more details in Sec.~\ref{sec:phasestructure}.
We reproduce the well-known picture in $d=3$, where the IFP is stable at low $N=M$, and the DFP is stable at large $N$. Inbetween, the BFP is the stable fixed point. These three fixed points trade stability in fixed-point collisions, which entails that only one of them can be stable at any given value of $N=M$. 
This reflects the 
usual expectation that the universality class is determined solely by the dimensionality, symmetries and long-range degrees of freedom. 
Towards lower $d$, a new, stable, bicritical fixed point appears, cf.~Fig.~\ref{fig:stabilityregions}. It trades stability with the IFP. Due to its bicritical nature, it \emph{cannot} trade stability with the DFP
 whose stability trading partner is still the BFP. We obtain regions of two simultaneously stable fixed points with different signs of $\Delta$.
 This tweaks the usual expectation of universality in bosonic models. 
 We conjecture that the sign of $\Delta$ in the microscopic potential decides about which universality class is realized.

Towards $d=2$, the stability regions of the DFP and the IFP begin to overlap. Comparing our results with the stability region of the DFP as inferred from the scaling solution and the Onsager solution, 
this overlapping stability region should be shifted to $N=M \approx 1$. This  would open the interesting possibility, that the universality class of two coupled Ising models might exhibit either a tetracritical, decoupled fixed point, or instead be a BKT-type phase transitions.

\section{Functional Renormalization Group tools}
\label{sec:FRG}
To evaluate the beta functions in $2\leq d \leq 3$, 
we employ the nonperturbative functional  RG, for reviews see \cite{Berges:2000ew,Pawlowski:2005xe,Gies:2006wv,Delamotte:2007pf,Bervillier:2007rc,Braun:2011pp}. Readers familiar with the FRG can skip this section.

\subsection{Functional Renormalization Group equation}
The central object of the functional  RG is a scale-dependent effective action, or free-energy functional, that is obtained from the generating functional (defined with a UV cutoff $\Lambda$) by imposing an infrared cutoff at the momentum scale $k$, and taking a modified Legendre transform. For a single scalar field $\varphi$ with expectation value $\phi$, the definition reads \cite{Wetterich:1992yh}
\bea
\Gamma_k[\phi] &=& \underset{J}{\rm sup} \Bigl( \int d^dx \, J(x)\phi(x) \nonumber\\
&{}&- {\rm ln} \int_{\Lambda}\mathcal{D}\varphi\, e^{-S[\varphi] - \frac{1}{2}\int \frac{d^dp}{(2\pi)^d}\varphi(-p)R_k(p^2)\varphi(p)}\Bigr)\nonumber\\
&{}& - \frac{1}{2} \int \frac{d^dp}{(2\pi)^d}\phi(-p)R_k(p^2)\phi(p),
\eea
and generalizes appropriately to the case of several fields.
Herein, the regulator function $R_k(p^2)$ depends on the cutoff scale $k$ and on the momenta of the fluctuating fields in such a way that it suppresses low-momentum modes with $p^2<k^2$ in the path integral. By contrast, high-momentum modes are integrated out, $R_k(p^2)=0$ for $p^2>k^2$. This implements the Wilsonian idea of momentum-shell-wise integration of the path integral \cite{Kadanoff:1966wm,Wilson:1971bg,Wegner:1972ih,Wilson:1973jj}. 
This procedure generates all interactions that are compatible with the symmetries of the model. The effective action can then be expanded in a derivative expansion, see, e.g., \cite{Morris:1994au,Morris:1994ie,Morris:1994ki,Morris:1996kn}, where all terms with a given number of derivatives, but arbitrary powers of the field are grouped into the same order. In particular in the vicinity of 
$d=2$, this is advantageous, as all interaction terms with zero derivatives have canonical dimensionality 2, i.e., they are canonically relevant. We thus expand 
\bea
\Gamma_k &=& \frac{Z_{\phi}(k)}{2}\int d^dx \partial_{\mu}\phi^I \partial_{\mu}\phi^I + \frac{Z_{\chi}(k)}{2}\int d^dx \partial_{\mu}\chi^J \partial_{\mu}\chi^J \nonumber\\
&{}& + \int d^dx\, U(\bar{\rho}_{\phi}, \bar{\rho}_{\chi}),\label{eq:derexp}
\eea
where we have neglected that radial modes and Goldstone modes in $\phi^I$ can have different wave-function renormalizations. \eqref{eq:derexp} is known as the local potential approximation (LPA) if $Z_{\phi}= Z_{\chi}=1$, and as LPA' if the anomalous dimensions 
\begin{equation}
 \eta_\phi = -\partial_t \ln Z_\phi \qquad \text{and} \qquad \eta_\chi = -\partial_t \ln Z_\chi,
\end{equation}
with $\partial_t = k\, \partial_k$ are taken into account.

The effective action obeys an RG flow equation that is exact at the formal level, but becomes approximate once we introduce a truncation as in \eqref{eq:derexp}. It reads \cite{Wetterich:1992yh}
\be
\partial_t \Gamma_k = \frac{1}{2} {\rm Tr} \left(\Gamma_k^{(2)}+R_k \right)^{-1} \partial_t R_k,\label{eq:floweq}
\ee
see also \cite{Morris:1993qb,Ellwanger:1993mw}.
By the superscript $(2)$ we denote the second functional derivative with respect to the fields, cf.~\eqref{eq:notation2}. Thus, $(\Gamma_k^{(2)}+R_k)$ is a matrix in field space.
The trace is over all terms on the right-hand-side and is a trace over the eigenmodes of the generalized, field-dependent propagator $(\Gamma_k^{(2)}+R_k)^{-1}$. In our case it 
translates into a momentum integration and a trace in field space. To exemplify the structure, we show the case of a field  $\phi$ with one component, where the trace in field space is absent. We further set $\phi=\rm const$, after taking the derivative in $\Gamma_k^{(2)}$:
\be
\partial_t \Gamma_k = \frac{1}{2} \int \frac{d^dp}{(2\pi)^d} \left(Z_{\phi} p^2 + U^{(2)}+ R_k(p^2) \right)^{-1} \partial_t R_k(p^2).
\ee
The momentum integral is IR finite due to the presence of the mass-like regulator $R_k(p^2)$, and UV finite due to the presence of $\partial_t (R_k) \rightarrow 0$ for $p^2>k^2$, as required for the regulator. 

For the regulator function, we choose a form that satisfies optimization criteria \cite{Litim:2000ci,Litim:2001up} within the LPA 
\be
R_k(p^2) = Z_{\phi,\chi}(k^2-p^2)\theta(k^2-p^2).
\ee
\subsection{Truncation and flow of the effective potential}
\label{sec:flow}
The flow equation for the dimensionless effective potential
expressed in dimensionless field variables reads
\bea
\partial_t u&=&-d u+\!(d-2+\eta_\phi)\rho_\phi u^{(1,0)} + \!(d-2+\eta_\chi)\rho_\chi u^{(0,1)}\nonumber\\
	 &&+I_{R,\phi}^d(\omega_\chi,\omega_\phi,\omega_{\phi\chi})+(N-1)I_{G,\phi}^d(u^{(1,0)})\nonumber\\
	 &&+I_{R,\chi}^d(\omega_\phi,\omega_\chi,\omega_{\phi\chi})+(M-1)I_{G,\chi}^d(u^{(0,1)})\,.\label{eq:potflow}
\eea
The contributions in the first line are the canonical and anomalous scaling of the fields and the potential.
The second and third line in \eqref{eq:potflow} arise from non-perturbative loop contributions of the massive radial and the Goldstone modes with factors $(N-1)$ and $(M-1)$. We use the threshold functions,
\bea
I_{R,i}^{d}(x,y,z)&=&\frac{4v_d}{d}\left(1-\frac{\eta_i}{d+2}\right)\frac{1+x}{(1+x)(1+y)-z},\nonumber\\
I_{G,i}^d(x)&=&\frac{4v_d}{d}\left(1-\frac{\eta_i}{d+2}\right)\frac{1}{(1+x)},
\eea
with the volume factor $v_d^{-1}=2^{d+1}\pi^{d/2}\Gamma(\frac{d}{2})$. The arguments in the flow equation \eqref{eq:potflow} read
\bea
\omega_\phi&=&u^{(1,0)}+2\rho_\phi u^{(2,0)},\\
\omega_\chi&=&u^{(0,1)}+2\rho_\chi u^{(0,2)},\\
\omega_{\phi\chi}&=&4\rho_\phi \rho_\chi \big(u^{(1,1)}\big)^2.
\eea

From \eqref{eq:potflow}, the beta functions of the couplings $\lambda_{i,j}$ can be inferred.
However, we mostly solve \eqref{eq:potflow} without further approximations.
From the fixed-point equation \eqref{eq:potflow}, a linearized equation for small perturbations around the fixed point can be derived, cf. \eqref{eq:eigenvalueproblem}, which corresponds to the functional form of the eigenvalue problem \eqref{eq:critExp} and provides the critical exponents.
 
The equations for the anomalous dimensions read
\bea
\eta_{\phi}&=& \frac{16 v_d}{d} \bigg[\big(u^{(2,0)}\big)^2\kappa_\phi\big(1+2u^{(0,2)}\kappa_\chi\big)^2 \nonumber \\
&{}&-4u^{(2,0)}\big(u^{(1,1)}\big)^2\kappa_\chi\kappa_\phi\big(1+2u^{(0,2)}\kappa_\chi\big) \label{eq:etaphi} \\
&{}&+\big(u^{(1,1)}\big)^2\kappa_\chi\big(1+4\big(u^{(1,1)}\big)^2\kappa_\chi\kappa_\phi\big)\bigg]/ \nonumber \\
&{}&\bigg[\big(1+2u^{(0,2)}\kappa_\chi\big)\big(1+2u^{(2,0)}\kappa_\phi\big)-4\big(u^{(1,1)}\big)^2\kappa_\phi\kappa_\chi\bigg]^2, \nonumber
\\
\eta_{\chi}&=&\eta_\phi\big\{\phi\leftrightarrow \chi, u^{(i,j)}\leftrightarrow u^{(j,i)}\big\}, \label{eq:etachi}
\eea
which are projected onto the non trivial minimum $(\kappa_\phi,\kappa_\chi)$ for tetracritical fixed points. 
For bicritical fixed points, we evaluate $\eta_\phi$ at  $(\kappa_\phi,0)$ and $\eta_\chi$ at  $(0,\kappa_\chi)$, which corresponds to the minima of the potential.

The LPA' gives rise to an ambiguity for the third critical exponent of the DFP, see \cite{Boettcher:2015pja}:
Whereas the scaling relation \eqref{eq:scalerelation} is fulfilled within the LPA, in LPA' it is only satisfied if the anomalous dimensions are held fixed in the computation of the critical exponents. This ambiguity might be resolved in a further extension of the truncation beyond LPA'.
In order for the value of $N_{\rm crit,\, DFP}$ to correspond to the value where the BFP loses stability in our truncation, we compute the critical exponents by taking the variations of the anomalous dimensions into account if not stated differently.

A second major difference between the LPA and the LPA' lies in the relation of the RD(B)FP to the D(B)FP: Whereas the rotation \eqref{eq:coordrot} is exact in the LPA, it is only approximate in the LPA'. Thus, rotating the D(B)FP at $N=M=1$ by $\pi/4$ in field space results in a potential that is close, but not exactly equal to, the RD(B)FP. 

In this paper, we focus on the case $N=M$. Moreover, we concentrate on fixed points with an explicit $\phi \leftrightarrow \chi$ exchange symmetry and do not explore possible fixed points which do not share that symmetry.

\subsection{Pseudo-spectral methods}
\label{sec:numerics}

In $d<3$, an increasing number of couplings (within the Taylor expansion) is needed to obtain reliable results, cf.~App.~\ref{app:localbreakdown}.
Hence, a more efficient method resolving the potential on a whole interval or even globally is required.
In the present case, we have to deal with a partial differential equation (PDE) \eqref{eq:potflow} for the potential
and two algebraic equations \eqref{eq:etaphi} and \eqref{eq:etachi} for the anomalous dimensions.
We solve this system via pseudo-spectral methods.

The potential $u$ is represented as a non-local expansion of a set of orthogonal basis functions \cite{Boyd:ChebyFourier}.
Different choices for this set of functions include Hermite or Laguerre polynomials defined on $[0,\infty)$, Legendre or Chebyshev polynomials defined on $[-1,1]$.
Laguerre or Hermite polynomials are not suitable for our problem since the asymptotic behavior of the potential would depend on the order of expansion. In particular, questions of global stability would be difficult to decide in this way.
As they show the best convergence properties for the expansion coefficients $a_{ij}$, we employ Chebyshev polynomials $T_i(x)$.
For the expansion in both field directions on the domain $[0,\rho^{\rm max}_\phi] \times [0,\rho^{\rm max}_\chi]$, we use
\begin{equation}
 u(\rho_\phi,\rho_\chi) = \sum_{i=0}^{n^{\rm max}_\phi} \sum_{j=0}^{n^{\rm max}_\chi} a_{ij} T_i\left(2\frac{\rho_\phi}{\rho^{\rm max}_\phi}-1\right) T_j\left(2\frac{\rho_\chi}{\rho^{\rm max}_\chi}-1\right),
 \label{eq:expansion}
\end{equation}
where $n^{\rm max}_\phi$ and $n^{\rm max}_\chi$ denote the maximum order in each field direction.
We expand the potential on a finite domain, and do not observe any significant influence of the choice of $\rho^{\rm max}_\phi$ and $\rho^{\rm max}_\chi$ on  the results.
However, a global resolution in each field direction is possible as  shown in \cite{Borchardt:2015rxa,Borchardt:2016pif} for one field invariant.

The values of the function and its derivatives everywhere in field space are easily accessible via efficient, recursive algorithms.
Because of the fast convergence, only a small number of the coefficients $a_{ij}$ is needed to obtain highly accurate results. As a fast convergence implies that high-order coefficients approach zero very quickly,
the last coefficients of the expansion provide an error estimate 
\begin{equation}
 err=\sum_{i=0}^{n_\phi^{\rm max}} |a_{i n_\chi^{\rm max}}|+\sum_{i=0}^{n_\chi^{\rm max}} |a_{n_\phi^{\rm max} i}|,
\end{equation}
for the accuracy of that approximation. 

More specifically, we divide the whole domain $(\rho_\phi,\rho_\chi) \in [0,\rho^{\rm max}_\phi] \times [0,\rho^{\rm max}_\chi]$ into subdomains and employ the expansion \eqref{eq:expansion} on each patch.
This results in a more efficient solver algorithm and a higher resolution in the physically interesting regions of small values of the field. In particular, this allows us to only use a high resolution in regions of field space where it is required to resolve the details.
We insert \eqref{eq:expansion} into the PDE and evaluate it at a set of collocation points%
 \footnote{ These collocation points correspond to a special choice of a non-equidistant grid in field space.
This important property supports the fast convergence of the $a_{ij}$ calculated from the values of the function at these points.}
(collocation method).
The multiple domains are connected by additionally demanding smoothness of $u$ and its first derivatives across the boundaries of the domains.
The resulting algebraic, non-linear system of equations is solved by a Newton-Raphson iteration.

The application of this method to the  functional RG flow and fixed-point equations has recently been put forward 
\cite{Borchardt:2015rxa,Borchardt:2016pif,Litim:2003kf,Fischer:2004uk,Gneiting:2005,Zappala:2012wh,Heilmann:2014iga}.
For more details on pseudo-spectral methods and the particular implementation used here, we refer to \cite{Boyd:ChebyFourier} and \cite{Borchardt:2015rxa,Borchardt:2016pif}, respectively.

\section{Interplay of fixed points and phase structure in $d\leq3$}\label{sec:phasestructure}

\subsection{Stability trading and allowed fixed-point collisions}

Fixed points trade their stability in collisions in the space of couplings that occur at particular values of $N$, see, e.g., \cite{Eichhorn:2014asa}, where one critical exponent vanishes. These are distinct from fixed-point annihilations in which the two fixed points collide and then disappear into the complex plane, and cease to be physically relevant. In a stability-trading collision, both fixed points continue to exist after the collision, and simply ``pass through" each other on the real line.
As stability trading must involve fixed-point collisions, a fixed point can only become (un)stable, if it approaches another fixed point. 
At the collision point, the fixed-point action of both fixed points is the same. 
Therefore, collisions between specific classes of fixed points cannot occur. 
In particular, tetracritical ($\Delta>0$) and bicritical ($\Delta<0$) fixed points cannot collide with each other due to the different location of the minima of the potential.
Both types of fixed points can however collide with the IFP. 
In that collision, they pass through the surface $\Delta=0$ and thereby change their nature from bi- to tetracritical and vice-versa. Note that it is not a contradiction that the RG flow for a given theory, at fixed $N, M$ cannot pass through the symmetry-enhanced surface $\Delta=0$, while the location of fixed points under variations of $N,M$, i.e., for different theories, can, of course, pass through that surface.

\subsubsection{Review of stability trading in $d=3$}
\begin{figure}[!t]
 \includegraphics[width=\columnwidth]{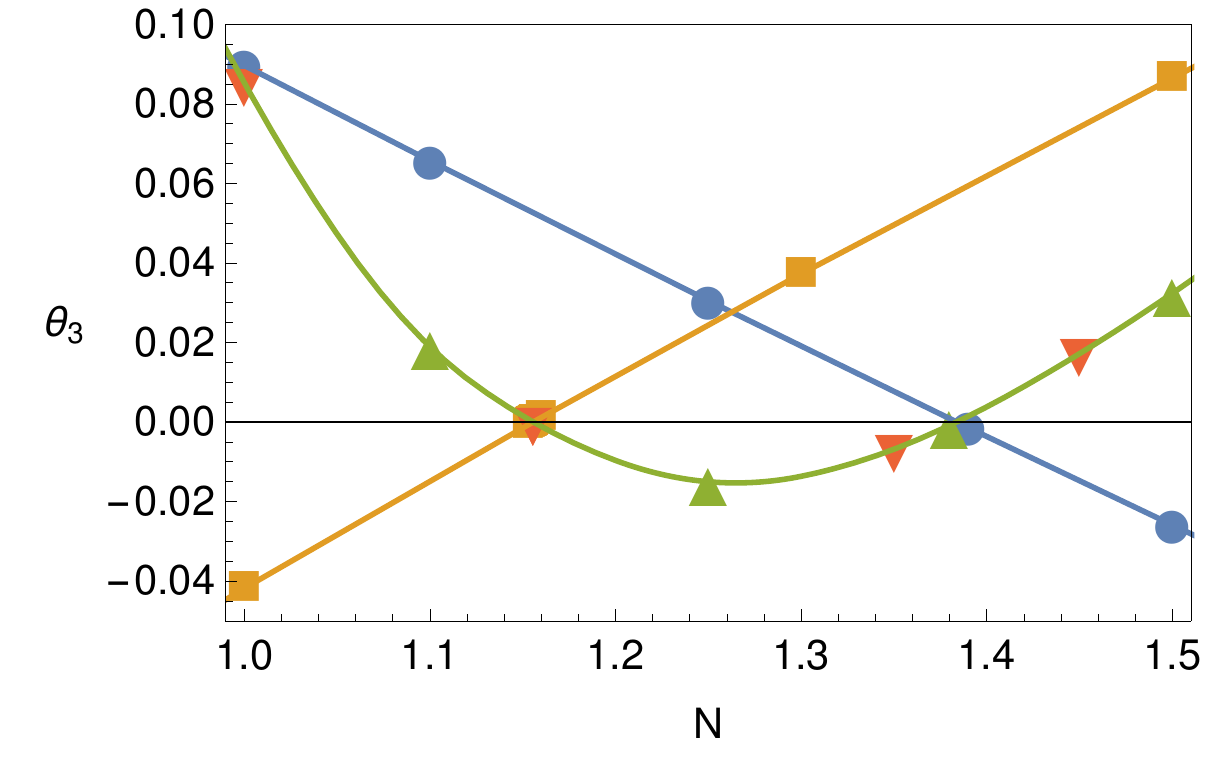}
 \caption{The third critical exponent of the DFP (blue dots), IFP (orange squares), BFP (green triangles/red inverted triangles) as a function of $N$
 for $d=3$.}
 \label{fig:FPstructured3}
\end{figure}
\begin{figure}[!t]
\includegraphics[width=\linewidth]{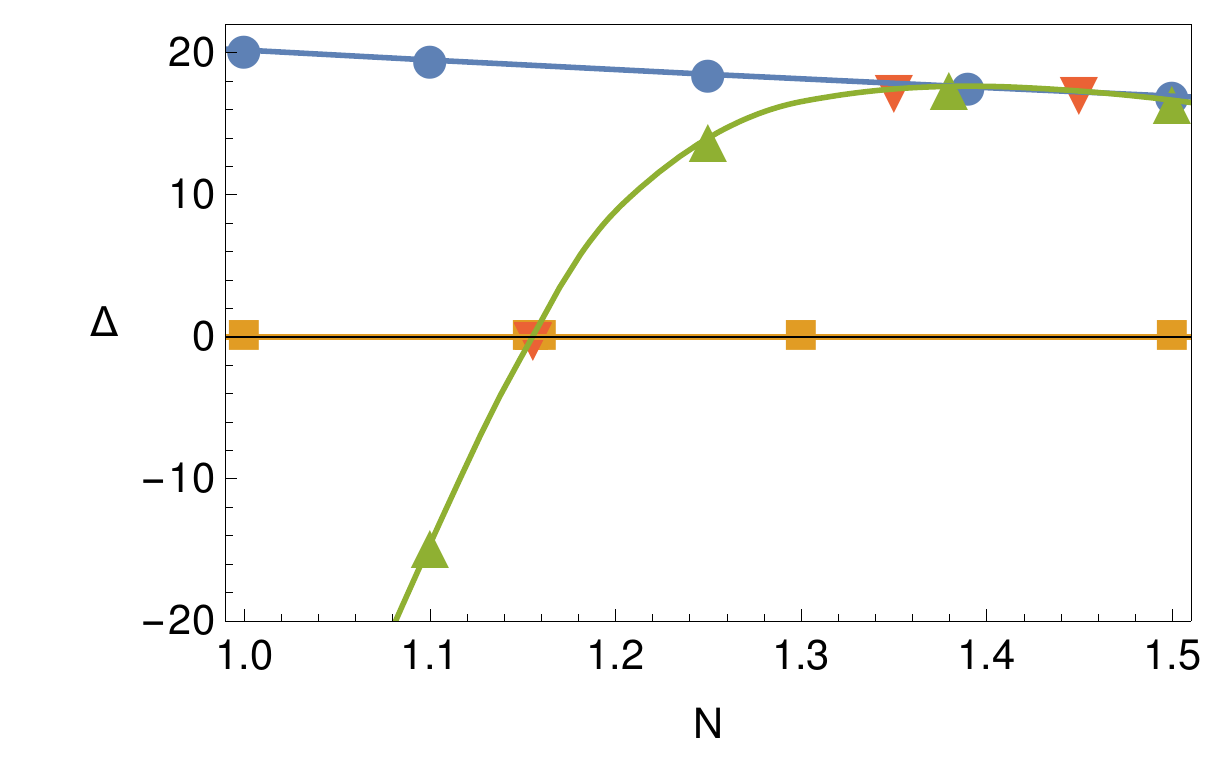}
\caption{\label{fig:Deltad3} We plot $\Delta$ as a function of $N=M$ for the IFP (orange squares), which divides the bicritical region ($\Delta<0$) from the tetracritical region ($\Delta>0$). The BFP (green triangles/red inverted triangles) collides with the DFP (blue dots) at $N=N_{\rm crit,\, DFP}=1.38$ and with the IFP at $N=N_{\rm crit, \,IFP}=1.15$. }
\end{figure}
The case of $d=3$ has been explored extensively \cite{Calabrese:2002bm,Calabrese:2002gv,Aharony:2002,Pelissetto:2000ek,Folk2008}, also with functional RG methods \cite{Eichhorn:2013zza,Boettcher:2015pja,Pawlowski:2015mlf}, where stability-trading was discussed in detail \cite{Eichhorn:2014asa}. Here, we briefly review the mechanism, and benchmark our method by comparing our LPA' results to previous results.

The DFP is stable at large $N$, and loses its stability in a collision with the tetracritical BFP at $N=N_{\rm crit,\, DFP}=1.4$, cf.~Fig.~\ref{fig:FPstructured3}.  This lies at a slightly higher value  than that obtained by perturbative methods, cf.~\cite{Folk2008}.
Towards smaller $N$, the BFP approaches the IFP, as its mixed couplings $\lambda_{i,j}$, $i,j \neq 0$, grow. It collides with the IFP at $N= N_{\rm crit,\, IFP}=1.15$.  Again, this estimate is slightly above that obtained in \cite{Folk2008}, and agrees -- as it should -- with the  functional RG result in \cite{Boettcher:2015pja} that employed the shooting method.
``Tunneling" through the IFP, cf.~Fig.~\ref{fig:Deltad3}, the BFP becomes bicritical, as its mixed couplings continue to grow. 
At $N=1$, the BFP is a bicritical fixed point, with the minima of the potential lying along the axes in field space. Rotating  this fixed point by $\pi/4$ in field space reproduces the DFP at $N=1$. 
In fact, as discussed in App.~\ref{rotatedsolutions}, this exact symmetry is only true in the LPA. 
However, within the LPA'  the rotation of the BFP by $\pi/4$ in field space  features a vanishing mixed coupling $\lambda_{1,1}=0$ at $N=1$, which is the defining property of the DFP. 
Thus, we conclude that the BFP and RDFP are degenerate in $d=3$.  In particular, they remain degenerate away from $N=1$, i.e., the existence of the BFP can be viewed as arising from the existence of the single-field Wilson-Fisher fixed point, combined with the rotational symmetry of the fixed-point equation.

This picture of stability trading and the degeneracy of the RDFP and the BFP persist down to $d\approx 2.8$, where additional fixed points appear.

\subsubsection{Appearence of new fixed points for $d<3$}
\begin{figure}[!t]
\includegraphics[width=\linewidth]{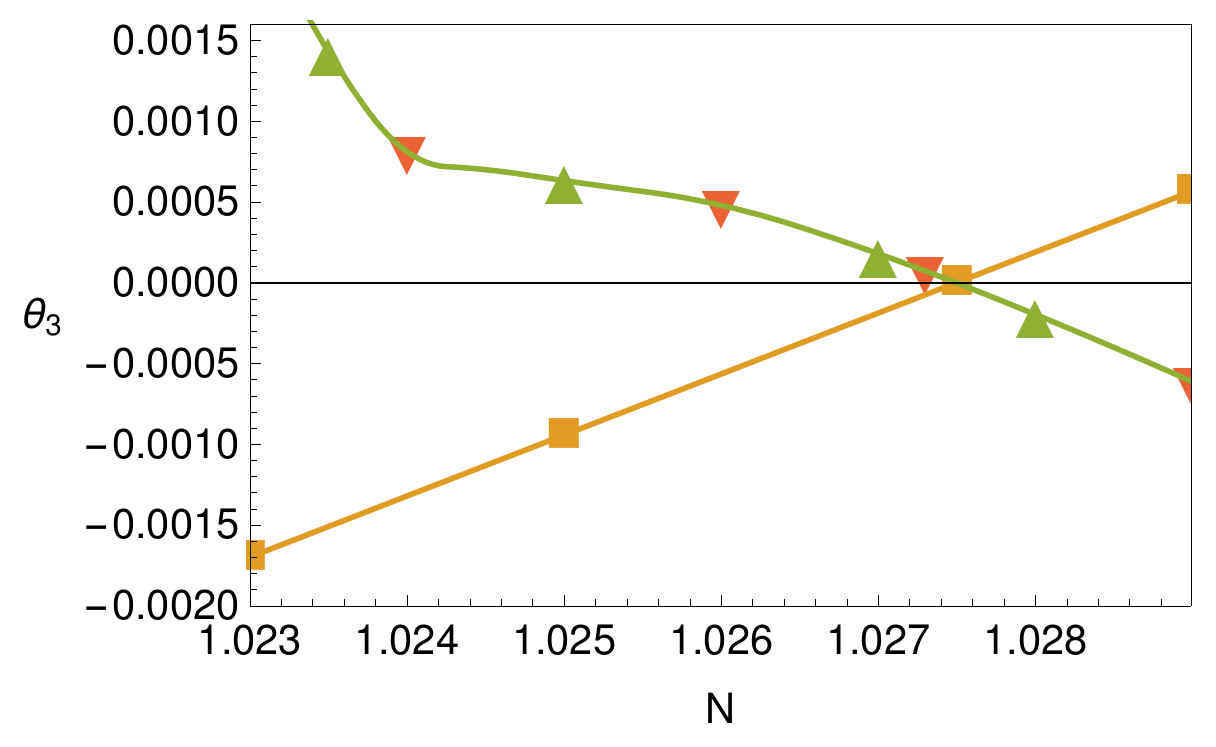}\\
\includegraphics[width=\linewidth]{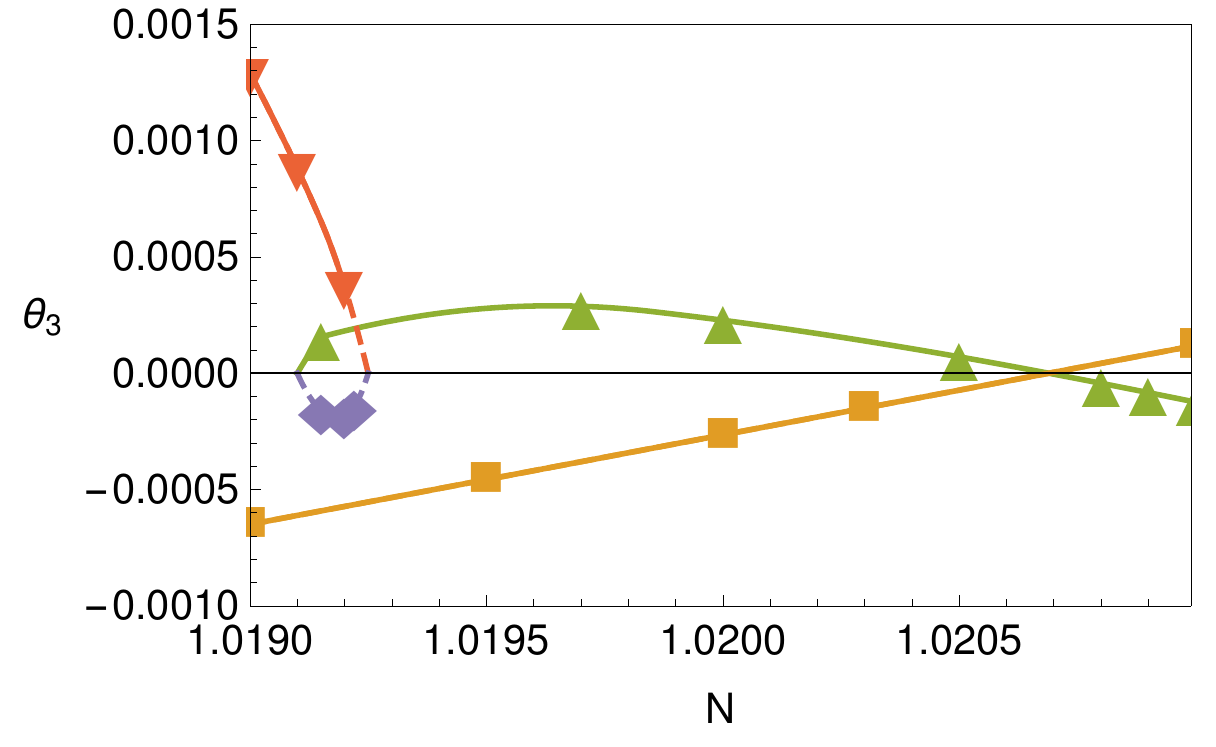}
\caption{\label{fig:newFPs}We show the third critical exponent of the IFP (orange squares) and the BFP/RDFP (green/red triangles) in $d=2.88$ within the LPA (upper panel) and $d=2.87$ (lower panel), where two additional bicritical fixed points (red inverted triangles and purple diamonds) exist. The second bicritical fixed point, indicated by red inverted triangles corresponds to the rotation of the DFP at $N=1$. In $d=2.88$, the BFP and RDFP are  still degenerate.}
\end{figure}
A pair of new fixed points appears below $d=3$: Its influence can already be detected in the upper panel of Fig.~\ref{fig:newFPs}: There, the third critical exponent of the BFP/RDFP exhibits a slight kink near $N\approx 1.024<N_{\rm crit, \, IFP}$. That kink is due to a pair of fixed points that still lies within the complex plane at that value of $d$, but already starts to approach the BFP. At a slightly lower value of $d$, that pair emerges from the complex plane  at $N=N_{\rm em}$, cf.~lower panel of Fig.~\ref{fig:newFPs}.
As soon as these new bicritical fixed points appear, the degeneracy between the BFP and the RDFP is lifted.
As $ N_{\rm crit, \, IFP}>N_{\rm em}$, the BFP is bicritical in that region as well, allowing it to collide with one of the new fixed points. At that collision point, they move off into the complex plane. Thus, one of the newly appearing fixed points has a rather short ``lifespan", emerging from the complex plane at $N_{\rm em}=1.01925$, and disappearing at  $N_{\rm ann}=1.0191$ (in LPA).  It serves as the annihilation partner of the BFP. In that process, the second new bicritical fixed point is left behind and continues to exist at lower $N$. At $N=1$, it can be rotated by $\pi/4$ in the space of fields, where it maps onto the DFP. Continuing to lower $d$, the new fixed points 
play a more important role, as they take part in stability-trading mechanisms.

\subsubsection{Separation of stability-trading mechanisms in $d\approx2.7$}
\begin{figure}[!t]
\includegraphics[width=\columnwidth]{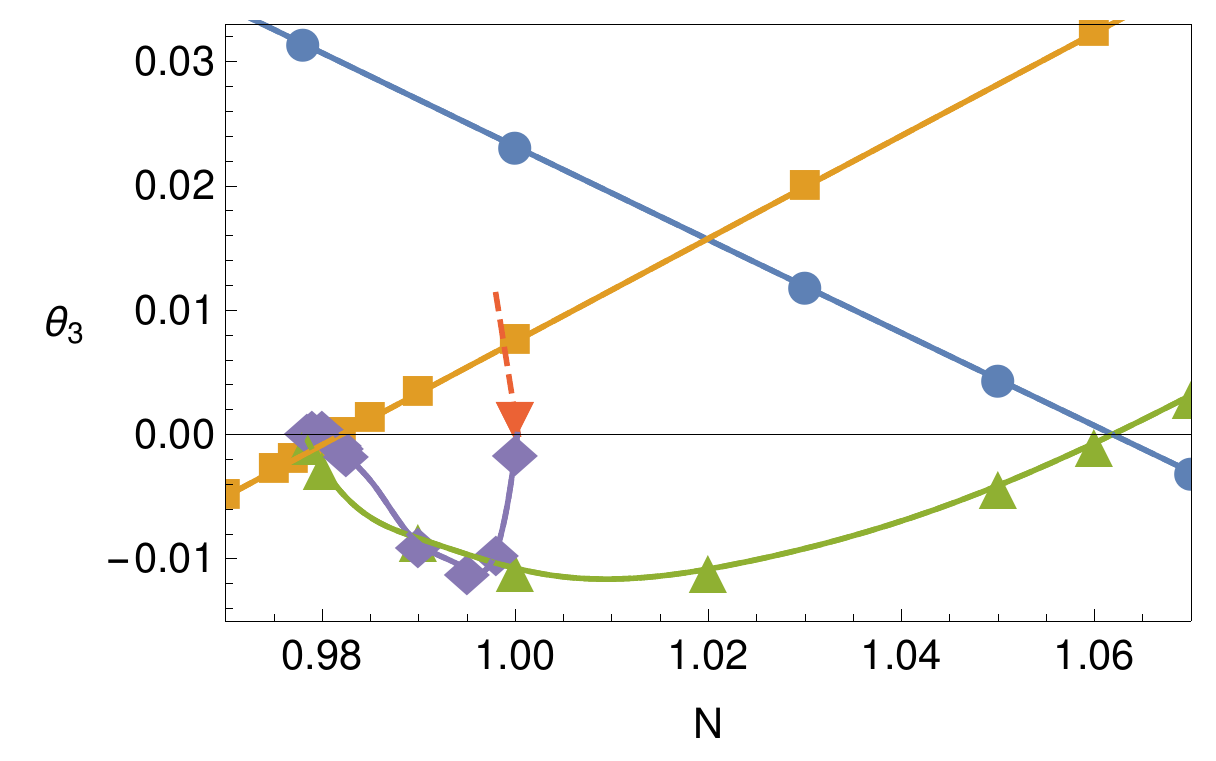}
 \caption{The third critical exponent of the DFP (blue dots), IFP (orange squares), BFP (green triangles), RBFP (purple diamonds) and RDFP (red triangles) as a function of $N$
 for $d=2.7$. The stability trading occurs between IFP and RBFP, as well as BFP and DFP, cf.~Fig.~\ref{fig:Deltad27}.}
 \label{fig:FPstructured27}
\end{figure}
\begin{figure}[!t]
\includegraphics[width=\linewidth]{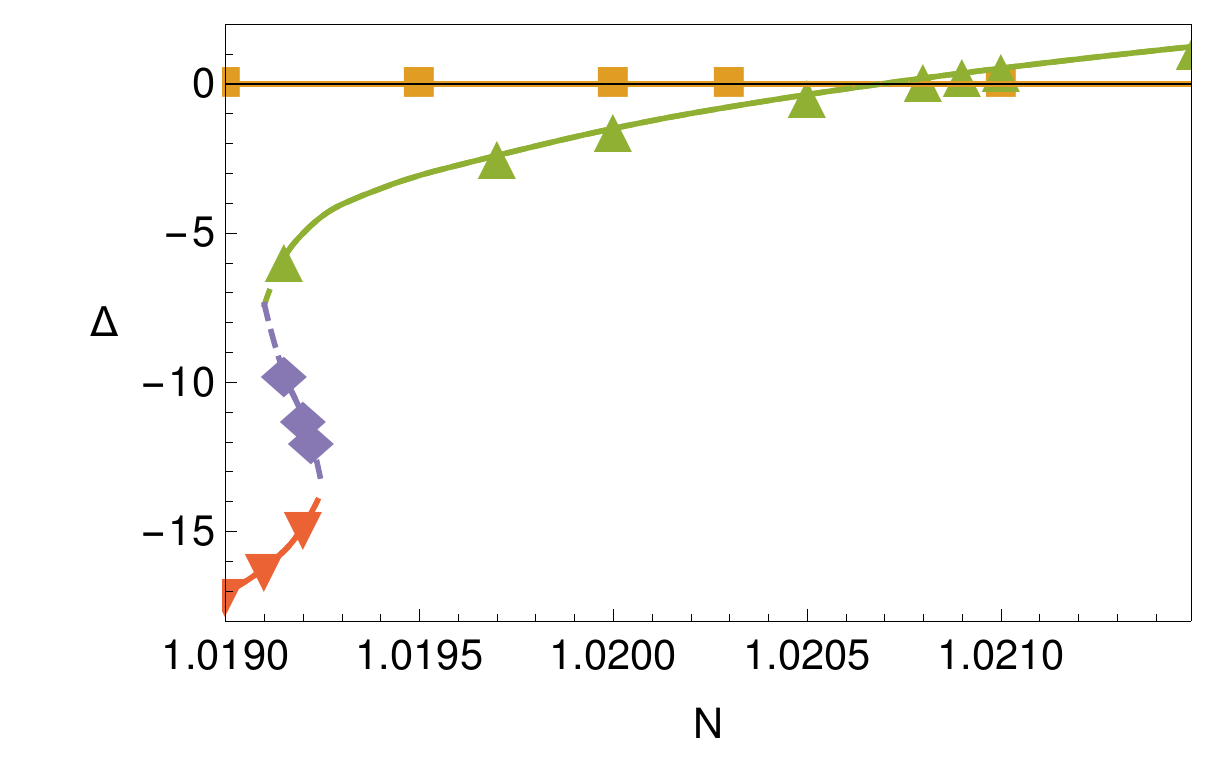}
\caption{\label{fig:Deltad287} We plot $\Delta$ in $d=2.87$ (within the LPA) as a function of $N$ for the IFP (orange squares), the BFP (green triangles), the RDFP (red inverted triangles) and the RBFP (purple diamonds). The IFP divides the bicritical region ($\Delta<0$) from the tetracritical region ($\Delta>0$).}
\end{figure}
\begin{figure}[!t]
\includegraphics[width=\linewidth]{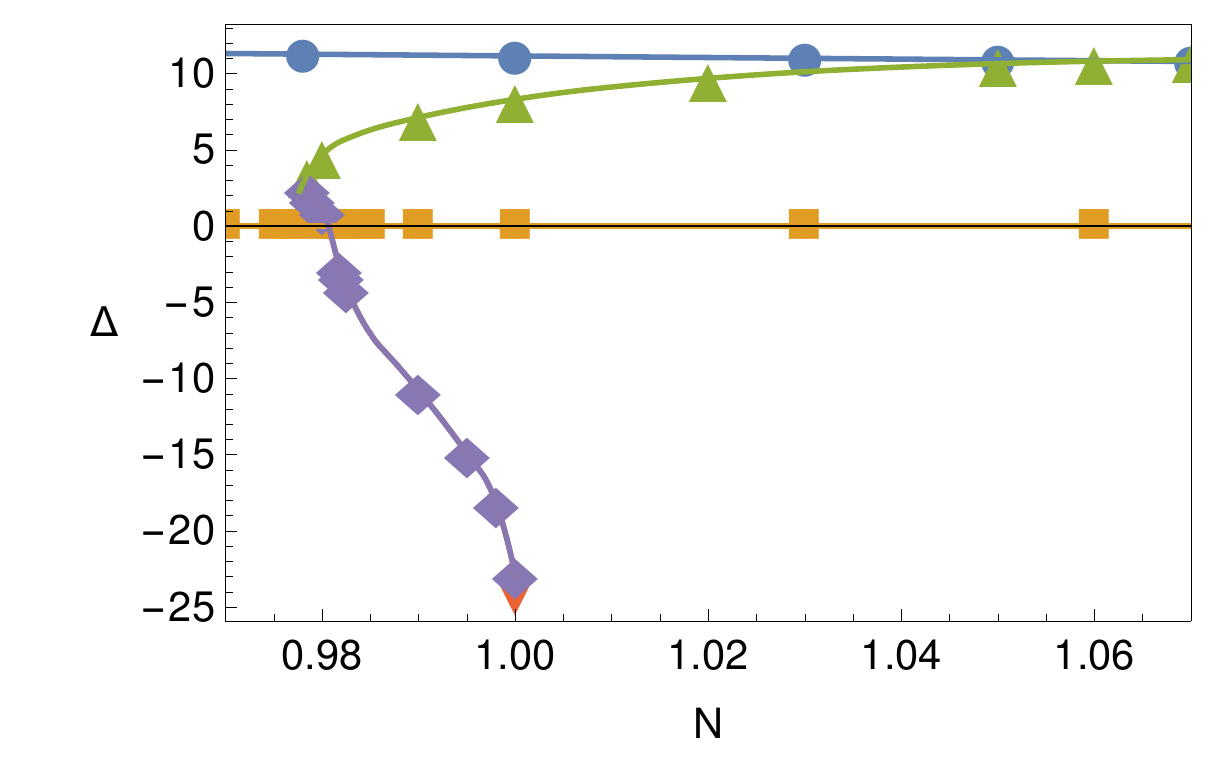}
\caption{\label{fig:Deltad27} We plot $\Delta$ in $d=2.7$ as a function of $N$ for the IFP (orange squares), which divides the bicritical region ($\Delta<0$) from the tetracritical region ($\Delta>0$). The BFP (green triangles) collides with the DFP (blue dots) at $N=N_{\rm crit, DFP}=1.06$. The IFP collides with the RBFP (purple diamonds) at $N=N_{\rm crit, IFP}=0.98$. }
\end{figure}

At $N_{\rm crit, \, DFP}$, the tetracritical BFP trades stability with the DFP for all $d\leq 3$. However, it ceases to trade stability with the IFP, which instead finds a new trading partner. Hence, the stability trading of IFP and DFP becomes disconnected, i.e., the stability is no longer transmitted between the two by a single fixed point, cf.~Fig.~\ref{fig:FPstructured27}. Instead, the trading partner of the IFP  never approaches the DFP.

Towards lower $d$, $N_{\rm crit, \, IFP}$ decreases. The major change in the stability-trading mechanism occurs when $N_{\rm crit, \, IFP}\lesssim 1$, where the IFP finds a new trading partner. 
The dynamics underlying this change of stability-trading is clearly visible by comparing Figs.~\ref{fig:Deltad287} and \ref{fig:Deltad27}: When the new bicritical fixed point first appears, it annihilates with the BFP at $N_{\rm ann}$ at a point in the space of couplings, where $\Delta<0$. Towards lower $d$, we observe that $\Delta(N_{\rm ann})$ \emph{increases}, i.e., the annihilation of the BFP occurs at larger values of $\Delta$.
Thus the annihilation point of the BFP starts to move closer to its collision point with the IFP. Finally, $\Delta(N_{\rm ann})>0$, and therefore the BFP remains tetracritical over its entire lifespan. As soon as $\Delta>0$ at $N_{\rm ann}$, the collision partner of the IFP must change. 
Hence, the new bicritical fixed point collides with the IFP, trades stability, thereby becoming tetracritical, and can 
then annihilate the BFP at $N_{\rm ann}$, where $\Delta>0$. 
For $N=1$, these two annihilation partners can actually be related by an (approximate) $\pi/4$ rotation in field space, i.e., the new collision partner of the IFP is the RBFP.

Note that within the LPA', additional fixed points appear for tiny ranges of $N$, close to the collision points of the other fixed points. These are not present within the LPA. We thus tentatively discard them as truncation artifacts, and discuss the mechanisms that we observe within the LPA, even when referring to numerical values for $N_{\rm crit}$ etc. from the LPA'.

\subsubsection{Coexisting universality classes}
In $d=2.7$, we first observe the new property of coexisting universality classes: 
Usually, one might expect that bosonic universality classes only depend on the long-range degrees of freedom, the symmetries and the dimensionality.
Here, there are two fixed points that are simultaneously stable for $0.98<N=M\lesssim1$, and underly possible continuous phase transitions. These two are the BFP,  and the newly generated bicritical fixed point, the RBFP, cf.~Fig.~\ref{fig:FPstructured27}.

These two imply very distinct phase diagrams in the vicinity of the multicritical point: The tetracritical BFP implies the existence of a fourth, mixed phase, however the other fixed point is bicricitcal, preventing the formation of a mixed phase. 
 To decide which of the two stable fixed points is  dominant for low-energy physics, 
 we conjecture that  the sign of $\Delta$ at the extremum/saddle point is all the additional information that is required in this case.  
We assume that microscopic models with $\Delta>0$  most likely flow towards the BFP, and exhibit tetracritical behavior. Microscopic models with $\Delta<0$ conversely flow towards the RBFP and exhibit bicritical behavior. 
It would be interesting to understand whether $\Delta$ corresponds to a microscopic parameter in realistic models, or whether it can be related to a macroscopic parameter, just as the mass-like couplings in these models can be related to the temperature or magnetic field.

A related property, namely that universality classes can depend on the presence of unbroken ``spectator symmetries" has been discussed in \cite{Gehring:2015vja}. In fermionic system, coexisting universality classes are a common phenomenon, see, e.g., \cite{Gies:2003dp,Gies:2010st,Braun:2014wja,Esbensen:2015cjw}. In these cases, there is one unique fixed point with the smallest number of relevant directions, but there exist several fixed points with one relevant direction.
To our best knowledge, ours is the first example of coexisting universality classes in bosonic systems. Moreover, the coexisting universality classes are \emph{stable} fixed points.

\subsubsection{Stability trading in $d=2.5$}
\begin{figure}[!t]
\includegraphics[width=\linewidth]{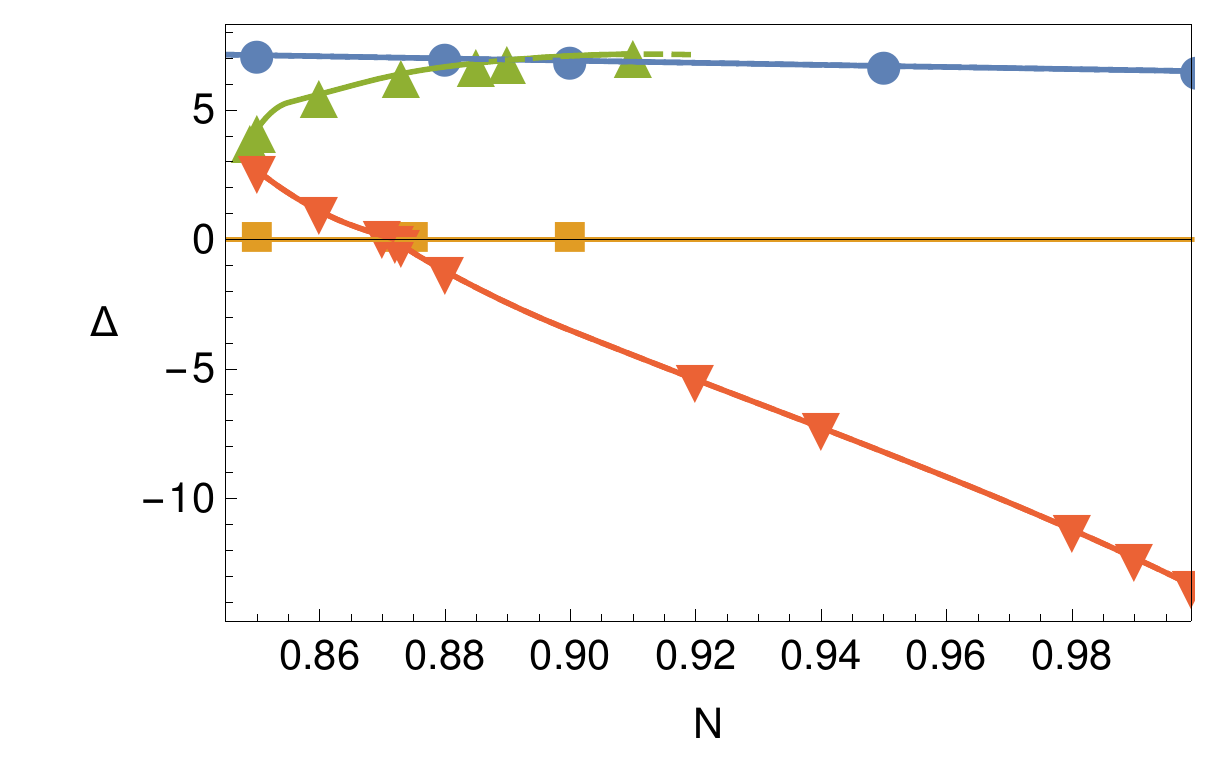}
\caption{\label{fig:Deltad25} We plot $\Delta$ in $d=2.5$ as a function of $N$ for the IFP (orange squares), which divides the bicritical region ($\Delta<0$) from the tetracritical region ($\Delta>0$). The BFP (green triangles) collides with the DFP (blue dots) at $N=N_{\rm crit, DFP}=0.89$. The IFP collides with the RDFP (red inverted triangles) at $N=N_{\rm crit, IFP}=0.87$. }
\end{figure}
\begin{figure}[!t]
\includegraphics[width=\columnwidth]{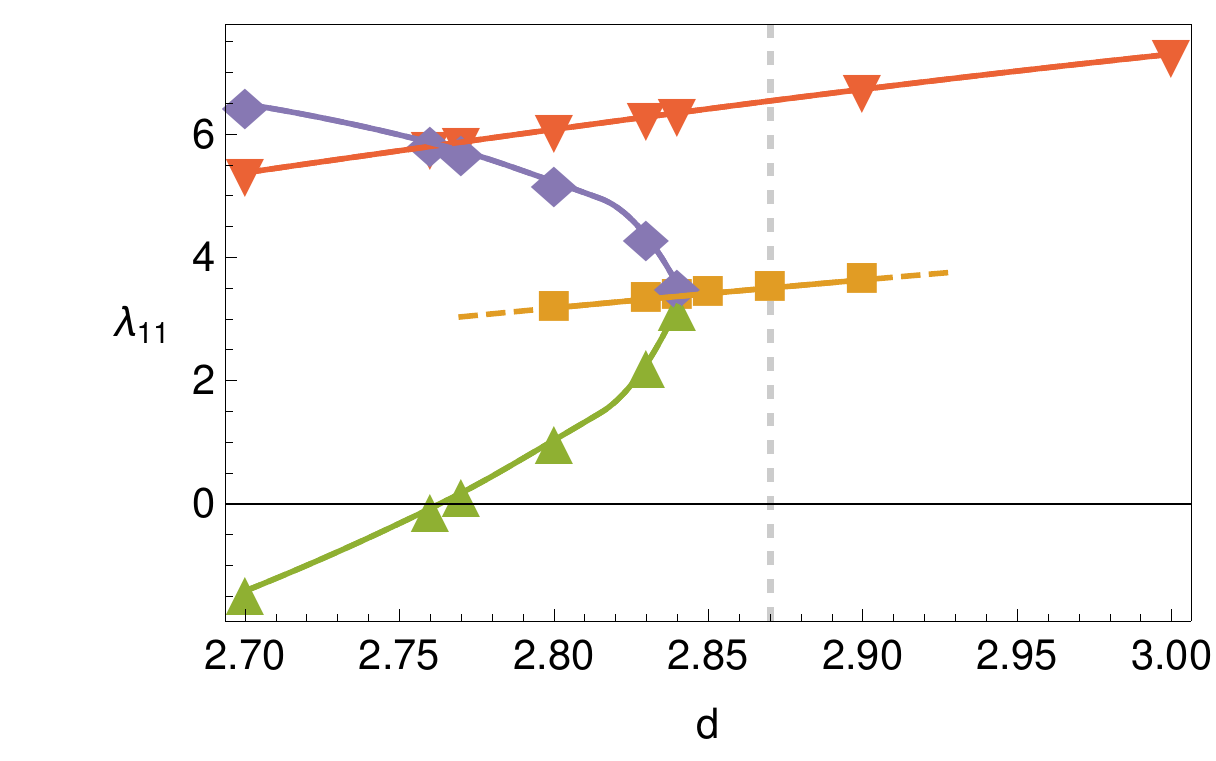}
 \caption{The coupling $\lambda_{1,1}$ of the IFP (orange squares), BFP (green triangles), RBFP (purple diamonds) and RDFP (red triangles) as a function of $d$
 for $N=1$ (within the LPA). 
The dashed line shows the point at which the two bicritical fixed points become complex. The plot summarizes the main changes in the stability trading: As $N_{\rm ann}= N_{\rm crit, \, IFP}=1$, the stability-trading-partner of the IFP changes from the BFP to the RBFP. When the RBFP and the RDFP become degenerate, the IFP again changes the stability-trading partner from the RBFP to the RDFP. }
 \label{fig:lambda11}
\end{figure}

The only change between $d \approx 2.7$ and $d\lesssim 2.5$ lies in the properties of the IFP's collision partner: At $N=1$, the collision partner of the IFP is always bicritical, cf.~Fig.~\ref{fig:Deltad25}, and can thus be related to one of the two tetracritical fixed points, the DFP or the BFP, by an (approximate) $\pi/4$ rotation in field space. In $d\approx 2.7$, that relation is with the BFP, i.e., the collision partner of the IFP is the RBFP. Towards $d\approx 2.5$, $N_{\rm crit, \, DFP}$ approaches 1, i.e., the BFP and the DFP become more similar to each other at $N=1$. Accordingly, so do their rotated counterparts, the RBFP and the RDFP. For $d=2.63$ (within the LPA'), $N_{\rm crit, \, DFP}=1$, i.e., the DFP and BFP lie on top of each other. At that point, the rotated counterparts must be degenerate as well\footnote{Actually, this happens slightly before $N_{\rm crit, \, DFP}=1$ at $d\approx2.69$ in LPA', since the rotation in field space is not exact.}. Thus, at lower $d$, the 
bicritical stability-trading partner of the IFP is the RDFP, cf.~Fig.~\ref{fig:lambda11}.

\begin{figure}[!t]
 \includegraphics[width=\columnwidth]{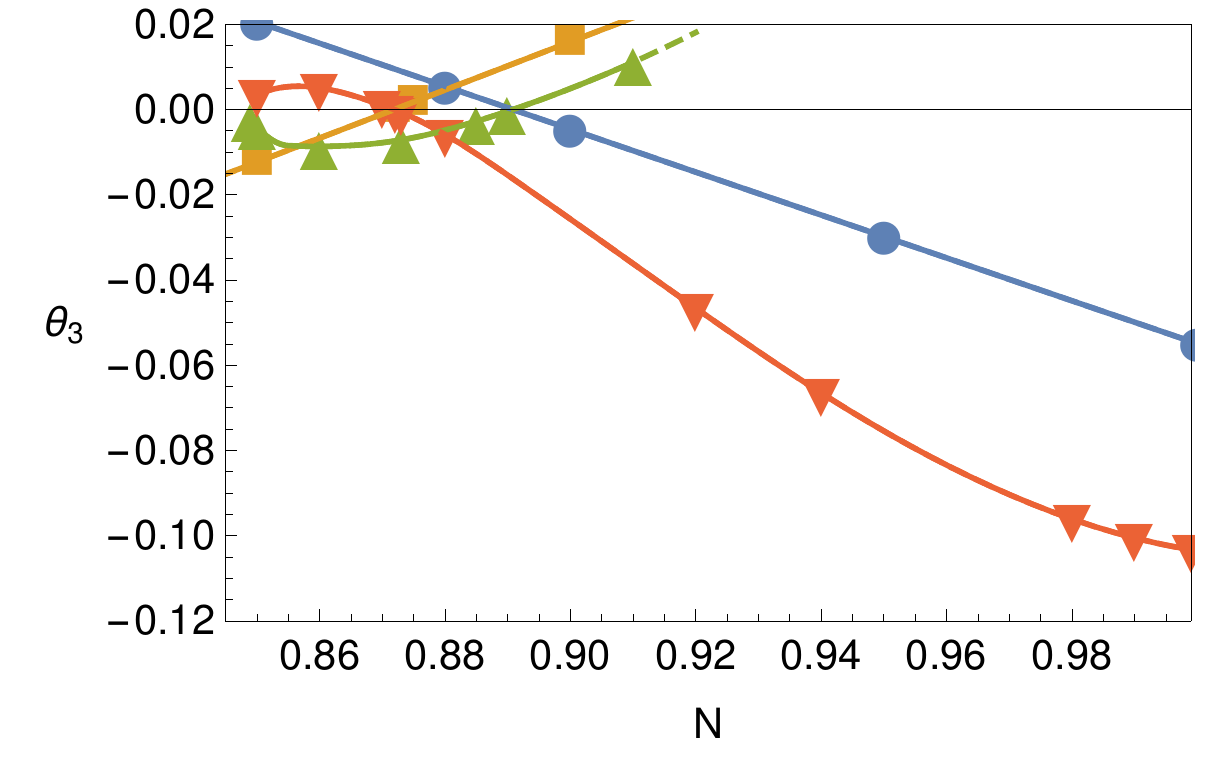}
 \caption{The third critical exponent of the DFP (blue dots), IFP (orange squares), BFP (green triangles) and RDFP (red inverted triangles) as a function of $N$
 for $d=2.5$.}
 \label{fig:FPstructured25}
\end{figure}

There is a significant region in which more than one fixed point is stable, cf.~Fig.~\ref{fig:FPstructured25}. In particular, there are ranges of $N$ for which the DFP is stable simultaneously with the RDFP.  Moreover, the IFP is stable simultaneously with the BFP. Again, there is never more than one stable fixed point
in each of the three separate regions of the theory space, $\Delta>0$, $\Delta=0$ and $\Delta<0$ (imposed as global conditions). 
In the case of the  IFP being stable, more information is required than only the sign of $\Delta$ to determine which of the possible universality classes dominates low-energy physics since
the hypersurface $\Delta=0$ may be attractive also for theories starting at $\Delta \gtrless 0$.
The coexistence of two universality classes, that we have first observed in $d \approx 2.7$, continues to the case $d =2.5$. In particular, it now includes the region $N=M=1$, at least in our approximation.

\subsubsection{Stability trading towards $d=2$: Overlapping stability regions of the DFP and the IFP}

\begin{figure}[!t]
\includegraphics[width=\linewidth]{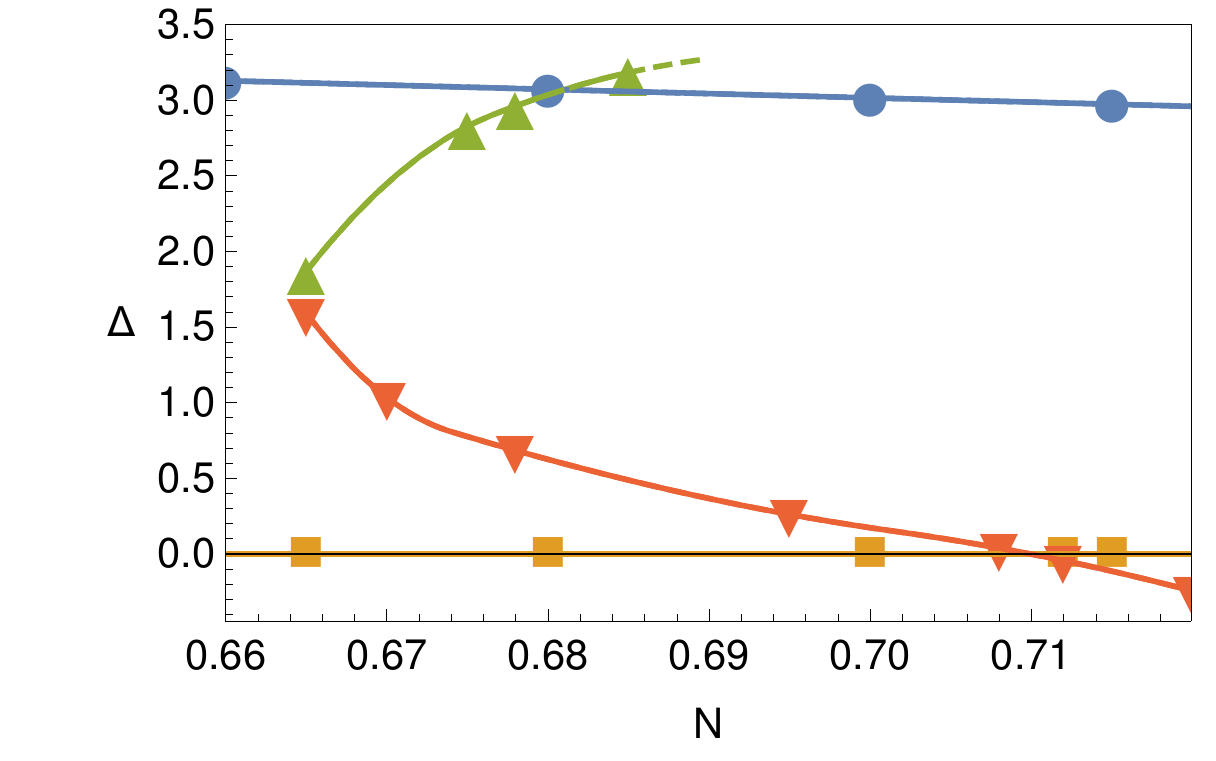}
\caption{\label{fig:Deltad22} We plot $\Delta$ in $d=2.2$ as a function of $N$ for the IFP (orange squares), which divides the bicritical region ($\Delta<0$) from the tetracritical region ($\Delta>0$). The BFP (green triangles) collides with the DFP (blue dots) at $N=N_{\rm crit, DFP}=0.68$. The IFP collides with the RDFP (red inverted triangles) at $N=N_{\rm crit, IFP}=0.71$.}
\end{figure}

Towards lower $d$,  
$N_{\rm crit,\, IFP}$ and $N_{\rm crit,\, DFP}$ approach each other, cf.~Fig.~\ref{fig:stabilityregions}. 
Finally, both stability regions touch and then start to overlap.
Note that the DFP and the IFP can never be degenerate. Whereas the IFP still lies in the plane of enhanced symmetry $\Delta=0$, the DFP must stay tetracritical $\Delta>0$.
Thus, collisions between them are excluded. This also holds for the case, where both scaling solutions change stability at the same value of $N$. Viewed in the space of couplings, the two stability-trading fixed-point collisions occur at rather different positions,  similar to the case shown in Fig.~\ref{fig:Deltad22}, even if they accidentally occur at the same value of $N$.

\begin{figure}[!t]
 \includegraphics[width=\columnwidth]{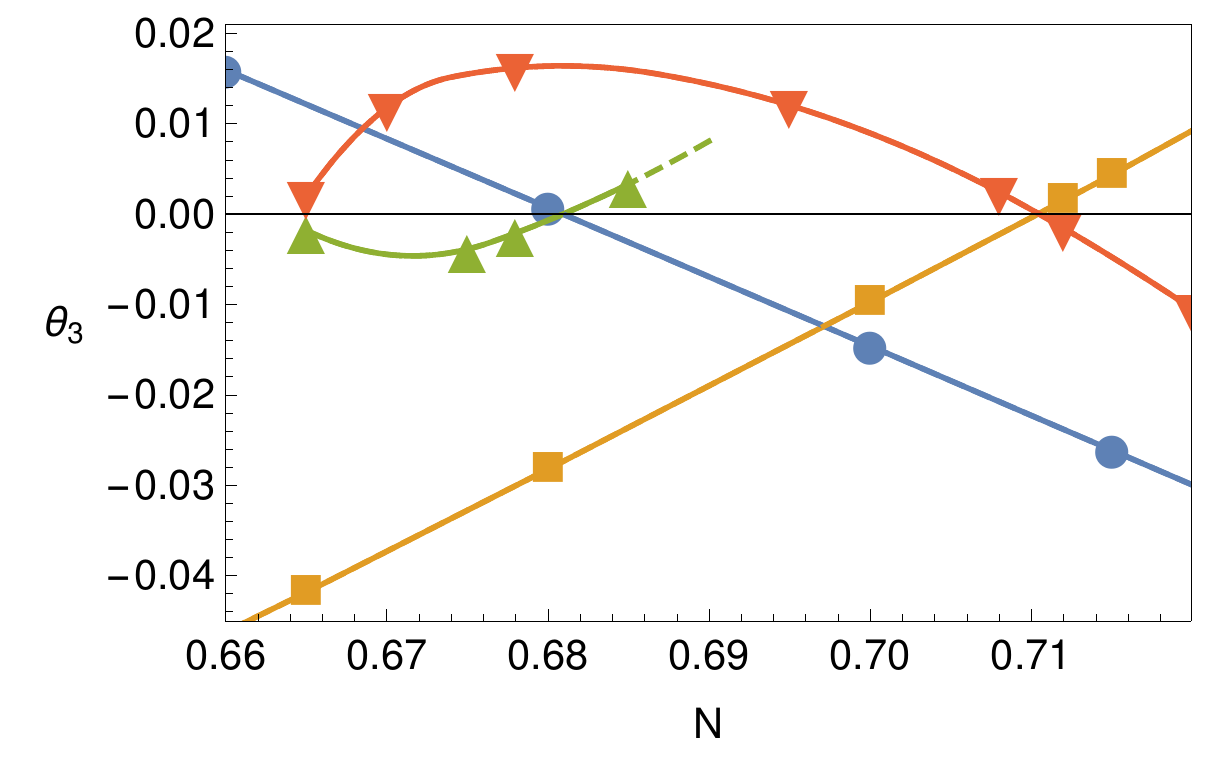}
 \caption{The third critical exponent of the DFP (blue dots), IFP (orange squares), BFP (green triangles), and RDFP (red inverted triangles) as a function of $N$
 for $d=2.2$. }
 \label{fig:FPstructured22}
\end{figure}

The collision partners of the IFP and DFP are the RDFP and BFP, respectively.
Fig.~\ref{fig:FPstructured22} depicts the situation where both stability regions overlap $N_{\rm crit,\, DFP}<N_{\rm crit, \,IFP}<1$ at $d=2.2$.
Thus, the IFP and the DFP feature an overlapping region of stability, cf.~Fig.~\ref{fig:stabilityregions}, which is another  instance of two coexisting universality classes. Note that our estimates for $N_{\rm crit,\, I/DFP}$ are not yet quantitatively exact, cf.~Sec.~\ref{sec:check}.

Fig.~\ref{fig:stabilityregions} summarizes the stability regions of the IFP, DFP and BFP in LPA and LPA'.
 Where it exists, the stability region of the new, stable, bicritical fixed point is depicted as well. Note that this bicritical fixed point is called RBFP at first, and later RDFP, as it corresponds to a rotation of the BFP, or DFP at $N=1$, respectively.
For $N_{\rm crit,\, IFP}>1$, the stabilization of the IFP and the destabilization of the BFP occur simultaneously. Close to $N_{\rm crit,\, IFP}=1$,
the degeneracy of the BFP and RDFP is lifted (appearence of mauve region), and regions of simultaneously stable fixed points appear. For even lower $d$, the BFP does not interact with the IFP anymore, thus the two lines separate. 

\subsection{Testing the quantitative reliability of our results}\label{sec:check}
\begin{figure}[!t]
 \includegraphics[width=\columnwidth]{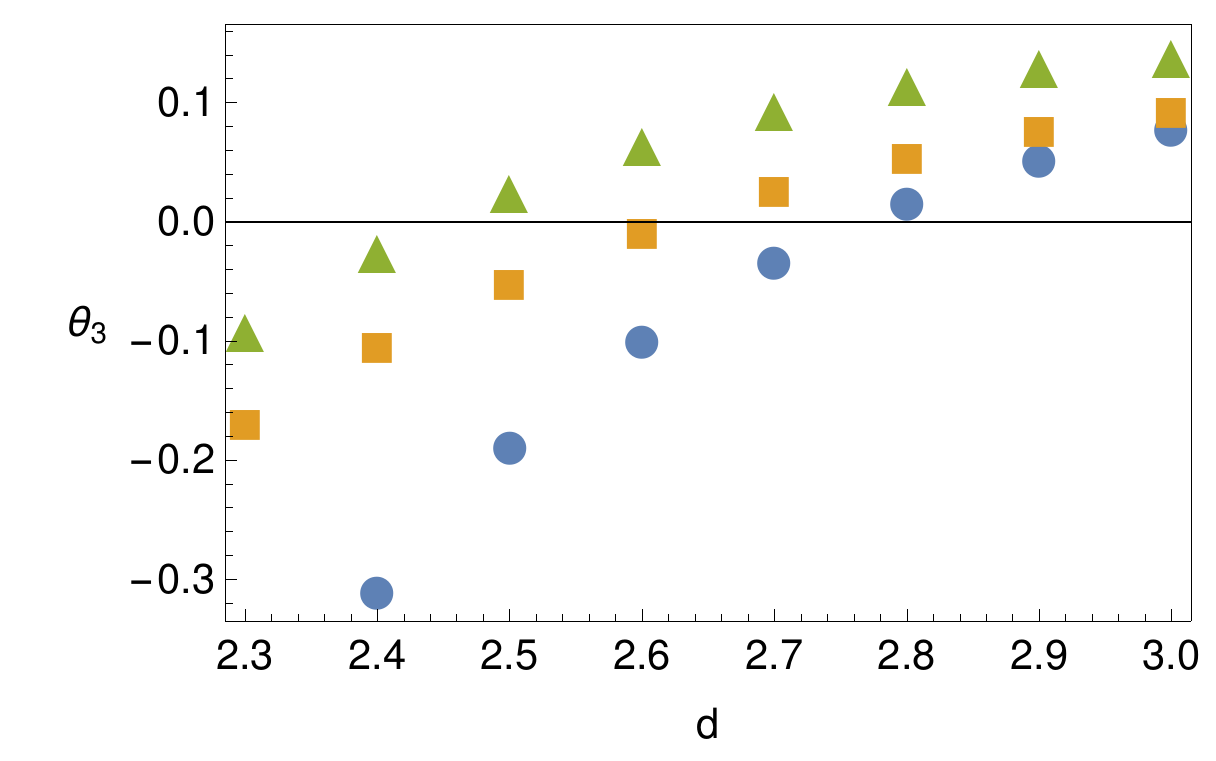}
 \caption{\label{fig:theta3DFP}The third largest critical exponent of the DFP for $N=M=1$ as a function of $d$ is depicted.
 The symbols (colors) indicate different truncations: The blue dots correspond to LPA, the orange squares to LPA' and the green triangles to LPA' but employing the scaling relation \eqref{eq:scalerelation}.}
\end{figure}

The LPA and LPA', combined with a Taylor expansion, give rather good results in $d=3$, \cite{Eichhorn:2013zza}, in comparison to high orders of the loop expansion \cite{Folk2008, Calabrese:2002bm}, at a very manageable computational complexity. 
Towards $d=2$, the Taylor expansion breaks down and we therefore resort to global methods.
Moreover, momentum-dependence is becoming more important, indicated, e.g., by the growth of the anomalous dimension.
Thus, we expect that our estimates for $N_{\rm crit,\, IFP}$ and $N_{\rm crit,\, DFP}$ are not accurate in the limit $d\rightarrow 2$. 

Comparing LPA and LPA' in Fig.~\ref{fig:stabilityregions}, the stability boundaries are shifted to lower values of $d$ in LPA'.
The point where the DFP and IFP stability lines intersect lies at a different value of $d$, but at a similar value of $N$.
Taking these observations into account, it is an interesting question, how far the stability lines are shifted to lower dimensions when the order of the derivative expansion is increased.

To judge the quantitative reliability of our results, we use the Onsager solution for the Ising model in $d=2$.
We can combine the scaling relation \eqref{eq:scalerelation} for the DFP with the Onsager solution in $d=2$ \cite{PhysRev.65.117}
\be
\nu_{\rm Onsager} =1,
\ee
to obtain
\be
\theta_{3}(d=2) = 1+1 -2 =0.\label{eq:theta3Onsager}
\ee
Thus, the DFP is on the verge of stability for the $\mathbb{Z}_2 + \mathbb{Z}_2$ model in $d=2$.
As $\theta_3>0$ for the DFP in $d=3$, and $\theta_3=0$ in $d=2$, a monotonic dependence on $d$ would suggest that $\theta_3>0$ for $2<d<3$.  
In our approximation, $\theta_3$ changes sign above $d=2$, cf.~Fig.~\ref{fig:theta3DFP}. 
As we improve the approximation from LPA to LPA' and employ the scaling relation, $d_{\rm crit}$ decreases, as expected. 
We find $d_{\rm crit} \approx 2.45$ and $\theta_3\approx 0.5$ at $d=2$, implying that our results are not yet quantitatively precise.
Extended truncations with momentum-dependent interactions are expected to improve these results.

\subsection{The $\eta$ conjecture}
\begin{table}[!top]
\begin{tabular}{c c c c c c}
$d$ & $N$ & fixed point & sign($\Delta$)&$\theta_3$& $\eta_{\phi/\chi}$\\\hline\hline
3 & 1&BFP(RDFP) & -& 0.0853 & 0.0243 \\\hline
3 & 1 & DFP &+ & 0.1407 (0.0897) & 0.0443 \\\hline
3 & 1&IFP & 0& -0.0418 & 0.0437 \\\hline\hline
3 &1.5& DFP &+& 0.0233 (-0.0259) & 0.0444 \\\hline
 3 &1.5& IFP& 0 & 0.0864 & 0.0409 \\ \hline
3 &1.5& BFP&+& 0.0324 & 0.0460 \\ \hline\hline\hline
2.7 &1& DFP &+& 0.0960 (0.0232) & 0.0841 \\\hline
2.7 &1& BFP&+& -0.0108 & 0.0789 \\ \hline
2.7 &1& RBFP &-& -0.00138 & 0.0491 \\\hline
2.7 &1& RDFP &-& 0.00147 & 0.0480 \\\hline
2.7 &1& IFP & 0& 0.00748 & 0.0801 \\\hline \hline\hline
2.5 &0.88& DFP &+ & 0.0906 (0.00546) & 0.119 \\\hline
2.5 &0.88& BFP &+ &-0.00483 & 0.118 \\\hline
2.5 & 0.88&RDFP & -&-0.00583 & 0.108\\\hline
2.5 & 0.88&IFP & 0&0.00459 & 0.113 \\\hline\hline
\end{tabular}
\caption{\label{tab:eta} We show the value of $\theta_3$ and $\eta$ at selected values of $N$ for the fixed points in different dimensions.
For the DFP, we give $\theta_3$ derived from the scaling relation \eqref{eq:scalerelation} and by taking the variation of $\eta$ into account (value in brackets).}
\end{table}

The $\eta$ conjecture has been put forward in \cite{Vicari:2006xr}: It states that there is always exactly one stable fixed point, which is the one with the largest value of the anomalous dimension. The following argument supports the conjecture: The beta function of a coupling $\lambda_{i,j}$ is of the form
\be
\beta_{\lambda_{i,j}} = d_{\bar{\lambda}_{i,j}} + \left(i \, \eta_{\phi}+ j \, \eta_{\chi} \right)\lambda_{i,j}+...\, .\label{eq:betafunctionschematic}
\ee
Additional terms in \eqref{eq:betafunctionschematic} arise from the evaluation of the flow equation and encode the effect of fluctuations at an interacting fixed point. While these terms are necessary for the existence of the fixed point, we concentrate on understanding the effect of the anomalous dimensions on the critical exponents. Ignoring that eigendirections of the stability matrix can be superpositions of couplings, the contribution of the anomalous dimensions to the critical exponent associated to $\lambda_{i,j}$ would be
\be
\theta_{\lambda_{i,j}}\Big|_{\eta}=- (i\, \eta_{\phi}+ j\, \eta_{\chi})+...\, \, .
\ee
Clearly, increasing $\eta_{\phi/\chi}$ lowers the value of $\theta_{\lambda_{i,j}}$, making it more likely that $\theta_{\lambda_{i,j}}<0$, and thus, indeed rendering the fixed point more likely to be stable. 

In $d=3$ dimensions, the existence of maximally one stable fixed point is true both for systems with two, as well as three or four order parameters. In fact, we observe that our results mostly support the $\eta$ conjecture in $d=3$. For instance, the IFP is stable at $N=1$, where it has a larger anomalous dimension than the BFP(RDFP), cf.~Tab.~\ref{tab:eta}. Comparing our results for the $\eta$s to those from Monte-Carlo simulations for the Wilson-Fisher fixed point, we observe deviations of about 25 \%. Thus, the difference between the value of $\eta$ for the IFP and the DFP in $d=3$ at $N=1$ is well within this systematic error.

For $d<3$, regions with more than one stable fixed point appear. Therefore, the $\eta$ conjecture can no longer hold in its simple form. On the other hand, in each subspace of the theory space defined by the sign of $\Delta$, i.e., in the region $\Delta>0$, $\Delta<0$ and $\Delta=0$, there is never more than one stable fixed point. Within our current truncation, we do not expect a sufficiently precise determination of the anomalous dimension. We thus leave the conjecture that each subspace of the theory space features never more than one stable fixed point, and that it is the one with the largest anomalous dimension, to be tested by future work.

\section{Conclusions}
\label{sec:conclusions}
\subsection{Summary}
We have shown that  functional RG methods, that allow to access fixed points in models with several interacting scalars with O($N$) $\oplus$ O($M$)-symmetry in $d\geq 3$, can also be extended to $d<3$ dimensions.
Here, the anomalous dimension becomes more important, and truncations at finite order in the effective potential are unreliable. This necessitates more advanced techniques, and
we apply pseudo-spectral methods to 
control 
the full effective potential. 
As both $d$ and $N=M$ can be treated as continuous parameters in the functional RG, we explore $2<d\leq 3$ and $N=M$ in the vicinity of $N=M \approx 1$.

The system is dominated by an interplay of several fixed points, which trade stability (and thus, physical relevance for the critical behavior of different systems) at fixed-point collisions.
Within our  functional RG formulation, we find a major difference between $d=3$ and $d=2$: Whereas in $d=3$ the destabilization of the isotropic, symmetry-enhanced IFP and the stabilization of the decoupled, tetracritical DFP occur in collisions with the same fixed point, the BFP, there are several fixed points involved in $d \lesssim 2.7$: While the stabilization of the DFP still occurs in a collision with the tetracritical BFP, the IFP is destabilized by a collision with the bicritical RDFP. As these two collisions are independent, the stability regions of IFP and DFP can shift towards each other. 
We observe regions of simultaneously stable fixed-point solutions, i.e., coexisting universality classes. For instance, in $d<2.4$, we find overlapping stability regions of the DFP and IFP. Interestingly, there is always only one stable fixed point in each of the separate regions of the theory space, defined by the sign of $\Delta$. 
To explore, whether the sign of $\Delta$ in the UV is sufficient to determine which of the different possible universality classes is realized in the IR, studies of RG trajectories are necessary. This is technically more involved than the search for fixed-point solutions that we have conducted here, and is thus left for future work.

Our main results are shown in Fig.~\ref{fig:stabilityregions}, where the DFP is stable at $N=M=1$, suggesting tetracritical behavior and the existence of a mixed phase for two coupled Ising models.
Comparing our results with the stability region for the DFP as implied by the exact Onsager solution for the Ising model and Aharony's scaling relation, we observe that our estimate for $N_{\rm crit,\, DFP}$, where  the DFP is stabilized, is off by about 0.45. Thus, our truncation is insufficient to obtain quantitative precision. On the other hand, we conjecture that the mechanisms that we observe here might only be shifted to lower $d$, however, the coexisting universality classes could remain. The region of overlapping stability of the IFP and the DFP, located at $0.55<N=M<0.6$ in our case, might also persist but be shifted to $N=M \approx 1$. Alternatively, the stability regions could be shifted independently, and the overlapping stability of IFP and DFP  might not persist.
In the former case, the system of two coupled Ising models in two dimensions might either exhibit a tetracritical phase diagram with a mixed phase, or a multicritical point associated to a Kosterlitz-Thouless type phase transition. Note that exactly in $d=2$, the description of BKT-like behavior within the functional RG is possible, but requires some care \cite{Grater:1994qx,VonGersdorff:2000kp,Jakubczyk:2014isa,Jakubczyk:2016sul}. For the DFP, as $\theta_3=0$, generically a third parameter would have to be tuned in order to approach criticality. If the coexistence of several universality classes persists, the RDFP or the RBFP might potentially provide further candidates for stable fixed points.

Our conjecture could be tested employing, e.g., lattice simulations.

\subsection{Outlook}

To reach quantitative precision, momentum-dependent interactions should be taken into account. As a first step, the next order in a derivative expansion, $\mathcal{O}(\partial^2)$ should be added, resulting in
two coupled differential equations for the potential $u(\rho_{\phi}, \rho_{\chi})$ and the momentum-dependent interaction, $y(\rho_{\phi}, \rho_{\chi})$.

We have established that the functional RG can be used to probe the physics of coupled order parameters in $d=2$.
Thus, the case $N=1, M=2$, corresponding to anisotropic antiferromagnets, can now be studied with this method, and the nature of the multicritical point, explored in \cite{2007PhRvB..76b4436P,2007PhRvB..76v0405H}, see also references therein, can be further clarified, and connected to experimental results in quasi-two-dimensional systems \cite{1982JAP....53.7963D,1985SSCom..53..737D,1986PhyBC.141....1D,1986JPhC...19.4503R,1993ZPhyB..93....5C,1997PhyB..241..570V,2001PhRvB..63n0401C}.

For $d<3$, the single-sector O($N$) models feature multicritical, i.e., unstable points which can be discovered with the functional RG, see \cite{Morris:1994jc} as well as \cite{Codello:2012sc,Codello:2014yfa,Codello:2012ec,Borchardt:2015rxa}. The $i$th multicritical point appears in $d=d_{c,i}$. Within the two-field models in $d<3$, these scaling solutions can be combined to give decoupled fixed points, where at least one of the two separate sectors approaches one of the multicritical single-field fixed points. Moreover, a ``multicritical" generalization of the isotropic fixed point also exists.
In particular, we anticipate that there 
 are
new fixed points, such as generalizations of the BFP, which are unique to two-sector models. Thus, stability-trading mechanisms as the ones that we have discussed here, could also be relevant for each of the multicritical sectors. It would be interesting to explore this conjecture further, and it is technically possible with the method that we have further developed here.

\begin{acknowledgements}
We are indebted to I.~Boettcher and M.~M.~Scherer for collaboration during the initial stages of this project.
Furthermore, we thank B.~Knorr and H.~Gies for valuable discussions and H.~Gies, I.~Boettcher and M.~M.~Scherer for helpful comments on the draft.
 Parts of the pseudo-spectral code were co-produced by B.~Knorr.
J.~B.~acknowledges support by the DFG under Grants No.~GRK1523/2 and No.~Gi328/7-1. The work of A.~E.~is supported by an Imperial College Junior Research Fellowship.
\end{acknowledgements}

\appendix
\section{$\pi/4$ rotational symmetry of the fixed-point equation for $N=M=1$ for the LPA}\label{rotatedsolutions}

In this section, we discuss the $\pi/4$ rotational symmetry for the functional RG fixed-point equation \eqref{eq:potflow}.
For $N=M$, it gives rise to an exchange symmetry under $\phi\leftrightarrow \chi$. 
We find solutions exhibiting this symmetry as well as solutions which do not, e.g., the DGFPs.
Such solutions emerge in pairs which transform into each other under $\phi\leftrightarrow \chi$.
Thus, the complete spectrum of solutions is invariant under the exchange symmetry.

Now, we set $\eta_\phi,\eta_\chi=0$ and specialize to $N=M=1$.
Let us assume that $u(\rho_\phi,\rho_\chi)_*$ is a solution of \eqref{eq:potflow}.
Inserting the $\pi/4$-rotation \eqref{eq:coordrot} of $u(\rho_\phi,\rho_\chi)_*$,
\be
\tilde u(\rho_\phi,\rho_\chi)_{*} = u\left(\frac{\rho_\phi+\rho_\chi+2\sqrt{\rho_\phi\rho_\chi}}{2},\frac{\rho_\phi+\rho_\chi-2\sqrt{\rho_\phi\rho_\chi}}{2}\right)_*,\label{appeq:rotatedsoln}
\ee
into \eqref{eq:potflow}, it  becomes clear that $\tilde u(\rho_\phi,\rho_\chi)_{*}$ also satisfies the fixed-point equation.
As the IFP is invariant under such a transformation, it is rotated into itself.
For the BFP and the DFP the transformation \eqref{eq:coordrot} turns a tetracritical fixed point  with $\Delta>0$ into a bicritical one resulting in two distinct solutions.

Let us take a closer look at those solutions which do not respect the exchange symmetry, $u(\rho_\phi,\rho_\chi)_* \neq u(\rho_\chi,\rho_\phi)_*$.
Here, $\tilde u(\rho_\phi,\rho_\chi)_{*}$ denotes the formal rotation of $u(\rho_\phi,\rho_\chi)_{*}$.
 For any solution of \eqref{eq:potflow}, the first derivatives $\partial_\phi u(\rho_\phi,\rho_\chi)_{*}$ and $\partial_\chi u(\rho_\phi,\rho_\chi)_{*}$ have to vanish at the boundaries $\phi=0$ and $\chi=0$. 
This is required by the  $\mathbb{Z}_2$ reflection symmetry in $\phi$ and $\chi$, respectively: For any smooth solution that is symmetric under a reflection in $\phi$, the derivative must vanish at $\phi$=0. 
Rotations of solutions which do not respect the $\phi \leftrightarrow \chi$ exchange symmetry violate that boundary condition:
For the rotated function $\tilde{u}$, the boundary condition, using \eqref{appeq:rotatedsoln}, becomes
\bea
\partial_{\phi} \tilde{u}(\rho_{\phi}, \rho_{\chi})|_{\rho_\phi=0}&=& \sqrt{\frac{\rho_{\chi}}{2}} \big(\partial_{x}u(x, y) \nonumber\\
&{}& -\partial_{y}u(x, y) \big)|_{x=\rho_\chi/2,y=\rho_\chi/2}\nonumber\\
&\overset{!}{=}& 0.\label{appeq:bc}
\eea
The exchange symmetry of $u(\rho_\phi,\rho_\chi)_*$ would imply that $\partial_{\rho_\phi} u(\rho_\phi,\rho_\chi)_{*}|_{\rho_\phi=\rho_\chi} = \partial_{\rho_\chi} u(\rho_\phi,\rho_\chi)_{*}|_{\rho_\phi=\rho_\chi}$. Using this condition in \eqref{appeq:bc} allows us to conclude that
\be
\partial_{\phi}\tilde{u}(\rho_{\phi}, \rho_{\chi})= 0,
\ee
 if and only if $u$ preserves the exchange symmetry.
Thus, $\tilde u(\rho_\phi,\rho_\chi)_{*}$ cannot be a solution of \eqref{eq:potflow}, unless the original solution $u(\rho_{\phi}, \rho_{\chi})$ satisfies the exchange symmetry.
By contrast, rotating the linear combination $u(\rho_\phi,\rho_\chi)_* + u(\rho_\chi,\rho_\phi)_*$ gives a solution.

\begin{table}[!t]
\begin{tabular}{c| c c c c c c c}
 & $\eta_{\phi,\chi}$ & $\theta_1$ & $\theta_2$ & $\theta_3$ & $\theta_4$ & $\theta_5$ & $\theta_6$ \\ \hline
 \multicolumn{7}{c}{$d=2.7$}\\ \hline
 DFP & 0.0841 & 1.398 & 1.398 & 0.0232 & -0.916 & -0.916 & -2.051 \\
 RDFP & 0.0480 & 1.952 & 1.371 & 0.0015 & -0.102 & -0.864 & -1.952 \\ \hline
 BFP & 0.0789 & 1.522 & 1.296 & -0.0108 & -0.877 & -0.915 & -2.035\\
 RBFP & 0.0491 & 1.945 & 1.362 & -0.0014 & -0.123 & -0.866 & -1.963\\ \hline
 \multicolumn{7}{c}{$d=2.5$}\\ \hline
 DFP & 0.119 & 1.264 & 1.264 & -0.0549 & -1.038 & -1.038 &  -2.048\\
 RDFP & 0.0694 & 1.931 & 1.221 & -0.104 & -0.134 & -0.973 & -1.931 \\ \hline
\end{tabular}
\caption{Anomalous dimensions and first critical exponents of the DFP and BFP and the rotated counterparts in LPA', where the symmetry \eqref{eq:coordrot} is slightly broken.
The dimensions $d=2.7$ is chosen as a representative. 
For the DFP we additionally give the values for $d=2.5$ to show that we do not obtain $d$-independent critical exponents anymore.}
\label{tab:critExpLPAprime}
\end{table}

Now, let us assume that $u(\rho_\phi,\rho_\chi)_*$ is invariant under $\phi\leftrightarrow \chi$.
From the considerations above, 
one can infer that the eigenvalue spectra of $u(\rho_\phi,\rho_\chi)_*$ and its rotated counterpart $\tilde u(\rho_\phi,\rho_\chi)_{*}$ are related to each other.
The linearized equation describing small perturbations around the fixed point  reads
\begin{equation}
 -\theta \, \delta u = \sum_{i,j=0} \left. \frac{\partial (\partial_t u)}{\partial u^{(i,j)}}\right|_{u=u_*} \delta u^{(i,j)},
 \label{eq:eigenvalueproblem}
\end{equation}
where $\delta u$ is the eigenperturbation and $\theta$ the critical exponent. 
As \eqref{eq:potflow} preserves the $\pi/4$ rotational symmetry and \eqref{eq:eigenvalueproblem} is linear in $\delta u$, it preserves that symmetry as well. 
According to the line of argument for the fixed point solutions,
only those eigenperturbations $\delta u$  that preserve the $\phi\leftrightarrow \chi$ symmetry are also eigenperturbations of the rotated solution $\tilde{u}$,
cf.~table \ref{tab:critExpLPA}. 
The rotation of an eigenperturbation not exhibiting $\phi \leftrightarrow \chi$ exchange symmetry are not a solution of \eqref{eq:eigenvalueproblem}.

We emphasize that the decoupled fixed points are an exceptional case.
For the decoupled solutions, some of the eigenvalues are degenerate.
The corresponding eigenperturbations separately break the exchange symmetry.
However, the linear combination of both eigendirections results in a $\phi\leftrightarrow \chi$ invariant perturbation.
Thus, the corresponding critical exponent is also contained in the spectrum of the rotated fixed points, cf.~table \ref{tab:critExpLPA}.

Let us now take the anomalous dimensions \eqref{eq:etaphi} and \eqref{eq:etachi} into account.
They are evaluated at the global minimum of the fixed-point potential.
For bicritical fixed points, $\eta_\phi$ is evaluated at the minimum in field direction $\rho_\phi$ and $\eta_\chi$ in field direction $\rho_\chi$. Thus, we evaluate $\eta_{\phi}$ at the point $(\kappa_\phi \neq 0, \kappa_{\chi}=0)$, and conversely, $\eta_{\chi}$ at the point $(\kappa_{\phi}=0, \kappa_\chi\neq 0)$.
Table \ref{tab:critExpLPAprime} shows that the anomalous dimensions are not invariant under \eqref{eq:coordrot}.
In fact, the difference between the anomalous dimensions of $u(\rho_\phi,\rho_\chi)_{*}$ and $\tilde u(\rho_\phi,\rho_\chi)_{*}$ may be large.
Thus, the $\pi/4$ rotational symmetry is broken in LPA'.  Note that this could change in a more extensive truncation, where a field-dependent wave function renormalization is taken into account.
However, that does not affect the existence of $\tilde u(\rho_\phi,\rho_\chi)_{*}$.
 Moreover, those critical exponents that are exactly equal for the solution $u$ and its rotation $\tilde{u}$ in the LPA, are still close to each other in the LPA', cf~Tab.~\ref{tab:critExpLPAprime}.

Now, we consider general values of $N=M$.
Besides the radial mode, the Goldstone modes additionally contribute to the flow \eqref{eq:potflow}.
It can be easily seen that they violate the $\pi/4$-rotational symmetry
\begin{equation*}
 \frac{1}{1+\tilde u_*^{(1,0)}}+\frac{1}{1+\tilde u_*^{(0,1)}} \rightarrow 
 \frac{2 \tilde\rho_\phi (1+ u_*^{(1,0)}) - 2 \tilde\rho_\chi (1+ u_*^{(0,1)})}{\tilde\rho_\phi (1+ u_*^{(1,0)})^2 - \tilde\rho_\chi (1+ u_*^{(0,1)})^2} \, ,
\end{equation*}
where $\tilde \rho_\phi = (\rho_\phi+\rho_\chi+2\sqrt{\rho_\phi\rho_\chi})/2, \tilde \rho_\chi = (\rho_\phi+\rho_\chi+-2\sqrt{\rho_\phi\rho_\chi})/2$.
This is already clear since the radial part contains derivatives with respect to both fields whereas the Goldstone terms are fully decoupled. The transformation \eqref{eq:coordrot} may generally couple both sectors.
Similar to the LPA' case, for small deviations from $N=M=1$, the symmetry is only broken slightly.
Hence, we observe that $\tilde u(\rho_\phi,\rho_\chi)_{*}$ may still exist for larger and smaller $N=M$.
Moreover, for $N=M$ far away from $N=M=1$ we still may find $\tilde u(\rho_\phi,\rho_\chi)_{*}$, which becomes fully independent from $u(\rho_\phi,\rho_\chi)_{*}$.

\section{Breakdown of local expansions}\label{app:localbreakdown}
Here, we review the convergence properties of the LPA' in a local expansion. We focus on the DFP as a simple example, cf.~Tab.~\ref{tab:convergence}. The breakdown of convergence towards $d=2$ is related to the canonical dimensionality of the couplings: As the dimensionality of the fields vanishes in $d=2$, all couplings $\lambda_{i,j}$ have the same dimensionality $[\lambda_{i,j}]=2$ in $d=2$ dimensions. Our results clearly  confirm
the necessity to go beyond local expansions, and instead use a method that allows us to solve the complete fixed-point potential.

\begin{table}[!t]
\begin{tabular}{c c c c c c}
$d$& $\theta_3(n=4)$& $\theta_3(n=8)$& $\theta_3(n=12)$& $\theta_3(n=16)$ & $\theta_3(n=\infty)$  \\ \hline \hline
3 & 0.8026 &0.1274 &0.1412&0.1408&0.1407\\\hline
2.9 &0.9919& 0.1317&0.1298& 0.1327&0.1321\\\hline
2.8&1.1976&0.1644& 0.1033& 0.1208&0.1175\\\hline
\end{tabular}
\caption{\label{tab:convergence} We give the third largest critical exponent $\theta_3$ at the DFP for $N=M=1$ in $d$ dimensions in LPA' $n$ using the scaling relation \eqref{eq:scalerelation}. $n=4$ includes all couplings up to four powers in the fields (i.e., two powers in the invariants $\rho_{\phi}$, $\rho_{\chi}$) and correspondingly for higher $n$.
Results from the solution computed via pseudo-spectral methods ($n=\infty$) are given for comparison. Whereas in $d=3$ the difference between LPA' 12 and LPA' 16 is at the level of 0.3 \%, it is of order 2\% in $d=2.9$, 17 \% in $d=2.8$. Reaching comparably high orders in the  LPA' is much more challenging  for coupled fixed points such as the BFP, as they do not follow from the Wilson-Fisher fixed point, and thus require the simultaneous solution of a much larger set of fixed-point equations for all the couplings.}
\end{table}

\bibliography{onombib}

\begin{thebibliography}{104}%
\makeatletter
\providecommand \@ifxundefined [1]{%
 \@ifx{#1\undefined}
}%
\providecommand \@ifnum [1]{%
 \ifnum #1\expandafter \@firstoftwo
 \else \expandafter \@secondoftwo
 \fi
}%
\providecommand \@ifx [1]{%
 \ifx #1\expandafter \@firstoftwo
 \else \expandafter \@secondoftwo
 \fi
}%
\providecommand \natexlab [1]{#1}%
\providecommand \enquote  [1]{``#1''}%
\providecommand \bibnamefont  [1]{#1}%
\providecommand \bibfnamefont [1]{#1}%
\providecommand \citenamefont [1]{#1}%
\providecommand \href@noop [0]{\@secondoftwo}%
\providecommand \href [0]{\begingroup \@sanitize@url \@href}%
\providecommand \@href[1]{\@@startlink{#1}\@@href}%
\providecommand \@@href[1]{\endgroup#1\@@endlink}%
\providecommand \@sanitize@url [0]{\catcode `\\12\catcode `\$12\catcode
  `\&12\catcode `\#12\catcode `\^12\catcode `\_12\catcode `\%12\relax}%
\providecommand \@@startlink[1]{}%
\providecommand \@@endlink[0]{}%
\providecommand \url  [0]{\begingroup\@sanitize@url \@url }%
\providecommand \@url [1]{\endgroup\@href {#1}{\urlprefix }}%
\providecommand \urlprefix  [0]{URL }%
\providecommand \Eprint [0]{\href }%
\providecommand \doibase [0]{http://dx.doi.org/}%
\providecommand \selectlanguage [0]{\@gobble}%
\providecommand \bibinfo  [0]{\@secondoftwo}%
\providecommand \bibfield  [0]{\@secondoftwo}%
\providecommand \translation [1]{[#1]}%
\providecommand \BibitemOpen [0]{}%
\providecommand \bibitemStop [0]{}%
\providecommand \bibitemNoStop [0]{.\EOS\space}%
\providecommand \EOS [0]{\spacefactor3000\relax}%
\providecommand \BibitemShut  [1]{\csname bibitem#1\endcsname}%
\let\auto@bib@innerbib\@empty
\bibitem [{\citenamefont {Fisher}\ and\ \citenamefont
  {Nelson}(1974)}]{Fisher:1974zz}%
  \BibitemOpen
  \bibfield  {author} {\bibinfo {author} {\bibfnamefont {M.~E.}\ \bibnamefont
  {Fisher}}\ and\ \bibinfo {author} {\bibfnamefont {D.~R.}\ \bibnamefont
  {Nelson}},\ }\href {\doibase 10.1103/PhysRevLett.32.1350} {\bibfield
  {journal} {\bibinfo  {journal} {Phys. Rev. Lett.}\ }\textbf {\bibinfo
  {volume} {32}},\ \bibinfo {pages} {1350} (\bibinfo {year}
  {1974})}\BibitemShut {NoStop}%
\bibitem [{\citenamefont {Kosterlitz}\ \emph {et~al.}(1976)\citenamefont
  {Kosterlitz}, \citenamefont {Nelson},\ and\ \citenamefont
  {Fisher}}]{Kosterlitz:1976zza}%
  \BibitemOpen
  \bibfield  {author} {\bibinfo {author} {\bibfnamefont {J.~M.}\ \bibnamefont
  {Kosterlitz}}, \bibinfo {author} {\bibfnamefont {D.~R.}\ \bibnamefont
  {Nelson}}, \ and\ \bibinfo {author} {\bibfnamefont {M.~E.}\ \bibnamefont
  {Fisher}},\ }\href {\doibase 10.1103/PhysRevB.13.412} {\bibfield  {journal}
  {\bibinfo  {journal} {Phys. Rev.}\ }\textbf {\bibinfo {volume} {B13}},\
  \bibinfo {pages} {412} (\bibinfo {year} {1976})}\BibitemShut {NoStop}%
\bibitem [{\citenamefont {Liu}\ and\ \citenamefont {Fisher}(1972)}]{FisherLiu}%
  \BibitemOpen
  \bibfield  {author} {\bibinfo {author} {\bibfnamefont {K.}~\bibnamefont
  {Liu}}\ and\ \bibinfo {author} {\bibfnamefont {M.}~\bibnamefont {Fisher}},\
  }\href@noop {} {\bibfield  {journal} {\bibinfo  {journal} {J. Low. Temp.
  Phys.}\ }\textbf {\bibinfo {volume} {10}},\ \bibinfo {pages} {655} (\bibinfo
  {year} {1972})}\BibitemShut {NoStop}%
\bibitem [{\citenamefont {Nelson}\ \emph {et~al.}(1974)\citenamefont {Nelson},
  \citenamefont {Kosterlitz},\ and\ \citenamefont
  {Fisher}}]{PhysRevLett.33.813}%
  \BibitemOpen
  \bibfield  {author} {\bibinfo {author} {\bibfnamefont {D.~R.}\ \bibnamefont
  {Nelson}}, \bibinfo {author} {\bibfnamefont {J.~M.}\ \bibnamefont
  {Kosterlitz}}, \ and\ \bibinfo {author} {\bibfnamefont {M.~E.}\ \bibnamefont
  {Fisher}},\ }\href {\doibase 10.1103/PhysRevLett.33.813} {\bibfield
  {journal} {\bibinfo  {journal} {Phys. Rev. Lett.}\ }\textbf {\bibinfo
  {volume} {33}},\ \bibinfo {pages} {813} (\bibinfo {year} {1974})}\BibitemShut
  {NoStop}%
\bibitem [{\citenamefont {Rohrer}(1975)}]{Rohrer1975}%
  \BibitemOpen
  \bibfield  {author} {\bibinfo {author} {\bibfnamefont {H.}~\bibnamefont
  {Rohrer}},\ }\href@noop {} {\bibfield  {journal} {\bibinfo  {journal} {Phys.
  Rev. Lett}\ }\textbf {\bibinfo {volume} {34}},\ \bibinfo {pages} {1638}
  (\bibinfo {year} {1975})}\BibitemShut {NoStop}%
\bibitem [{\citenamefont {Rohrer}(1977)}]{Rohrer1977}%
  \BibitemOpen
  \bibfield  {author} {\bibinfo {author} {\bibfnamefont {H.}~\bibnamefont
  {Rohrer}},\ }\href@noop {} {\bibfield  {journal} {\bibinfo  {journal} {Phys.
  Rev. Lett}\ }\textbf {\bibinfo {volume} {38}},\ \bibinfo {pages} {909}
  (\bibinfo {year} {1977})}\BibitemShut {NoStop}%
\bibitem [{\citenamefont {King}\ and\ \citenamefont
  {Rohrer}(1979)}]{KingRohrer}%
  \BibitemOpen
  \bibfield  {author} {\bibinfo {author} {\bibfnamefont {A.}~\bibnamefont
  {King}}\ and\ \bibinfo {author} {\bibfnamefont {H.}~\bibnamefont {Rohrer}},\
  }\href@noop {} {\bibfield  {journal} {\bibinfo  {journal} {Phys. Rev.}\
  }\textbf {\bibinfo {volume} {19}},\ \bibinfo {pages} {5864} (\bibinfo {year}
  {1979})}\BibitemShut {NoStop}%
\bibitem [{\citenamefont {Oliveira~Jr.}\ \emph {et~al.}(1978)\citenamefont
  {Oliveira~Jr.}, \citenamefont {Paduan~Filho}, \citenamefont {Salinas},\ and\
  \citenamefont {Becerra}}]{Oliveira}%
  \BibitemOpen
  \bibfield  {author} {\bibinfo {author} {\bibfnamefont {N.}~\bibnamefont
  {Oliveira~Jr.}}, \bibinfo {author} {\bibfnamefont {A.}~\bibnamefont
  {Paduan~Filho}}, \bibinfo {author} {\bibfnamefont {S.}~\bibnamefont
  {Salinas}}, \ and\ \bibinfo {author} {\bibfnamefont {C.}~\bibnamefont
  {Becerra}},\ }\href@noop {} {\bibfield  {journal} {\bibinfo  {journal} {Phys.
  Rev. B}\ }\textbf {\bibinfo {volume} {18}},\ \bibinfo {pages} {6165}
  (\bibinfo {year} {1978})}\BibitemShut {NoStop}%
\bibitem [{\citenamefont {Butera}\ \emph {et~al.}(1981)\citenamefont {Butera},
  \citenamefont {Corliss}, \citenamefont {Hastings}, \citenamefont {Thomas},\
  and\ \citenamefont {Mukamel}}]{Butera}%
  \BibitemOpen
  \bibfield  {author} {\bibinfo {author} {\bibfnamefont {R.~A.}\ \bibnamefont
  {Butera}}, \bibinfo {author} {\bibfnamefont {L.~M.}\ \bibnamefont {Corliss}},
  \bibinfo {author} {\bibfnamefont {J.~M.}\ \bibnamefont {Hastings}}, \bibinfo
  {author} {\bibfnamefont {R.}~\bibnamefont {Thomas}}, \ and\ \bibinfo {author}
  {\bibfnamefont {D.}~\bibnamefont {Mukamel}},\ }\href {\doibase
  10.1103/PhysRevB.24.1244} {\bibfield  {journal} {\bibinfo  {journal} {Phys.
  Rev. B}\ }\textbf {\bibinfo {volume} {24}},\ \bibinfo {pages} {1244}
  (\bibinfo {year} {1981})}\BibitemShut {NoStop}%
\bibitem [{\citenamefont {Ohgushi}\ and\ \citenamefont {Ueda}(2005)}]{Ohgushi}%
  \BibitemOpen
  \bibfield  {author} {\bibinfo {author} {\bibfnamefont {K.}~\bibnamefont
  {Ohgushi}}\ and\ \bibinfo {author} {\bibfnamefont {Y.}~\bibnamefont {Ueda}},\
  }\href@noop {} {\bibfield  {journal} {\bibinfo  {journal} {Phys. Rev. Lett.}\
  }\textbf {\bibinfo {volume} {95}},\ \bibinfo {pages} {217202} (\bibinfo
  {year} {2005})}\BibitemShut {NoStop}%
\bibitem [{\citenamefont {Becerra}\ \emph {et~al.}(1988)\citenamefont
  {Becerra}, \citenamefont {Oliveira~Jr.}, \citenamefont {Filho}, \citenamefont
  {Figueiredo},\ and\ \citenamefont {Souza}}]{beccera88}%
  \BibitemOpen
  \bibfield  {author} {\bibinfo {author} {\bibfnamefont {C.}~\bibnamefont
  {Becerra}}, \bibinfo {author} {\bibfnamefont {N.}~\bibnamefont
  {Oliveira~Jr.}}, \bibinfo {author} {\bibfnamefont {P.}~\bibnamefont {Filho}},
  \bibinfo {author} {\bibfnamefont {W.}~\bibnamefont {Figueiredo}}, \ and\
  \bibinfo {author} {\bibfnamefont {M.}~\bibnamefont {Souza}},\ }\href@noop {}
  {\bibfield  {journal} {\bibinfo  {journal} {Phys. Rev. B}\ }\textbf {\bibinfo
  {volume} {38}},\ \bibinfo {pages} {6887} (\bibinfo {year}
  {1988})}\BibitemShut {NoStop}%
\bibitem [{\citenamefont {Basten}\ \emph {et~al.}(1980)\citenamefont {Basten},
  \citenamefont {Frikkee},\ and\ \citenamefont {de~Jonge}}]{basten80}%
  \BibitemOpen
  \bibfield  {author} {\bibinfo {author} {\bibfnamefont {J.}~\bibnamefont
  {Basten}}, \bibinfo {author} {\bibfnamefont {E.}~\bibnamefont {Frikkee}}, \
  and\ \bibinfo {author} {\bibfnamefont {W.}~\bibnamefont {de~Jonge}},\
  }\href@noop {} {\bibfield  {journal} {\bibinfo  {journal} {Phys. Rev. B}\
  }\textbf {\bibinfo {volume} {22}},\ \bibinfo {pages} {1429} (\bibinfo {year}
  {1980})}\BibitemShut {NoStop}%
\bibitem [{\citenamefont {Zhang}(1997)}]{Zhang1997}%
  \BibitemOpen
  \bibfield  {author} {\bibinfo {author} {\bibfnamefont {S.-C.}\ \bibnamefont
  {Zhang}},\ }\href {\doibase 10.1126/science.275.5303.1089} {\bibfield
  {journal} {\bibinfo  {journal} {Science}\ }\textbf {\bibinfo {volume}
  {275}},\ \bibinfo {pages} {1089} (\bibinfo {year} {1997})}\BibitemShut
  {NoStop}%
\bibitem [{\citenamefont {Roy}(2011)}]{Roy:2011pg}%
  \BibitemOpen
  \bibfield  {author} {\bibinfo {author} {\bibfnamefont {B.}~\bibnamefont
  {Roy}},\ }\href {\doibase 10.1103/PhysRevB.84.113404} {\bibfield  {journal}
  {\bibinfo  {journal} {Phys. Rev.}\ }\textbf {\bibinfo {volume} {B84}},\
  \bibinfo {pages} {113404} (\bibinfo {year} {2011})},\ \Eprint
  {http://arxiv.org/abs/1106.1419} {arXiv:1106.1419 [cond-mat.str-el]}
  \BibitemShut {NoStop}%
\bibitem [{\citenamefont {Classen}\ \emph {et~al.}(2015)\citenamefont
  {Classen}, \citenamefont {Herbut}, \citenamefont {Janssen},\ and\
  \citenamefont {Scherer}}]{Classen:2015ssa}%
  \BibitemOpen
  \bibfield  {author} {\bibinfo {author} {\bibfnamefont {L.}~\bibnamefont
  {Classen}}, \bibinfo {author} {\bibfnamefont {I.~F.}\ \bibnamefont {Herbut}},
  \bibinfo {author} {\bibfnamefont {L.}~\bibnamefont {Janssen}}, \ and\
  \bibinfo {author} {\bibfnamefont {M.~M.}\ \bibnamefont {Scherer}},\ }\href
  {\doibase 10.1103/PhysRevB.92.035429} {\bibfield  {journal} {\bibinfo
  {journal} {Phys. Rev.}\ }\textbf {\bibinfo {volume} {B92}},\ \bibinfo {pages}
  {035429} (\bibinfo {year} {2015})},\ \Eprint
  {http://arxiv.org/abs/1503.05002} {arXiv:1503.05002 [cond-mat.str-el]}
  \BibitemShut {NoStop}%
\bibitem [{\citenamefont {Classen}\ \emph {et~al.}(2016)\citenamefont
  {Classen}, \citenamefont {Herbut}, \citenamefont {Janssen},\ and\
  \citenamefont {Scherer}}]{Classen:2015mar}%
  \BibitemOpen
  \bibfield  {author} {\bibinfo {author} {\bibfnamefont {L.}~\bibnamefont
  {Classen}}, \bibinfo {author} {\bibfnamefont {I.~F.}\ \bibnamefont {Herbut}},
  \bibinfo {author} {\bibfnamefont {L.}~\bibnamefont {Janssen}}, \ and\
  \bibinfo {author} {\bibfnamefont {M.~M.}\ \bibnamefont {Scherer}},\ }\href
  {\doibase 10.1103/PhysRevB.93.125119} {\bibfield  {journal} {\bibinfo
  {journal} {Phys. Rev.}\ }\textbf {\bibinfo {volume} {B93}},\ \bibinfo {pages}
  {125119} (\bibinfo {year} {2016})},\ \Eprint
  {http://arxiv.org/abs/1510.09003} {arXiv:1510.09003 [cond-mat.str-el]}
  \BibitemShut {NoStop}%
\bibitem [{\citenamefont {Aharony}()}]{Aharony:2002}%
  \BibitemOpen
  \bibfield  {author} {\bibinfo {author} {\bibfnamefont {A.}~\bibnamefont
  {Aharony}},\ }\href {\doibase 10.1023/A:1022103717585} {\bibfield  {journal}
  {\bibinfo  {journal} {Journal of Statistical Physics}\ }\textbf {\bibinfo
  {volume} {110}},\ \bibinfo {pages} {659}}\BibitemShut {NoStop}%
\bibitem [{\citenamefont {Aharony}(2002)}]{Aharony:2002zz}%
  \BibitemOpen
  \bibfield  {author} {\bibinfo {author} {\bibfnamefont {A.}~\bibnamefont
  {Aharony}},\ }\href {\doibase 10.1103/PhysRevLett.88.059703} {\bibfield
  {journal} {\bibinfo  {journal} {Phys. Rev. Lett.}\ }\textbf {\bibinfo
  {volume} {88}},\ \bibinfo {pages} {059703} (\bibinfo {year}
  {2002})}\BibitemShut {NoStop}%
\bibitem [{\citenamefont {Aharony}(1976)}]{Aharony:1976book}%
  \BibitemOpen
  \bibfield  {author} {\bibinfo {author} {\bibfnamefont {A.}~\bibnamefont
  {Aharony}},\ }\href@noop {} {\emph {\bibinfo {title} {Phase transitions and
  critical phenomena}}},\ Vol.~\bibinfo {volume} {6}\ (\bibinfo  {publisher}
  {Academic Press},\ \bibinfo {year} {1976})\ p.\ \bibinfo {pages}
  {357}\BibitemShut {NoStop}%
\bibitem [{\citenamefont {Aharony}(1973)}]{Aharony:1973zz}%
  \BibitemOpen
  \bibfield  {author} {\bibinfo {author} {\bibfnamefont {A.}~\bibnamefont
  {Aharony}},\ }\href {\doibase 10.1103/PhysRevB.8.4270} {\bibfield  {journal}
  {\bibinfo  {journal} {Phys. Rev.}\ }\textbf {\bibinfo {volume} {B8}},\
  \bibinfo {pages} {4270} (\bibinfo {year} {1973})}\BibitemShut {NoStop}%
\bibitem [{\citenamefont {Calabrese}\ \emph
  {et~al.}(2003{\natexlab{a}})\citenamefont {Calabrese}, \citenamefont
  {Pelissetto},\ and\ \citenamefont {Vicari}}]{Calabrese:2002bm}%
  \BibitemOpen
  \bibfield  {author} {\bibinfo {author} {\bibfnamefont {P.}~\bibnamefont
  {Calabrese}}, \bibinfo {author} {\bibfnamefont {A.}~\bibnamefont
  {Pelissetto}}, \ and\ \bibinfo {author} {\bibfnamefont {E.}~\bibnamefont
  {Vicari}},\ }\href {\doibase 10.1103/PhysRevB.67.054505} {\bibfield
  {journal} {\bibinfo  {journal} {Phys. Rev.}\ }\textbf {\bibinfo {volume}
  {B67}},\ \bibinfo {pages} {054505} (\bibinfo {year} {2003}{\natexlab{a}})},\
  \Eprint {http://arxiv.org/abs/cond-mat/0209580} {arXiv:cond-mat/0209580
  [cond-mat]} \BibitemShut {NoStop}%
\bibitem [{\citenamefont {Calabrese}\ \emph {et~al.}(2002)\citenamefont
  {Calabrese}, \citenamefont {Pelissetto},\ and\ \citenamefont
  {Vicari}}]{Calabrese:2002bq}%
  \BibitemOpen
  \bibfield  {author} {\bibinfo {author} {\bibfnamefont {P.}~\bibnamefont
  {Calabrese}}, \bibinfo {author} {\bibfnamefont {A.}~\bibnamefont
  {Pelissetto}}, \ and\ \bibinfo {author} {\bibfnamefont {E.}~\bibnamefont
  {Vicari}},\ }\href@noop {} {\  (\bibinfo {year} {2002})},\ \Eprint
  {http://arxiv.org/abs/cond-mat/203533} {arXiv:cond-mat/203533 [cond-mat]}
  \BibitemShut {NoStop}%
\bibitem [{\citenamefont {Vicari}(2007)}]{Vicari:2007ma}%
  \BibitemOpen
  \bibfield  {author} {\bibinfo {author} {\bibfnamefont {E.}~\bibnamefont
  {Vicari}},\ }\href@noop {} {\bibfield  {journal} {\bibinfo  {journal} {PoS}\
  }\textbf {\bibinfo {volume} {LAT2007}},\ \bibinfo {pages} {023} (\bibinfo
  {year} {2007})},\ \Eprint {http://arxiv.org/abs/0709.1014} {arXiv:0709.1014
  [hep-lat]} \BibitemShut {NoStop}%
\bibitem [{\citenamefont {{Folk}}\ \emph {et~al.}(2008)\citenamefont {{Folk}},
  \citenamefont {{Holovatch}},\ and\ \citenamefont {{Moser}}}]{Folk2008}%
  \BibitemOpen
  \bibfield  {author} {\bibinfo {author} {\bibfnamefont {R.}~\bibnamefont
  {{Folk}}}, \bibinfo {author} {\bibfnamefont {Y.}~\bibnamefont {{Holovatch}}},
  \ and\ \bibinfo {author} {\bibfnamefont {G.}~\bibnamefont {{Moser}}},\ }\href
  {\doibase 10.1103/PhysRevE.78.041124} {\bibfield  {journal} {\bibinfo
  {journal} {\pre}\ }\textbf {\bibinfo {volume} {78}},\ \bibinfo {eid} {041124}
  (\bibinfo {year} {2008})},\ \Eprint {http://arxiv.org/abs/0808.0314}
  {arXiv:0808.0314 [cond-mat.stat-mech]} \BibitemShut {NoStop}%
\bibitem [{\citenamefont {Eichhorn}\ \emph {et~al.}(2013)\citenamefont
  {Eichhorn}, \citenamefont {Mesterházy},\ and\ \citenamefont
  {Scherer}}]{Eichhorn:2013zza}%
  \BibitemOpen
  \bibfield  {author} {\bibinfo {author} {\bibfnamefont {A.}~\bibnamefont
  {Eichhorn}}, \bibinfo {author} {\bibfnamefont {D.}~\bibnamefont
  {Mesterházy}}, \ and\ \bibinfo {author} {\bibfnamefont {M.~M.}\ \bibnamefont
  {Scherer}},\ }\href {\doibase 10.1103/PhysRevE.88.042141} {\bibfield
  {journal} {\bibinfo  {journal} {Phys. Rev.}\ }\textbf {\bibinfo {volume}
  {E88}},\ \bibinfo {pages} {042141} (\bibinfo {year} {2013})},\ \Eprint
  {http://arxiv.org/abs/1306.2952} {arXiv:1306.2952 [cond-mat.stat-mech]}
  \BibitemShut {NoStop}%
\bibitem [{\citenamefont {Eichhorn}\ \emph {et~al.}(2014)\citenamefont
  {Eichhorn}, \citenamefont {Mesterházy},\ and\ \citenamefont
  {Scherer}}]{Eichhorn:2014asa}%
  \BibitemOpen
  \bibfield  {author} {\bibinfo {author} {\bibfnamefont {A.}~\bibnamefont
  {Eichhorn}}, \bibinfo {author} {\bibfnamefont {D.}~\bibnamefont
  {Mesterházy}}, \ and\ \bibinfo {author} {\bibfnamefont {M.~M.}\ \bibnamefont
  {Scherer}},\ }\href {\doibase 10.1103/PhysRevE.90.052129} {\bibfield
  {journal} {\bibinfo  {journal} {Phys. Rev.}\ }\textbf {\bibinfo {volume}
  {E90}},\ \bibinfo {pages} {052129} (\bibinfo {year} {2014})},\ \Eprint
  {http://arxiv.org/abs/1407.7442} {arXiv:1407.7442 [cond-mat.stat-mech]}
  \BibitemShut {NoStop}%
\bibitem [{\citenamefont {Boettcher}(2015)}]{Boettcher:2015pja}%
  \BibitemOpen
  \bibfield  {author} {\bibinfo {author} {\bibfnamefont {I.}~\bibnamefont
  {Boettcher}},\ }\href {\doibase 10.1103/PhysRevE.91.062112} {\bibfield
  {journal} {\bibinfo  {journal} {Phys. Rev.}\ }\textbf {\bibinfo {volume}
  {E91}},\ \bibinfo {pages} {062112} (\bibinfo {year} {2015})},\ \Eprint
  {http://arxiv.org/abs/1503.07817} {arXiv:1503.07817 [cond-mat.stat-mech]}
  \BibitemShut {NoStop}%
\bibitem [{\citenamefont {Eichhorn}\ \emph
  {et~al.}(2016{\natexlab{a}})\citenamefont {Eichhorn}, \citenamefont {Helfer},
  \citenamefont {Mesterházy},\ and\ \citenamefont
  {Scherer}}]{Eichhorn:2015woa}%
  \BibitemOpen
  \bibfield  {author} {\bibinfo {author} {\bibfnamefont {A.}~\bibnamefont
  {Eichhorn}}, \bibinfo {author} {\bibfnamefont {T.}~\bibnamefont {Helfer}},
  \bibinfo {author} {\bibfnamefont {D.}~\bibnamefont {Mesterházy}}, \ and\
  \bibinfo {author} {\bibfnamefont {M.~M.}\ \bibnamefont {Scherer}},\ }\href
  {\doibase 10.1140/epjc/s10052-016-3921-3} {\bibfield  {journal} {\bibinfo
  {journal} {Eur. Phys. J.}\ }\textbf {\bibinfo {volume} {C76}},\ \bibinfo
  {pages} {88} (\bibinfo {year} {2016}{\natexlab{a}})},\ \Eprint
  {http://arxiv.org/abs/1510.04807} {arXiv:1510.04807 [cond-mat.stat-mech]}
  \BibitemShut {NoStop}%
\bibitem [{\citenamefont {Mermin}\ and\ \citenamefont
  {Wagner}(1966)}]{PhysRevLett.17.1133}%
  \BibitemOpen
  \bibfield  {author} {\bibinfo {author} {\bibfnamefont {N.~D.}\ \bibnamefont
  {Mermin}}\ and\ \bibinfo {author} {\bibfnamefont {H.}~\bibnamefont
  {Wagner}},\ }\href {\doibase 10.1103/PhysRevLett.17.1133} {\bibfield
  {journal} {\bibinfo  {journal} {Phys. Rev. Lett.}\ }\textbf {\bibinfo
  {volume} {17}},\ \bibinfo {pages} {1133} (\bibinfo {year}
  {1966})}\BibitemShut {NoStop}%
\bibitem [{\citenamefont {Kosterlitz}\ and\ \citenamefont
  {Thouless}(1973)}]{Kosterlitz:1973xp}%
  \BibitemOpen
  \bibfield  {author} {\bibinfo {author} {\bibfnamefont {J.~M.}\ \bibnamefont
  {Kosterlitz}}\ and\ \bibinfo {author} {\bibfnamefont {D.~J.}\ \bibnamefont
  {Thouless}},\ }\href@noop {} {\bibfield  {journal} {\bibinfo  {journal} {J.
  Phys.}\ }\textbf {\bibinfo {volume} {C6}},\ \bibinfo {pages} {1181} (\bibinfo
  {year} {1973})}\BibitemShut {NoStop}%
\bibitem [{\citenamefont {Bishop}\ and\ \citenamefont
  {Reppy}(1978)}]{Bishop:1978zz}%
  \BibitemOpen
  \bibfield  {author} {\bibinfo {author} {\bibfnamefont {D.~J.}\ \bibnamefont
  {Bishop}}\ and\ \bibinfo {author} {\bibfnamefont {J.~D.}\ \bibnamefont
  {Reppy}},\ }\href {\doibase 10.1103/PhysRevLett.40.1727} {\bibfield
  {journal} {\bibinfo  {journal} {Phys. Rev. Lett.}\ }\textbf {\bibinfo
  {volume} {40}},\ \bibinfo {pages} {1727} (\bibinfo {year}
  {1978})}\BibitemShut {NoStop}%
\bibitem [{\citenamefont {Maps}\ and\ \citenamefont
  {Hallock}(1981)}]{PhysRevLett.47.1533}%
  \BibitemOpen
  \bibfield  {author} {\bibinfo {author} {\bibfnamefont {J.}~\bibnamefont
  {Maps}}\ and\ \bibinfo {author} {\bibfnamefont {R.~B.}\ \bibnamefont
  {Hallock}},\ }\href {\doibase 10.1103/PhysRevLett.47.1533} {\bibfield
  {journal} {\bibinfo  {journal} {Phys. Rev. Lett.}\ }\textbf {\bibinfo
  {volume} {47}},\ \bibinfo {pages} {1533} (\bibinfo {year}
  {1981})}\BibitemShut {NoStop}%
\bibitem [{\citenamefont {Hadzibabic}\ \emph {et~al.}(2006)\citenamefont
  {Hadzibabic}, \citenamefont {Kruger}, \citenamefont {Cheneau}, \citenamefont
  {Battelier},\ and\ \citenamefont {Dalibard}}]{Nature05851}%
  \BibitemOpen
  \bibfield  {author} {\bibinfo {author} {\bibfnamefont {Z.}~\bibnamefont
  {Hadzibabic}}, \bibinfo {author} {\bibfnamefont {P.}~\bibnamefont {Kruger}},
  \bibinfo {author} {\bibfnamefont {M.}~\bibnamefont {Cheneau}}, \bibinfo
  {author} {\bibfnamefont {B.}~\bibnamefont {Battelier}}, \ and\ \bibinfo
  {author} {\bibfnamefont {J.}~\bibnamefont {Dalibard}},\ }\href
  {http://dx.doi.org/10.1038/nature04851} {\bibfield  {journal} {\bibinfo
  {journal} {Nature}\ }\textbf {\bibinfo {volume} {441}},\ \bibinfo {pages}
  {1118} (\bibinfo {year} {2006})}\BibitemShut {NoStop}%
\bibitem [{\citenamefont {Tung}\ \emph {et~al.}(2010)\citenamefont {Tung},
  \citenamefont {Lamporesi}, \citenamefont {Lobser}, \citenamefont {Xia},\ and\
  \citenamefont {Cornell}}]{PhysRevLett.105.230408}%
  \BibitemOpen
  \bibfield  {author} {\bibinfo {author} {\bibfnamefont {S.}~\bibnamefont
  {Tung}}, \bibinfo {author} {\bibfnamefont {G.}~\bibnamefont {Lamporesi}},
  \bibinfo {author} {\bibfnamefont {D.}~\bibnamefont {Lobser}}, \bibinfo
  {author} {\bibfnamefont {L.}~\bibnamefont {Xia}}, \ and\ \bibinfo {author}
  {\bibfnamefont {E.~A.}\ \bibnamefont {Cornell}},\ }\href {\doibase
  10.1103/PhysRevLett.105.230408} {\bibfield  {journal} {\bibinfo  {journal}
  {Phys. Rev. Lett.}\ }\textbf {\bibinfo {volume} {105}},\ \bibinfo {pages}
  {230408} (\bibinfo {year} {2010})}\BibitemShut {NoStop}%
\bibitem [{\citenamefont {{Desbuquois}}\ \emph {et~al.}(2012)\citenamefont
  {{Desbuquois}}, \citenamefont {{Chomaz}}, \citenamefont {{Yefsah}},
  \citenamefont {{L{\'e}onard}}, \citenamefont {{Beugnon}}, \citenamefont
  {{Weitenberg}},\ and\ \citenamefont {{Dalibard}}}]{2012NatPh...8..645D}%
  \BibitemOpen
  \bibfield  {author} {\bibinfo {author} {\bibfnamefont {R.}~\bibnamefont
  {{Desbuquois}}}, \bibinfo {author} {\bibfnamefont {L.}~\bibnamefont
  {{Chomaz}}}, \bibinfo {author} {\bibfnamefont {T.}~\bibnamefont {{Yefsah}}},
  \bibinfo {author} {\bibfnamefont {J.}~\bibnamefont {{L{\'e}onard}}}, \bibinfo
  {author} {\bibfnamefont {J.}~\bibnamefont {{Beugnon}}}, \bibinfo {author}
  {\bibfnamefont {C.}~\bibnamefont {{Weitenberg}}}, \ and\ \bibinfo {author}
  {\bibfnamefont {J.}~\bibnamefont {{Dalibard}}},\ }\href {\doibase
  10.1038/nphys2378} {\bibfield  {journal} {\bibinfo  {journal} {Nature
  Physics}\ }\textbf {\bibinfo {volume} {8}},\ \bibinfo {pages} {645} (\bibinfo
  {year} {2012})},\ \Eprint {http://arxiv.org/abs/1205.4536} {arXiv:1205.4536
  [cond-mat.quant-gas]} \BibitemShut {NoStop}%
\bibitem [{\citenamefont {Murthy}\ \emph {et~al.}(2015)\citenamefont {Murthy},
  \citenamefont {Boettcher}, \citenamefont {Bayha}, \citenamefont {Holzmann},
  \citenamefont {Kedar}, \citenamefont {Neidig}, \citenamefont {Ries},
  \citenamefont {Wenz}, \citenamefont {Z\"urn},\ and\ \citenamefont
  {Jochim}}]{PhysRevLett.115.010401}%
  \BibitemOpen
  \bibfield  {author} {\bibinfo {author} {\bibfnamefont {P.~A.}\ \bibnamefont
  {Murthy}}, \bibinfo {author} {\bibfnamefont {I.}~\bibnamefont {Boettcher}},
  \bibinfo {author} {\bibfnamefont {L.}~\bibnamefont {Bayha}}, \bibinfo
  {author} {\bibfnamefont {M.}~\bibnamefont {Holzmann}}, \bibinfo {author}
  {\bibfnamefont {D.}~\bibnamefont {Kedar}}, \bibinfo {author} {\bibfnamefont
  {M.}~\bibnamefont {Neidig}}, \bibinfo {author} {\bibfnamefont {M.~G.}\
  \bibnamefont {Ries}}, \bibinfo {author} {\bibfnamefont {A.~N.}\ \bibnamefont
  {Wenz}}, \bibinfo {author} {\bibfnamefont {G.}~\bibnamefont {Z\"urn}}, \ and\
  \bibinfo {author} {\bibfnamefont {S.}~\bibnamefont {Jochim}},\ }\href
  {\doibase 10.1103/PhysRevLett.115.010401} {\bibfield  {journal} {\bibinfo
  {journal} {Phys. Rev. Lett.}\ }\textbf {\bibinfo {volume} {115}},\ \bibinfo
  {pages} {010401} (\bibinfo {year} {2015})}\BibitemShut {NoStop}%
\bibitem [{\citenamefont {Morris}(1995{\natexlab{a}})}]{Morris:1994jc}%
  \BibitemOpen
  \bibfield  {author} {\bibinfo {author} {\bibfnamefont {T.~R.}\ \bibnamefont
  {Morris}},\ }\href {\doibase 10.1016/0370-2693(94)01603-A} {\bibfield
  {journal} {\bibinfo  {journal} {Phys. Lett.}\ }\textbf {\bibinfo {volume}
  {B345}},\ \bibinfo {pages} {139} (\bibinfo {year} {1995}{\natexlab{a}})},\
  \Eprint {http://arxiv.org/abs/hep-th/9410141} {arXiv:hep-th/9410141 [hep-th]}
  \BibitemShut {NoStop}%
\bibitem [{\citenamefont {Canet}\ \emph {et~al.}(2003)\citenamefont {Canet},
  \citenamefont {Delamotte}, \citenamefont {Mouhanna},\ and\ \citenamefont
  {Vidal}}]{Canet:2003qd}%
  \BibitemOpen
  \bibfield  {author} {\bibinfo {author} {\bibfnamefont {L.}~\bibnamefont
  {Canet}}, \bibinfo {author} {\bibfnamefont {B.}~\bibnamefont {Delamotte}},
  \bibinfo {author} {\bibfnamefont {D.}~\bibnamefont {Mouhanna}}, \ and\
  \bibinfo {author} {\bibfnamefont {J.}~\bibnamefont {Vidal}},\ }\href
  {\doibase 10.1103/PhysRevB.68.064421} {\bibfield  {journal} {\bibinfo
  {journal} {Phys. Rev.}\ }\textbf {\bibinfo {volume} {B68}},\ \bibinfo {pages}
  {064421} (\bibinfo {year} {2003})},\ \Eprint
  {http://arxiv.org/abs/hep-th/0302227} {arXiv:hep-th/0302227 [hep-th]}
  \BibitemShut {NoStop}%
\bibitem [{\citenamefont {Bervillier}\ \emph {et~al.}(2007)\citenamefont
  {Bervillier}, \citenamefont {Juttner},\ and\ \citenamefont
  {Litim}}]{Bervillier:2007rc}%
  \BibitemOpen
  \bibfield  {author} {\bibinfo {author} {\bibfnamefont {C.}~\bibnamefont
  {Bervillier}}, \bibinfo {author} {\bibfnamefont {A.}~\bibnamefont {Juttner}},
  \ and\ \bibinfo {author} {\bibfnamefont {D.~F.}\ \bibnamefont {Litim}},\
  }\href {\doibase 10.1016/j.nuclphysb.2007.03.036} {\bibfield  {journal}
  {\bibinfo  {journal} {Nucl. Phys.}\ }\textbf {\bibinfo {volume} {B783}},\
  \bibinfo {pages} {213} (\bibinfo {year} {2007})},\ \Eprint
  {http://arxiv.org/abs/hep-th/0701172} {arXiv:hep-th/0701172 [hep-th]}
  \BibitemShut {NoStop}%
\bibitem [{\citenamefont {Litim}\ and\ \citenamefont
  {Zappala}(2011)}]{Litim:2010tt}%
  \BibitemOpen
  \bibfield  {author} {\bibinfo {author} {\bibfnamefont {D.~F.}\ \bibnamefont
  {Litim}}\ and\ \bibinfo {author} {\bibfnamefont {D.}~\bibnamefont
  {Zappala}},\ }\href {\doibase 10.1103/PhysRevD.83.085009} {\bibfield
  {journal} {\bibinfo  {journal} {Phys. Rev.}\ }\textbf {\bibinfo {volume}
  {D83}},\ \bibinfo {pages} {085009} (\bibinfo {year} {2011})},\ \Eprint
  {http://arxiv.org/abs/1009.1948} {arXiv:1009.1948 [hep-th]} \BibitemShut
  {NoStop}%
\bibitem [{\citenamefont {Codello}\ and\ \citenamefont
  {D'Odorico}(2013)}]{Codello:2012ec}%
  \BibitemOpen
  \bibfield  {author} {\bibinfo {author} {\bibfnamefont {A.}~\bibnamefont
  {Codello}}\ and\ \bibinfo {author} {\bibfnamefont {G.}~\bibnamefont
  {D'Odorico}},\ }\href {\doibase 10.1103/PhysRevLett.110.141601} {\bibfield
  {journal} {\bibinfo  {journal} {Phys.Rev.Lett.}\ }\textbf {\bibinfo {volume}
  {110}},\ \bibinfo {pages} {141601} (\bibinfo {year} {2013})},\ \Eprint
  {http://arxiv.org/abs/1210.4037} {arXiv:1210.4037 [hep-th]} \BibitemShut
  {NoStop}%
\bibitem [{\citenamefont {Codello}(2012)}]{Codello:2012sc}%
  \BibitemOpen
  \bibfield  {author} {\bibinfo {author} {\bibfnamefont {A.}~\bibnamefont
  {Codello}},\ }\href {\doibase 10.1088/1751-8113/45/46/465006} {\bibfield
  {journal} {\bibinfo  {journal} {J.Phys.}\ }\textbf {\bibinfo {volume}
  {A45}},\ \bibinfo {pages} {465006} (\bibinfo {year} {2012})},\ \Eprint
  {http://arxiv.org/abs/1204.3877} {arXiv:1204.3877 [hep-th]} \BibitemShut
  {NoStop}%
\bibitem [{\citenamefont {Delamotte}\ \emph {et~al.}(2016)\citenamefont
  {Delamotte}, \citenamefont {Tissier},\ and\ \citenamefont
  {Wschebor}}]{Delamotte:2015aaa}%
  \BibitemOpen
  \bibfield  {author} {\bibinfo {author} {\bibfnamefont {B.}~\bibnamefont
  {Delamotte}}, \bibinfo {author} {\bibfnamefont {M.}~\bibnamefont {Tissier}},
  \ and\ \bibinfo {author} {\bibfnamefont {N.}~\bibnamefont {Wschebor}},\
  }\href {\doibase 10.1103/PhysRevE.93.012144} {\bibfield  {journal} {\bibinfo
  {journal} {Phys. Rev.}\ }\textbf {\bibinfo {volume} {E93}},\ \bibinfo {pages}
  {012144} (\bibinfo {year} {2016})},\ \Eprint
  {http://arxiv.org/abs/1501.01776} {arXiv:1501.01776 [cond-mat.stat-mech]}
  \BibitemShut {NoStop}%
\bibitem [{\citenamefont {Grater}\ and\ \citenamefont
  {Wetterich}(1995)}]{Grater:1994qx}%
  \BibitemOpen
  \bibfield  {author} {\bibinfo {author} {\bibfnamefont {M.}~\bibnamefont
  {Grater}}\ and\ \bibinfo {author} {\bibfnamefont {C.}~\bibnamefont
  {Wetterich}},\ }\href {\doibase 10.1103/PhysRevLett.75.378} {\bibfield
  {journal} {\bibinfo  {journal} {Phys. Rev. Lett.}\ }\textbf {\bibinfo
  {volume} {75}},\ \bibinfo {pages} {378} (\bibinfo {year} {1995})},\ \Eprint
  {http://arxiv.org/abs/hep-ph/9409459} {arXiv:hep-ph/9409459 [hep-ph]}
  \BibitemShut {NoStop}%
\bibitem [{\citenamefont {Von~Gersdorff}\ and\ \citenamefont
  {Wetterich}(2001)}]{VonGersdorff:2000kp}%
  \BibitemOpen
  \bibfield  {author} {\bibinfo {author} {\bibfnamefont {G.}~\bibnamefont
  {Von~Gersdorff}}\ and\ \bibinfo {author} {\bibfnamefont {C.}~\bibnamefont
  {Wetterich}},\ }\href {\doibase 10.1103/PhysRevB.64.054513} {\bibfield
  {journal} {\bibinfo  {journal} {Phys. Rev.}\ }\textbf {\bibinfo {volume}
  {B64}},\ \bibinfo {pages} {054513} (\bibinfo {year} {2001})},\ \Eprint
  {http://arxiv.org/abs/hep-th/0008114} {arXiv:hep-th/0008114 [hep-th]}
  \BibitemShut {NoStop}%
\bibitem [{\citenamefont {Jakubczyk}\ \emph {et~al.}(2014)\citenamefont
  {Jakubczyk}, \citenamefont {Dupuis},\ and\ \citenamefont
  {Delamotte}}]{Jakubczyk:2014isa}%
  \BibitemOpen
  \bibfield  {author} {\bibinfo {author} {\bibfnamefont {P.}~\bibnamefont
  {Jakubczyk}}, \bibinfo {author} {\bibfnamefont {N.}~\bibnamefont {Dupuis}}, \
  and\ \bibinfo {author} {\bibfnamefont {B.}~\bibnamefont {Delamotte}},\ }\href
  {\doibase 10.1103/PhysRevE.90.062105} {\bibfield  {journal} {\bibinfo
  {journal} {Phys. Rev.}\ }\textbf {\bibinfo {volume} {E90}},\ \bibinfo {pages}
  {062105} (\bibinfo {year} {2014})},\ \Eprint {http://arxiv.org/abs/1409.1374}
  {arXiv:1409.1374 [cond-mat.stat-mech]} \BibitemShut {NoStop}%
\bibitem [{\citenamefont {Jakubczyk}\ and\ \citenamefont
  {Eberlein}(2016)}]{Jakubczyk:2016sul}%
  \BibitemOpen
  \bibfield  {author} {\bibinfo {author} {\bibfnamefont {P.}~\bibnamefont
  {Jakubczyk}}\ and\ \bibinfo {author} {\bibfnamefont {A.}~\bibnamefont
  {Eberlein}},\ }\href@noop {} {\  (\bibinfo {year} {2016})},\ \Eprint
  {http://arxiv.org/abs/1604.06470} {arXiv:1604.06470 [cond-mat.stat-mech]}
  \BibitemShut {NoStop}%
\bibitem [{\citenamefont {Borchardt}\ and\ \citenamefont
  {Knorr}(2015)}]{Borchardt:2015rxa}%
  \BibitemOpen
  \bibfield  {author} {\bibinfo {author} {\bibfnamefont {J.}~\bibnamefont
  {Borchardt}}\ and\ \bibinfo {author} {\bibfnamefont {B.}~\bibnamefont
  {Knorr}},\ }\href {\doibase 10.1103/PhysRevD.91.105011} {\bibfield  {journal}
  {\bibinfo  {journal} {Phys. Rev.}\ }\textbf {\bibinfo {volume} {D91}},\
  \bibinfo {pages} {105011} (\bibinfo {year} {2015})},\ \Eprint
  {http://arxiv.org/abs/1502.07511} {arXiv:1502.07511 [hep-th]} \BibitemShut
  {NoStop}%
\bibitem [{\citenamefont {Borchardt}\ and\ \citenamefont
  {Knorr}(2016)}]{Borchardt:2016pif}%
  \BibitemOpen
  \bibfield  {author} {\bibinfo {author} {\bibfnamefont {J.}~\bibnamefont
  {Borchardt}}\ and\ \bibinfo {author} {\bibfnamefont {B.}~\bibnamefont
  {Knorr}},\ }\href@noop {} {\  (\bibinfo {year} {2016})},\ \Eprint
  {http://arxiv.org/abs/1603.06726} {arXiv:1603.06726 [hep-th]} \BibitemShut
  {NoStop}%
\bibitem [{\citenamefont {Gies}\ \emph {et~al.}(2014)\citenamefont {Gies},
  \citenamefont {Gneiting},\ and\ \citenamefont {Sondenheimer}}]{Gies:2013fua}%
  \BibitemOpen
  \bibfield  {author} {\bibinfo {author} {\bibfnamefont {H.}~\bibnamefont
  {Gies}}, \bibinfo {author} {\bibfnamefont {C.}~\bibnamefont {Gneiting}}, \
  and\ \bibinfo {author} {\bibfnamefont {R.}~\bibnamefont {Sondenheimer}},\
  }\href {\doibase 10.1103/PhysRevD.89.045012} {\bibfield  {journal} {\bibinfo
  {journal} {Phys. Rev.}\ }\textbf {\bibinfo {volume} {D89}},\ \bibinfo {pages}
  {045012} (\bibinfo {year} {2014})},\ \Eprint {http://arxiv.org/abs/1308.5075}
  {arXiv:1308.5075 [hep-ph]} \BibitemShut {NoStop}%
\bibitem [{\citenamefont {Gies}\ and\ \citenamefont
  {Sondenheimer}(2015)}]{Gies:2014xha}%
  \BibitemOpen
  \bibfield  {author} {\bibinfo {author} {\bibfnamefont {H.}~\bibnamefont
  {Gies}}\ and\ \bibinfo {author} {\bibfnamefont {R.}~\bibnamefont
  {Sondenheimer}},\ }\href {\doibase 10.1140/epjc/s10052-015-3284-1} {\bibfield
   {journal} {\bibinfo  {journal} {Eur. Phys. J.}\ }\textbf {\bibinfo {volume}
  {C75}},\ \bibinfo {pages} {68} (\bibinfo {year} {2015})},\ \Eprint
  {http://arxiv.org/abs/1407.8124} {arXiv:1407.8124 [hep-ph]} \BibitemShut
  {NoStop}%
\bibitem [{\citenamefont {Eichhorn}\ \emph {et~al.}(2015)\citenamefont
  {Eichhorn}, \citenamefont {Gies}, \citenamefont {Jaeckel}, \citenamefont
  {Plehn}, \citenamefont {Scherer},\ and\ \citenamefont
  {Sondenheimer}}]{Eichhorn:2015kea}%
  \BibitemOpen
  \bibfield  {author} {\bibinfo {author} {\bibfnamefont {A.}~\bibnamefont
  {Eichhorn}}, \bibinfo {author} {\bibfnamefont {H.}~\bibnamefont {Gies}},
  \bibinfo {author} {\bibfnamefont {J.}~\bibnamefont {Jaeckel}}, \bibinfo
  {author} {\bibfnamefont {T.}~\bibnamefont {Plehn}}, \bibinfo {author}
  {\bibfnamefont {M.~M.}\ \bibnamefont {Scherer}}, \ and\ \bibinfo {author}
  {\bibfnamefont {R.}~\bibnamefont {Sondenheimer}},\ }\href {\doibase
  10.1007/JHEP04(2015)022} {\bibfield  {journal} {\bibinfo  {journal} {JHEP}\
  }\textbf {\bibinfo {volume} {04}},\ \bibinfo {pages} {022} (\bibinfo {year}
  {2015})},\ \Eprint {http://arxiv.org/abs/1501.02812} {arXiv:1501.02812
  [hep-ph]} \BibitemShut {NoStop}%
\bibitem [{\citenamefont {Borchardt}\ \emph {et~al.}(2016)\citenamefont
  {Borchardt}, \citenamefont {Gies},\ and\ \citenamefont
  {Sondenheimer}}]{Borchardt:2016xju}%
  \BibitemOpen
  \bibfield  {author} {\bibinfo {author} {\bibfnamefont {J.}~\bibnamefont
  {Borchardt}}, \bibinfo {author} {\bibfnamefont {H.}~\bibnamefont {Gies}}, \
  and\ \bibinfo {author} {\bibfnamefont {R.}~\bibnamefont {Sondenheimer}},\
  }\href@noop {} {\  (\bibinfo {year} {2016})},\ \Eprint
  {http://arxiv.org/abs/1603.05861} {arXiv:1603.05861 [hep-ph]} \BibitemShut
  {NoStop}%
\bibitem [{\citenamefont {Eichhorn}\ and\ \citenamefont
  {Scherer}(2014)}]{Eichhorn:2014qka}%
  \BibitemOpen
  \bibfield  {author} {\bibinfo {author} {\bibfnamefont {A.}~\bibnamefont
  {Eichhorn}}\ and\ \bibinfo {author} {\bibfnamefont {M.~M.}\ \bibnamefont
  {Scherer}},\ }\href {\doibase 10.1103/PhysRevD.90.025023} {\bibfield
  {journal} {\bibinfo  {journal} {Phys. Rev.}\ }\textbf {\bibinfo {volume}
  {D90}},\ \bibinfo {pages} {025023} (\bibinfo {year} {2014})},\ \Eprint
  {http://arxiv.org/abs/1404.5962} {arXiv:1404.5962 [hep-ph]} \BibitemShut
  {NoStop}%
\bibitem [{\citenamefont {Orlov}\ and\ \citenamefont
  {Sokolov}(2000)}]{Orlov:2000wn}%
  \BibitemOpen
  \bibfield  {author} {\bibinfo {author} {\bibfnamefont {E.~V.}\ \bibnamefont
  {Orlov}}\ and\ \bibinfo {author} {\bibfnamefont {A.~I.}\ \bibnamefont
  {Sokolov}},\ }\href@noop {} {\bibfield  {journal} {\bibinfo  {journal}
  {Submitted to: Sov. Phys. Solid State}\ } (\bibinfo {year} {2000})},\ \Eprint
  {http://arxiv.org/abs/hep-th/0003140} {arXiv:hep-th/0003140 [hep-th]}
  \BibitemShut {NoStop}%
\bibitem [{\citenamefont {{Pelissetto}}\ and\ \citenamefont
  {{Vicari}}(2007)}]{2007PhRvB..76b4436P}%
  \BibitemOpen
  \bibfield  {author} {\bibinfo {author} {\bibfnamefont {A.}~\bibnamefont
  {{Pelissetto}}}\ and\ \bibinfo {author} {\bibfnamefont {E.}~\bibnamefont
  {{Vicari}}},\ }\href {\doibase 10.1103/PhysRevB.76.024436} {\bibfield
  {journal} {\bibinfo  {journal} {\prb}\ }\textbf {\bibinfo {volume} {76}},\
  \bibinfo {eid} {024436} (\bibinfo {year} {2007})},\ \Eprint
  {http://arxiv.org/abs/cond-mat/0702273} {cond-mat/0702273} \BibitemShut
  {NoStop}%
\bibitem [{\citenamefont {Esbensen}\ \emph {et~al.}(2016)\citenamefont
  {Esbensen}, \citenamefont {Ryttov},\ and\ \citenamefont
  {Sannino}}]{Esbensen:2015cjw}%
  \BibitemOpen
  \bibfield  {author} {\bibinfo {author} {\bibfnamefont {J.~K.}\ \bibnamefont
  {Esbensen}}, \bibinfo {author} {\bibfnamefont {T.~A.}\ \bibnamefont
  {Ryttov}}, \ and\ \bibinfo {author} {\bibfnamefont {F.}~\bibnamefont
  {Sannino}},\ }\href {\doibase 10.1103/PhysRevD.93.045009} {\bibfield
  {journal} {\bibinfo  {journal} {Phys. Rev.}\ }\textbf {\bibinfo {volume}
  {D93}},\ \bibinfo {pages} {045009} (\bibinfo {year} {2016})},\ \Eprint
  {http://arxiv.org/abs/1512.04402} {arXiv:1512.04402 [hep-th]} \BibitemShut
  {NoStop}%
\bibitem [{\citenamefont {Eichhorn}(2012)}]{Eichhorn:2012va}%
  \BibitemOpen
  \bibfield  {author} {\bibinfo {author} {\bibfnamefont {A.}~\bibnamefont
  {Eichhorn}},\ }\href {\doibase 10.1103/PhysRevD.86.105021} {\bibfield
  {journal} {\bibinfo  {journal} {Phys. Rev.}\ }\textbf {\bibinfo {volume}
  {D86}},\ \bibinfo {pages} {105021} (\bibinfo {year} {2012})},\ \Eprint
  {http://arxiv.org/abs/1204.0965} {arXiv:1204.0965 [gr-qc]} \BibitemShut
  {NoStop}%
\bibitem [{\citenamefont {Eichhorn}\ \emph
  {et~al.}(2016{\natexlab{b}})\citenamefont {Eichhorn}, \citenamefont {Held},\
  and\ \citenamefont {Pawlowski}}]{Eichhorn:2016esv}%
  \BibitemOpen
  \bibfield  {author} {\bibinfo {author} {\bibfnamefont {A.}~\bibnamefont
  {Eichhorn}}, \bibinfo {author} {\bibfnamefont {A.}~\bibnamefont {Held}}, \
  and\ \bibinfo {author} {\bibfnamefont {J.~M.}\ \bibnamefont {Pawlowski}},\
  }\href@noop {} {\  (\bibinfo {year} {2016}{\natexlab{b}})},\ \Eprint
  {http://arxiv.org/abs/1604.02041} {arXiv:1604.02041 [hep-th]} \BibitemShut
  {NoStop}%
\bibitem [{\citenamefont {Bornholdt}\ \emph {et~al.}(1995)\citenamefont
  {Bornholdt}, \citenamefont {Tetradis},\ and\ \citenamefont
  {Wetterich}}]{Bornholdt:1994rf}%
  \BibitemOpen
  \bibfield  {author} {\bibinfo {author} {\bibfnamefont {S.}~\bibnamefont
  {Bornholdt}}, \bibinfo {author} {\bibfnamefont {N.}~\bibnamefont {Tetradis}},
  \ and\ \bibinfo {author} {\bibfnamefont {C.}~\bibnamefont {Wetterich}},\
  }\href {\doibase 10.1016/0370-2693(95)00045-M} {\bibfield  {journal}
  {\bibinfo  {journal} {Phys. Lett.}\ }\textbf {\bibinfo {volume} {B348}},\
  \bibinfo {pages} {89} (\bibinfo {year} {1995})},\ \Eprint
  {http://arxiv.org/abs/hep-th/9408132} {arXiv:hep-th/9408132 [hep-th]}
  \BibitemShut {NoStop}%
\bibitem [{\citenamefont {Bornholdt}\ \emph {et~al.}(1996)\citenamefont
  {Bornholdt}, \citenamefont {Tetradis},\ and\ \citenamefont
  {Wetterich}}]{Bornholdt:1995rn}%
  \BibitemOpen
  \bibfield  {author} {\bibinfo {author} {\bibfnamefont {S.}~\bibnamefont
  {Bornholdt}}, \bibinfo {author} {\bibfnamefont {N.}~\bibnamefont {Tetradis}},
  \ and\ \bibinfo {author} {\bibfnamefont {C.}~\bibnamefont {Wetterich}},\
  }\href {\doibase 10.1103/PhysRevD.53.4552} {\bibfield  {journal} {\bibinfo
  {journal} {Phys. Rev.}\ }\textbf {\bibinfo {volume} {D53}},\ \bibinfo {pages}
  {4552} (\bibinfo {year} {1996})},\ \Eprint
  {http://arxiv.org/abs/hep-ph/9503282} {arXiv:hep-ph/9503282 [hep-ph]}
  \BibitemShut {NoStop}%
\bibitem [{\citenamefont {Tissier}\ \emph {et~al.}(2002)\citenamefont
  {Tissier}, \citenamefont {Mouhanna}, \citenamefont {Vidal},\ and\
  \citenamefont {Delamotte}}]{Tissier:2002zz}%
  \BibitemOpen
  \bibfield  {author} {\bibinfo {author} {\bibfnamefont {M.}~\bibnamefont
  {Tissier}}, \bibinfo {author} {\bibfnamefont {D.}~\bibnamefont {Mouhanna}},
  \bibinfo {author} {\bibfnamefont {J.}~\bibnamefont {Vidal}}, \ and\ \bibinfo
  {author} {\bibfnamefont {B.}~\bibnamefont {Delamotte}},\ }\href {\doibase
  10.1103/PhysRevB.65.140402} {\bibfield  {journal} {\bibinfo  {journal} {Phys.
  Rev.}\ }\textbf {\bibinfo {volume} {B65}},\ \bibinfo {pages} {140402}
  (\bibinfo {year} {2002})}\BibitemShut {NoStop}%
\bibitem [{\citenamefont {Berges}\ \emph {et~al.}(2002)\citenamefont {Berges},
  \citenamefont {Tetradis},\ and\ \citenamefont {Wetterich}}]{Berges:2000ew}%
  \BibitemOpen
  \bibfield  {author} {\bibinfo {author} {\bibfnamefont {J.}~\bibnamefont
  {Berges}}, \bibinfo {author} {\bibfnamefont {N.}~\bibnamefont {Tetradis}}, \
  and\ \bibinfo {author} {\bibfnamefont {C.}~\bibnamefont {Wetterich}},\ }\href
  {\doibase 10.1016/S0370-1573(01)00098-9} {\bibfield  {journal} {\bibinfo
  {journal} {Phys.\ Rep.}\ }\textbf {\bibinfo {volume} {363}},\ \bibinfo
  {pages} {223} (\bibinfo {year} {2002})},\ \Eprint
  {http://arxiv.org/abs/hep-ph/0005122} {arXiv:hep-ph/0005122 [hep-ph]}
  \BibitemShut {NoStop}%
\bibitem [{\citenamefont {Pawlowski}(2007)}]{Pawlowski:2005xe}%
  \BibitemOpen
  \bibfield  {author} {\bibinfo {author} {\bibfnamefont {J.~M.}\ \bibnamefont
  {Pawlowski}},\ }\href {\doibase 10.1016/j.aop.2007.01.007} {\bibfield
  {journal} {\bibinfo  {journal} {Annals Phys.}\ }\textbf {\bibinfo {volume}
  {322}},\ \bibinfo {pages} {2831} (\bibinfo {year} {2007})},\ \Eprint
  {http://arxiv.org/abs/hep-th/0512261} {arXiv:hep-th/0512261 [hep-th]}
  \BibitemShut {NoStop}%
\bibitem [{\citenamefont {Gies}(2012)}]{Gies:2006wv}%
  \BibitemOpen
  \bibfield  {author} {\bibinfo {author} {\bibfnamefont {H.}~\bibnamefont
  {Gies}},\ }\href {\doibase 10.1007/978-3-642-27320-9_6} {\bibfield  {journal}
  {\bibinfo  {journal} {Lect.\ Notes Phys.}\ }\textbf {\bibinfo {volume}
  {852}},\ \bibinfo {pages} {287} (\bibinfo {year} {2012})},\ \Eprint
  {http://arxiv.org/abs/hep-ph/0611146} {arXiv:hep-ph/0611146 [hep-ph]}
  \BibitemShut {NoStop}%
\bibitem [{\citenamefont {Delamotte}(2012)}]{Delamotte:2007pf}%
  \BibitemOpen
  \bibfield  {author} {\bibinfo {author} {\bibfnamefont {B.}~\bibnamefont
  {Delamotte}},\ }\href {\doibase 10.1007/978-3-642-27320-9_2} {\bibfield
  {journal} {\bibinfo  {journal} {Lect.\ Notes Phys.}\ }\textbf {\bibinfo
  {volume} {852}},\ \bibinfo {pages} {49} (\bibinfo {year} {2012})},\ \Eprint
  {http://arxiv.org/abs/cond-mat/0702365} {arXiv:cond-mat/0702365 [cond-mat]}
  \BibitemShut {NoStop}%
\bibitem [{\citenamefont {Braun}(2012)}]{Braun:2011pp}%
  \BibitemOpen
  \bibfield  {author} {\bibinfo {author} {\bibfnamefont {J.}~\bibnamefont
  {Braun}},\ }\href {\doibase 10.1088/0954-3899/39/3/033001} {\bibfield
  {journal} {\bibinfo  {journal} {J.Phys.}\ }\textbf {\bibinfo {volume}
  {G39}},\ \bibinfo {pages} {033001} (\bibinfo {year} {2012})},\ \Eprint
  {http://arxiv.org/abs/1108.4449} {arXiv:1108.4449 [hep-ph]} \BibitemShut
  {NoStop}%
\bibitem [{\citenamefont {Wetterich}(1993)}]{Wetterich:1992yh}%
  \BibitemOpen
  \bibfield  {author} {\bibinfo {author} {\bibfnamefont {C.}~\bibnamefont
  {Wetterich}},\ }\href {\doibase 10.1016/0370-2693(93)90726-X} {\bibfield
  {journal} {\bibinfo  {journal} {Phys.\ Lett.}\ }\textbf {\bibinfo {volume}
  {B301}},\ \bibinfo {pages} {90} (\bibinfo {year} {1993})}\BibitemShut
  {NoStop}%
\bibitem [{\citenamefont {Kadanoff}(1966)}]{Kadanoff:1966wm}%
  \BibitemOpen
  \bibfield  {author} {\bibinfo {author} {\bibfnamefont {L.~P.}\ \bibnamefont
  {Kadanoff}},\ }\href@noop {} {\bibfield  {journal} {\bibinfo  {journal}
  {Physics}\ }\textbf {\bibinfo {volume} {2}},\ \bibinfo {pages} {263}
  (\bibinfo {year} {1966})}\BibitemShut {NoStop}%
\bibitem [{\citenamefont {Wilson}(1971)}]{Wilson:1971bg}%
  \BibitemOpen
  \bibfield  {author} {\bibinfo {author} {\bibfnamefont {K.~G.}\ \bibnamefont
  {Wilson}},\ }\href {\doibase 10.1103/PhysRevB.4.3174} {\bibfield  {journal}
  {\bibinfo  {journal} {Phys. Rev.}\ }\textbf {\bibinfo {volume} {B4}},\
  \bibinfo {pages} {3174} (\bibinfo {year} {1971})}\BibitemShut {NoStop}%
\bibitem [{\citenamefont {Wegner}\ and\ \citenamefont
  {Houghton}(1973)}]{Wegner:1972ih}%
  \BibitemOpen
  \bibfield  {author} {\bibinfo {author} {\bibfnamefont {F.~J.}\ \bibnamefont
  {Wegner}}\ and\ \bibinfo {author} {\bibfnamefont {A.}~\bibnamefont
  {Houghton}},\ }\href {\doibase 10.1103/PhysRevA.8.401} {\bibfield  {journal}
  {\bibinfo  {journal} {Phys. Rev.}\ }\textbf {\bibinfo {volume} {A8}},\
  \bibinfo {pages} {401} (\bibinfo {year} {1973})}\BibitemShut {NoStop}%
\bibitem [{\citenamefont {Wilson}\ and\ \citenamefont
  {Kogut}(1974)}]{Wilson:1973jj}%
  \BibitemOpen
  \bibfield  {author} {\bibinfo {author} {\bibfnamefont {K.~G.}\ \bibnamefont
  {Wilson}}\ and\ \bibinfo {author} {\bibfnamefont {J.~B.}\ \bibnamefont
  {Kogut}},\ }\href {\doibase 10.1016/0370-1573(74)90023-4} {\bibfield
  {journal} {\bibinfo  {journal} {Phys. Rept.}\ }\textbf {\bibinfo {volume}
  {12}},\ \bibinfo {pages} {75} (\bibinfo {year} {1974})}\BibitemShut {NoStop}%
\bibitem [{\citenamefont {Morris}(1995{\natexlab{b}})}]{Morris:1994au}%
  \BibitemOpen
  \bibfield  {author} {\bibinfo {author} {\bibfnamefont {T.~R.}\ \bibnamefont
  {Morris}},\ }\href {\doibase 10.1016/0920-5632(95)00389-Q} {\bibfield
  {journal} {\bibinfo  {journal} {Nucl.\ Phys.\ Proc.\ Suppl.}\ }\textbf
  {\bibinfo {volume} {42}},\ \bibinfo {pages} {811} (\bibinfo {year}
  {1995}{\natexlab{b}})},\ \Eprint {http://arxiv.org/abs/hep-lat/9411053}
  {arXiv:hep-lat/9411053 [hep-lat]} \BibitemShut {NoStop}%
\bibitem [{\citenamefont {Morris}(1994{\natexlab{a}})}]{Morris:1994ie}%
  \BibitemOpen
  \bibfield  {author} {\bibinfo {author} {\bibfnamefont {T.~R.}\ \bibnamefont
  {Morris}},\ }\href {\doibase 10.1016/0370-2693(94)90767-6} {\bibfield
  {journal} {\bibinfo  {journal} {Phys.\ Lett.}\ }\textbf {\bibinfo {volume}
  {B329}},\ \bibinfo {pages} {241} (\bibinfo {year} {1994}{\natexlab{a}})},\
  \Eprint {http://arxiv.org/abs/hep-ph/9403340} {arXiv:hep-ph/9403340 [hep-ph]}
  \BibitemShut {NoStop}%
\bibitem [{\citenamefont {Morris}(1994{\natexlab{b}})}]{Morris:1994ki}%
  \BibitemOpen
  \bibfield  {author} {\bibinfo {author} {\bibfnamefont {T.~R.}\ \bibnamefont
  {Morris}},\ }\href {\doibase 10.1016/0370-2693(94)90700-5} {\bibfield
  {journal} {\bibinfo  {journal} {Phys.\ Lett.}\ }\textbf {\bibinfo {volume}
  {B334}},\ \bibinfo {pages} {355} (\bibinfo {year} {1994}{\natexlab{b}})},\
  \Eprint {http://arxiv.org/abs/hep-th/9405190} {arXiv:hep-th/9405190 [hep-th]}
  \BibitemShut {NoStop}%
\bibitem [{\citenamefont {Morris}(1998)}]{Morris:1996kn}%
  \BibitemOpen
  \bibfield  {author} {\bibinfo {author} {\bibfnamefont {T.}~\bibnamefont
  {Morris}},\ }\href {\doibase 10.1142/S0217979298000752} {\bibfield  {journal}
  {\bibinfo  {journal} {Int.\ J.\ Mod.\ Phys.}\ }\textbf {\bibinfo {volume}
  {B12}},\ \bibinfo {pages} {1343} (\bibinfo {year} {1998})},\ \Eprint
  {http://arxiv.org/abs/hep-th/9610012} {arXiv:hep-th/9610012 [hep-th]}
  \BibitemShut {NoStop}%
\bibitem [{\citenamefont {Morris}(1994{\natexlab{c}})}]{Morris:1993qb}%
  \BibitemOpen
  \bibfield  {author} {\bibinfo {author} {\bibfnamefont {T.~R.}\ \bibnamefont
  {Morris}},\ }\href {\doibase 10.1142/S0217751X94000972} {\bibfield  {journal}
  {\bibinfo  {journal} {Int. J. Mod. Phys.}\ }\textbf {\bibinfo {volume}
  {A9}},\ \bibinfo {pages} {2411} (\bibinfo {year} {1994}{\natexlab{c}})},\
  \Eprint {http://arxiv.org/abs/hep-ph/9308265} {arXiv:hep-ph/9308265}
  \BibitemShut {NoStop}%
\bibitem [{\citenamefont {Ellwanger}(1994)}]{Ellwanger:1993mw}%
  \BibitemOpen
  \bibfield  {author} {\bibinfo {author} {\bibfnamefont {U.}~\bibnamefont
  {Ellwanger}},\ }\href {\doibase 10.1007/BF01555911} {\bibfield  {journal}
  {\bibinfo  {journal} {Z. Phys.}\ }\textbf {\bibinfo {volume} {C62}},\
  \bibinfo {pages} {503} (\bibinfo {year} {1994})},\ \bibinfo {note}
  {[,206(1993)]},\ \Eprint {http://arxiv.org/abs/hep-ph/9308260}
  {arXiv:hep-ph/9308260 [hep-ph]} \BibitemShut {NoStop}%
\bibitem [{\citenamefont {Litim}(2000)}]{Litim:2000ci}%
  \BibitemOpen
  \bibfield  {author} {\bibinfo {author} {\bibfnamefont {D.~F.}\ \bibnamefont
  {Litim}},\ }\href {\doibase 10.1016/S0370-2693(00)00748-6} {\bibfield
  {journal} {\bibinfo  {journal} {Phys.\ Lett.}\ }\textbf {\bibinfo {volume}
  {B486}},\ \bibinfo {pages} {92} (\bibinfo {year} {2000})},\ \Eprint
  {http://arxiv.org/abs/hep-th/0005245} {arXiv:hep-th/0005245 [hep-th]}
  \BibitemShut {NoStop}%
\bibitem [{\citenamefont {Litim}(2001)}]{Litim:2001up}%
  \BibitemOpen
  \bibfield  {author} {\bibinfo {author} {\bibfnamefont {D.~F.}\ \bibnamefont
  {Litim}},\ }\href {\doibase 10.1103/PhysRevD.64.105007} {\bibfield  {journal}
  {\bibinfo  {journal} {Phys.\ Rev.}\ }\textbf {\bibinfo {volume} {D64}},\
  \bibinfo {pages} {105007} (\bibinfo {year} {2001})},\ \Eprint
  {http://arxiv.org/abs/hep-th/0103195} {arXiv:hep-th/0103195 [hep-th]}
  \BibitemShut {NoStop}%
\bibitem [{\citenamefont {Boyd}(2000)}]{Boyd:ChebyFourier}%
  \BibitemOpen
  \bibfield  {author} {\bibinfo {author} {\bibfnamefont {J.~P.}\ \bibnamefont
  {Boyd}},\ }\href@noop {} {\emph {\bibinfo {title} {{Chebyshev and Fourier
  Spectral Methods}}}},\ \bibinfo {edition} {2nd}\ ed.\ (\bibinfo  {publisher}
  {Dover Publications},\ \bibinfo {year} {2000})\BibitemShut {NoStop}%
\bibitem [{\citenamefont {Litim}\ and\ \citenamefont
  {Vergara}(2004)}]{Litim:2003kf}%
  \BibitemOpen
  \bibfield  {author} {\bibinfo {author} {\bibfnamefont {D.~F.}\ \bibnamefont
  {Litim}}\ and\ \bibinfo {author} {\bibfnamefont {L.}~\bibnamefont
  {Vergara}},\ }\href {\doibase 10.1016/j.physletb.2003.11.047} {\bibfield
  {journal} {\bibinfo  {journal} {Phys.Lett.}\ }\textbf {\bibinfo {volume}
  {B581}},\ \bibinfo {pages} {263} (\bibinfo {year} {2004})},\ \Eprint
  {http://arxiv.org/abs/hep-th/0310101} {arXiv:hep-th/0310101 [hep-th]}
  \BibitemShut {NoStop}%
\bibitem [{\citenamefont {Fischer}\ and\ \citenamefont
  {Gies}(2004)}]{Fischer:2004uk}%
  \BibitemOpen
  \bibfield  {author} {\bibinfo {author} {\bibfnamefont {C.~S.}\ \bibnamefont
  {Fischer}}\ and\ \bibinfo {author} {\bibfnamefont {H.}~\bibnamefont {Gies}},\
  }\href {\doibase 10.1088/1126-6708/2004/10/048} {\bibfield  {journal}
  {\bibinfo  {journal} {JHEP}\ }\textbf {\bibinfo {volume} {0410}},\ \bibinfo
  {pages} {048} (\bibinfo {year} {2004})},\ \Eprint
  {http://arxiv.org/abs/hep-ph/0408089} {arXiv:hep-ph/0408089 [hep-ph]}
  \BibitemShut {NoStop}%
\bibitem [{\citenamefont {Gneiting}(2005)}]{Gneiting:2005}%
  \BibitemOpen
  \bibfield  {author} {\bibinfo {author} {\bibfnamefont {C.}~\bibnamefont
  {Gneiting}},\ }\emph {\bibinfo {title} {{}}},\ \href@noop {} {\bibinfo {type}
  {diploma thesis}},\ \bibinfo  {school} {Heidelberg} (\bibinfo {year}
  {2005})\BibitemShut {NoStop}%
\bibitem [{\citenamefont {Zappala}(2012)}]{Zappala:2012wh}%
  \BibitemOpen
  \bibfield  {author} {\bibinfo {author} {\bibfnamefont {D.}~\bibnamefont
  {Zappala}},\ }\href {\doibase 10.1103/PhysRevD.86.125003} {\bibfield
  {journal} {\bibinfo  {journal} {Phys. Rev.}\ }\textbf {\bibinfo {volume}
  {D86}},\ \bibinfo {pages} {125003} (\bibinfo {year} {2012})},\ \Eprint
  {http://arxiv.org/abs/1206.2480} {arXiv:1206.2480 [hep-th]} \BibitemShut
  {NoStop}%
\bibitem [{\citenamefont {Heilmann}\ \emph {et~al.}(2015)\citenamefont
  {Heilmann}, \citenamefont {Hellwig}, \citenamefont {Knorr}, \citenamefont
  {Ansorg},\ and\ \citenamefont {Wipf}}]{Heilmann:2014iga}%
  \BibitemOpen
  \bibfield  {author} {\bibinfo {author} {\bibfnamefont {M.}~\bibnamefont
  {Heilmann}}, \bibinfo {author} {\bibfnamefont {T.}~\bibnamefont {Hellwig}},
  \bibinfo {author} {\bibfnamefont {B.}~\bibnamefont {Knorr}}, \bibinfo
  {author} {\bibfnamefont {M.}~\bibnamefont {Ansorg}}, \ and\ \bibinfo {author}
  {\bibfnamefont {A.}~\bibnamefont {Wipf}},\ }\href {\doibase
  10.1007/JHEP02(2015)109} {\bibfield  {journal} {\bibinfo  {journal} {JHEP}\
  }\textbf {\bibinfo {volume} {02}},\ \bibinfo {pages} {109} (\bibinfo {year}
  {2015})},\ \Eprint {http://arxiv.org/abs/1409.5650} {arXiv:1409.5650
  [hep-th]} \BibitemShut {NoStop}%
\bibitem [{\citenamefont {Calabrese}\ \emph
  {et~al.}(2003{\natexlab{b}})\citenamefont {Calabrese}, \citenamefont
  {Pelissetto}, \citenamefont {Rossi},\ and\ \citenamefont
  {Vicari}}]{Calabrese:2002gv}%
  \BibitemOpen
  \bibfield  {author} {\bibinfo {author} {\bibfnamefont {P.}~\bibnamefont
  {Calabrese}}, \bibinfo {author} {\bibfnamefont {A.}~\bibnamefont
  {Pelissetto}}, \bibinfo {author} {\bibfnamefont {P.}~\bibnamefont {Rossi}}, \
  and\ \bibinfo {author} {\bibfnamefont {E.}~\bibnamefont {Vicari}},\ }\href
  {\doibase 10.1142/S0217979203023355} {\bibfield  {journal} {\bibinfo
  {journal} {Int. J. Mod. Phys.}\ }\textbf {\bibinfo {volume} {B17}},\ \bibinfo
  {pages} {5829} (\bibinfo {year} {2003}{\natexlab{b}})},\ \Eprint
  {http://arxiv.org/abs/hep-th/0212161} {arXiv:hep-th/0212161 [hep-th]}
  \BibitemShut {NoStop}%
\bibitem [{\citenamefont {Pelissetto}\ and\ \citenamefont
  {Vicari}(2002)}]{Pelissetto:2000ek}%
  \BibitemOpen
  \bibfield  {author} {\bibinfo {author} {\bibfnamefont {A.}~\bibnamefont
  {Pelissetto}}\ and\ \bibinfo {author} {\bibfnamefont {E.}~\bibnamefont
  {Vicari}},\ }\href {\doibase 10.1016/S0370-1573(02)00219-3} {\bibfield
  {journal} {\bibinfo  {journal} {Phys. Rept.}\ }\textbf {\bibinfo {volume}
  {368}},\ \bibinfo {pages} {549} (\bibinfo {year} {2002})},\ \Eprint
  {http://arxiv.org/abs/cond-mat/0012164} {arXiv:cond-mat/0012164 [cond-mat]}
  \BibitemShut {NoStop}%
\bibitem [{\citenamefont {Pawlowski}\ \emph {et~al.}(2015)\citenamefont
  {Pawlowski}, \citenamefont {Scherer}, \citenamefont {Schmidt},\ and\
  \citenamefont {Wetzel}}]{Pawlowski:2015mlf}%
  \BibitemOpen
  \bibfield  {author} {\bibinfo {author} {\bibfnamefont {J.~M.}\ \bibnamefont
  {Pawlowski}}, \bibinfo {author} {\bibfnamefont {M.~M.}\ \bibnamefont
  {Scherer}}, \bibinfo {author} {\bibfnamefont {R.}~\bibnamefont {Schmidt}}, \
  and\ \bibinfo {author} {\bibfnamefont {S.~J.}\ \bibnamefont {Wetzel}},\
  }\href@noop {} {\  (\bibinfo {year} {2015})},\ \Eprint
  {http://arxiv.org/abs/1512.03598} {arXiv:1512.03598 [hep-th]} \BibitemShut
  {NoStop}%
\bibitem [{\citenamefont {Gehring}\ \emph {et~al.}(2015)\citenamefont
  {Gehring}, \citenamefont {Gies},\ and\ \citenamefont
  {Janssen}}]{Gehring:2015vja}%
  \BibitemOpen
  \bibfield  {author} {\bibinfo {author} {\bibfnamefont {F.}~\bibnamefont
  {Gehring}}, \bibinfo {author} {\bibfnamefont {H.}~\bibnamefont {Gies}}, \
  and\ \bibinfo {author} {\bibfnamefont {L.}~\bibnamefont {Janssen}},\ }\href
  {\doibase 10.1103/PhysRevD.92.085046} {\bibfield  {journal} {\bibinfo
  {journal} {Phys. Rev.}\ }\textbf {\bibinfo {volume} {D92}},\ \bibinfo {pages}
  {085046} (\bibinfo {year} {2015})},\ \Eprint
  {http://arxiv.org/abs/1506.07570} {arXiv:1506.07570 [hep-th]} \BibitemShut
  {NoStop}%
\bibitem [{\citenamefont {Gies}\ \emph {et~al.}(2004)\citenamefont {Gies},
  \citenamefont {Jaeckel},\ and\ \citenamefont {Wetterich}}]{Gies:2003dp}%
  \BibitemOpen
  \bibfield  {author} {\bibinfo {author} {\bibfnamefont {H.}~\bibnamefont
  {Gies}}, \bibinfo {author} {\bibfnamefont {J.}~\bibnamefont {Jaeckel}}, \
  and\ \bibinfo {author} {\bibfnamefont {C.}~\bibnamefont {Wetterich}},\ }\href
  {\doibase 10.1103/PhysRevD.69.105008} {\bibfield  {journal} {\bibinfo
  {journal} {Phys. Rev.}\ }\textbf {\bibinfo {volume} {D69}},\ \bibinfo {pages}
  {105008} (\bibinfo {year} {2004})},\ \Eprint
  {http://arxiv.org/abs/hep-ph/0312034} {arXiv:hep-ph/0312034 [hep-ph]}
  \BibitemShut {NoStop}%
\bibitem [{\citenamefont {Gies}\ and\ \citenamefont
  {Janssen}(2010)}]{Gies:2010st}%
  \BibitemOpen
  \bibfield  {author} {\bibinfo {author} {\bibfnamefont {H.}~\bibnamefont
  {Gies}}\ and\ \bibinfo {author} {\bibfnamefont {L.}~\bibnamefont {Janssen}},\
  }\href {\doibase 10.1103/PhysRevD.82.085018} {\bibfield  {journal} {\bibinfo
  {journal} {Phys. Rev.}\ }\textbf {\bibinfo {volume} {D82}},\ \bibinfo {pages}
  {085018} (\bibinfo {year} {2010})},\ \Eprint {http://arxiv.org/abs/1006.3747}
  {arXiv:1006.3747 [hep-th]} \BibitemShut {NoStop}%
\bibitem [{\citenamefont {Braun}\ \emph {et~al.}(2014)\citenamefont {Braun},
  \citenamefont {Gies}, \citenamefont {Janssen},\ and\ \citenamefont
  {Roscher}}]{Braun:2014wja}%
  \BibitemOpen
  \bibfield  {author} {\bibinfo {author} {\bibfnamefont {J.}~\bibnamefont
  {Braun}}, \bibinfo {author} {\bibfnamefont {H.}~\bibnamefont {Gies}},
  \bibinfo {author} {\bibfnamefont {L.}~\bibnamefont {Janssen}}, \ and\
  \bibinfo {author} {\bibfnamefont {D.}~\bibnamefont {Roscher}},\ }\href
  {\doibase 10.1103/PhysRevD.90.036002} {\bibfield  {journal} {\bibinfo
  {journal} {Phys. Rev.}\ }\textbf {\bibinfo {volume} {D90}},\ \bibinfo {pages}
  {036002} (\bibinfo {year} {2014})},\ \Eprint {http://arxiv.org/abs/1404.1362}
  {arXiv:1404.1362 [hep-ph]} \BibitemShut {NoStop}%
\bibitem [{\citenamefont {Onsager}(1944)}]{PhysRev.65.117}%
  \BibitemOpen
  \bibfield  {author} {\bibinfo {author} {\bibfnamefont {L.}~\bibnamefont
  {Onsager}},\ }\href {\doibase 10.1103/PhysRev.65.117} {\bibfield  {journal}
  {\bibinfo  {journal} {Phys. Rev.}\ }\textbf {\bibinfo {volume} {65}},\
  \bibinfo {pages} {117} (\bibinfo {year} {1944})}\BibitemShut {NoStop}%
\bibitem [{\citenamefont {Vicari}\ and\ \citenamefont
  {Zinn-Justin}(2006)}]{Vicari:2006xr}%
  \BibitemOpen
  \bibfield  {author} {\bibinfo {author} {\bibfnamefont {E.}~\bibnamefont
  {Vicari}}\ and\ \bibinfo {author} {\bibfnamefont {J.}~\bibnamefont
  {Zinn-Justin}},\ }\href {\doibase 10.1088/1367-2630/8/12/321} {\bibfield
  {journal} {\bibinfo  {journal} {New J. Phys.}\ }\textbf {\bibinfo {volume}
  {8}},\ \bibinfo {pages} {321} (\bibinfo {year} {2006})},\ \Eprint
  {http://arxiv.org/abs/cond-mat/0611353} {arXiv:cond-mat/0611353 [cond-mat]}
  \BibitemShut {NoStop}%
\bibitem [{\citenamefont {{Holtschneider}}\ and\ \citenamefont
  {{Selke}}(2007)}]{2007PhRvB..76v0405H}%
  \BibitemOpen
  \bibfield  {author} {\bibinfo {author} {\bibfnamefont {M.}~\bibnamefont
  {{Holtschneider}}}\ and\ \bibinfo {author} {\bibfnamefont {W.}~\bibnamefont
  {{Selke}}},\ }\href {\doibase 10.1103/PhysRevB.76.220405} {\bibfield
  {journal} {\bibinfo  {journal} {\prb}\ }\textbf {\bibinfo {volume} {76}},\
  \bibinfo {eid} {220405} (\bibinfo {year} {2007})},\ \Eprint
  {http://arxiv.org/abs/0710.1251} {arXiv:0710.1251 [cond-mat.stat-mech]}
  \BibitemShut {NoStop}%
\bibitem [{\citenamefont {{de Jongh}}\ \emph {et~al.}(1982)\citenamefont {{de
  Jongh}}, \citenamefont {{Regnault}}, \citenamefont {{Rossat-Mignod}},\ and\
  \citenamefont {{Henry}}}]{1982JAP....53.7963D}%
  \BibitemOpen
  \bibfield  {author} {\bibinfo {author} {\bibfnamefont {L.~J.}\ \bibnamefont
  {{de Jongh}}}, \bibinfo {author} {\bibfnamefont {L.~P.}\ \bibnamefont
  {{Regnault}}}, \bibinfo {author} {\bibfnamefont {J.}~\bibnamefont
  {{Rossat-Mignod}}}, \ and\ \bibinfo {author} {\bibfnamefont {J.~Y.}\
  \bibnamefont {{Henry}}},\ }\href {\doibase 10.1063/1.330242} {\bibfield
  {journal} {\bibinfo  {journal} {Journal of Applied Physics}\ }\textbf
  {\bibinfo {volume} {53}},\ \bibinfo {pages} {7963} (\bibinfo {year}
  {1982})}\BibitemShut {NoStop}%
\bibitem [{\citenamefont {{de Jongh}}\ and\ \citenamefont {{de
  Groot}}(1985)}]{1985SSCom..53..737D}%
  \BibitemOpen
  \bibfield  {author} {\bibinfo {author} {\bibfnamefont {L.~J.}\ \bibnamefont
  {{de Jongh}}}\ and\ \bibinfo {author} {\bibfnamefont {H.~J.~M.}\ \bibnamefont
  {{de Groot}}},\ }\href {\doibase 10.1016/0038-1098(85)90210-8} {\bibfield
  {journal} {\bibinfo  {journal} {Solid State Communications}\ }\textbf
  {\bibinfo {volume} {53}},\ \bibinfo {pages} {737} (\bibinfo {year}
  {1985})}\BibitemShut {NoStop}%
\bibitem [{\citenamefont {{De Groot}}\ and\ \citenamefont {{De
  Jongh}}(1986)}]{1986PhyBC.141....1D}%
  \BibitemOpen
  \bibfield  {author} {\bibinfo {author} {\bibfnamefont {H.~J.~M.}\
  \bibnamefont {{De Groot}}}\ and\ \bibinfo {author} {\bibfnamefont {L.~J.}\
  \bibnamefont {{De Jongh}}},\ }\href {\doibase 10.1016/0378-4363(86)90346-3}
  {\bibfield  {journal} {\bibinfo  {journal} {Physica B+C}\ }\textbf {\bibinfo
  {volume} {141}},\ \bibinfo {pages} {1} (\bibinfo {year} {1986})}\BibitemShut
  {NoStop}%
\bibitem [{\citenamefont {{Rauh}}\ \emph {et~al.}(1986)\citenamefont {{Rauh}},
  \citenamefont {{Erkelens}}, \citenamefont {{Regnault}}, \citenamefont
  {{Rossat-Mignod}}, \citenamefont {{Kullman}},\ and\ \citenamefont
  {{Geick}}}]{1986JPhC...19.4503R}%
  \BibitemOpen
  \bibfield  {author} {\bibinfo {author} {\bibfnamefont {H.}~\bibnamefont
  {{Rauh}}}, \bibinfo {author} {\bibfnamefont {W.~A.~C.}\ \bibnamefont
  {{Erkelens}}}, \bibinfo {author} {\bibfnamefont {L.~P.}\ \bibnamefont
  {{Regnault}}}, \bibinfo {author} {\bibfnamefont {J.}~\bibnamefont
  {{Rossat-Mignod}}}, \bibinfo {author} {\bibfnamefont {W.}~\bibnamefont
  {{Kullman}}}, \ and\ \bibinfo {author} {\bibfnamefont {R.}~\bibnamefont
  {{Geick}}},\ }\href {\doibase 10.1088/0022-3719/19/23/013} {\bibfield
  {journal} {\bibinfo  {journal} {Journal of Physics C Solid State Physics}\
  }\textbf {\bibinfo {volume} {19}},\ \bibinfo {pages} {4503} (\bibinfo {year}
  {1986})}\BibitemShut {NoStop}%
\bibitem [{\citenamefont {{Cowley}}\ \emph {et~al.}(1993)\citenamefont
  {{Cowley}}, \citenamefont {{Aharony}}, \citenamefont {{Birgeneau}},
  \citenamefont {{Pelcovits}}, \citenamefont {{Shirane}},\ and\ \citenamefont
  {{Thurston}}}]{1993ZPhyB..93....5C}%
  \BibitemOpen
  \bibfield  {author} {\bibinfo {author} {\bibfnamefont {R.~A.}\ \bibnamefont
  {{Cowley}}}, \bibinfo {author} {\bibfnamefont {A.}~\bibnamefont {{Aharony}}},
  \bibinfo {author} {\bibfnamefont {R.~J.}\ \bibnamefont {{Birgeneau}}},
  \bibinfo {author} {\bibfnamefont {R.~A.}\ \bibnamefont {{Pelcovits}}},
  \bibinfo {author} {\bibfnamefont {G.}~\bibnamefont {{Shirane}}}, \ and\
  \bibinfo {author} {\bibfnamefont {T.~R.}\ \bibnamefont {{Thurston}}},\ }\href
  {\doibase 10.1007/BF01308802} {\bibfield  {journal} {\bibinfo  {journal}
  {Zeitschrift fur Physik B Condensed Matter}\ }\textbf {\bibinfo {volume}
  {93}},\ \bibinfo {pages} {5} (\bibinfo {year} {1993})}\BibitemShut {NoStop}%
\bibitem [{\citenamefont {{van de Kamp}}\ \emph {et~al.}(1997)\citenamefont
  {{van de Kamp}}, \citenamefont {{Steiner}},\ and\ \citenamefont
  {{Tietze-Jaensch}}}]{1997PhyB..241..570V}%
  \BibitemOpen
  \bibfield  {author} {\bibinfo {author} {\bibfnamefont {R.}~\bibnamefont {{van
  de Kamp}}}, \bibinfo {author} {\bibfnamefont {M.}~\bibnamefont {{Steiner}}},
  \ and\ \bibinfo {author} {\bibfnamefont {H.}~\bibnamefont
  {{Tietze-Jaensch}}},\ }\href {\doibase 10.1016/S0921-4526(97)00646-7}
  {\bibfield  {journal} {\bibinfo  {journal} {Physica B Condensed Matter}\
  }\textbf {\bibinfo {volume} {241}},\ \bibinfo {pages} {570} (\bibinfo {year}
  {1997})}\BibitemShut {NoStop}%
\bibitem [{\citenamefont {{Christianson}}\ \emph {et~al.}(2001)\citenamefont
  {{Christianson}}, \citenamefont {{Leheny}}, \citenamefont {{Birgeneau}},\
  and\ \citenamefont {{Erwin}}}]{2001PhRvB..63n0401C}%
  \BibitemOpen
  \bibfield  {author} {\bibinfo {author} {\bibfnamefont {R.~J.}\ \bibnamefont
  {{Christianson}}}, \bibinfo {author} {\bibfnamefont {R.~L.}\ \bibnamefont
  {{Leheny}}}, \bibinfo {author} {\bibfnamefont {R.~J.}\ \bibnamefont
  {{Birgeneau}}}, \ and\ \bibinfo {author} {\bibfnamefont {R.~W.}\ \bibnamefont
  {{Erwin}}},\ }\href {\doibase 10.1103/PhysRevB.63.140401} {\bibfield
  {journal} {\bibinfo  {journal} {\prb}\ }\textbf {\bibinfo {volume} {63}},\
  \bibinfo {eid} {140401} (\bibinfo {year} {2001})},\ \Eprint
  {http://arxiv.org/abs/cond-mat/0101097} {cond-mat/0101097} \BibitemShut
  {NoStop}%
\bibitem [{\citenamefont {Codello}\ \emph {et~al.}(2015)\citenamefont
  {Codello}, \citenamefont {Defenu},\ and\ \citenamefont
  {D’Odorico}}]{Codello:2014yfa}%
  \BibitemOpen
  \bibfield  {author} {\bibinfo {author} {\bibfnamefont {A.}~\bibnamefont
  {Codello}}, \bibinfo {author} {\bibfnamefont {N.}~\bibnamefont {Defenu}}, \
  and\ \bibinfo {author} {\bibfnamefont {G.}~\bibnamefont {D’Odorico}},\
  }\href {\doibase 10.1103/PhysRevD.91.105003} {\bibfield  {journal} {\bibinfo
  {journal} {Phys. Rev.}\ }\textbf {\bibinfo {volume} {D91}},\ \bibinfo {pages}
  {105003} (\bibinfo {year} {2015})},\ \Eprint {http://arxiv.org/abs/1410.3308}
  {arXiv:1410.3308 [hep-th]} \BibitemShut {NoStop}%
\end{thebibliography}%

\end{document}